\documentclass[twocolumn]{aastex6}

\begin{document}

\title{{\it HST} Imaging Of The Ionizing Radiation From A Star-forming Galaxy at z = 3.794} 

\author{Zhiyuan Ji\altaffilmark{1},
Mauro Giavalisco\altaffilmark{1}, 
Eros Vanzella\altaffilmark{2},
Brian Siana\altaffilmark{3},
Laura Pentericci\altaffilmark{4},
Anne Jaskot\altaffilmark{1,15},
Teng Liu \altaffilmark{1,5},
Mario Nonino\altaffilmark{6},
Henry C. Ferguson\altaffilmark{7},
Marco Castellano\altaffilmark{4}, 
Filippo Mannucci \altaffilmark{8},
Daniel Schaerer  \altaffilmark{9},
Johan Peter Uldall Fynbo \altaffilmark{10},
Casey Papovich \altaffilmark{11},
Adam C. Carnall \altaffilmark{12},
Ricardo Amorin \altaffilmark{13,14},
Raymond C. Simons \altaffilmark{7},
Nimish Hathi  \altaffilmark{7},
Fergus Cullen \altaffilmark{12},
Derek McLeod \altaffilmark{12}
}

\altaffiltext{1}{Department of Astronomy, University of Massachusetts Amherst, 710 N. Pleasant St., Amherst, MA 01003, USA; zhiyuanji@astro.umass.edu} 
\altaffiltext{2}{INAF -- OAS, Osservatorio di Astrofisica e Scienza dello Spazio di Bologna, via Gobetti 93/3, I-40129 Bologna, Italy} 
\altaffiltext{3}{Department of Physics and Astronomy, University of California, Riverside, CA 92507, USA} 
\altaffiltext{4}{INAF -- Osservatorio Astronomico di Roma, Via Frascati 33, I-00078 Monteporzio (RM), Italy} 
\altaffiltext{5}{University of Science and Technology of China, Hefei, Anhui 230026, China}
\altaffiltext{6}{INAF -- Osservatorio Astronomico di Trieste, Via G.B. Tiepolo 11, I-40131 Trieste, Italy} 
\altaffiltext{7}{Space Telescope Science Institute, 3700 San Martin Boulevard, Baltimore, MD 21218, USA} 
\altaffiltext{8}{INAF -- Osservatorio Astrofisico di Arcetri, Largo E. Fermi 5, I-50125 Firenze, Italy}
\altaffiltext{9}{Observatoire de Gen\`eve, Universit\`e de Gen\`eve, 51 Ch. des Maillettes, CH-1290 Versoix, Switzerland}
\altaffiltext{10}{The Cosmic Dawn Center, Niels Bohr Institute, University of Copenhagen, DK-2100 Copenhagen, Denmark}
\altaffiltext{11}{Department of Physics and Astronomy, Texas A\&M University, College Station, TX 77843-4242, USA}
\altaffiltext{12}{Institute for Astronomy, University of Edinburgh, Royal Observatory, Edinburgh EH9 3HJ, UK}
\altaffiltext{13}{Instituto de Investigaci\'on Multidisciplinar en Ciencia y Tecnolog\'ia, Universidad de La Serena, Ra\'ul Bitr\'an 1305, La Serena, Chile}
\altaffiltext{14}{Departamento de Astronomía, Universidad de La Serena, Av. Juan Cisternas 1200 Norte, La Serena, Chile}
\altaffiltext{15}{Astronomy Department, Williams College, Williamstown, MA 01267, USA}

\begin{abstract}
We report on the {\it HST} detection of the Lyman-continuum (LyC) radiation emitted by a galaxy at redshift $z=3.794$, dubbed Ion1 \citep{Vanzella2012}. The LyC from Ion1 is detected at $\lambda\lambda$ 820$\sim$890 \AA\ with {\it HST} WFC3/UVIS in the F410M band ($m_{410}=27.60\pm0.36$ m$\rm{_{AB}}$, peak SNR $=$ 4.17 in a r $=$ 0.12'' aperture) and at $\lambda\lambda$ 700$\sim$830 \AA\ with the VLT/VIMOS in the U-band ($m_U = 27.84\pm0.19$ m$\rm{_{AB}}$, peak SNR $=$ 6.7 with a r $=$ 0.6'' aperture). A 20-hr VLT/VIMOS spectrum shows low- and high- ionization interstellar metal absorption lines, the P-Cygni profile of \ion{C}{4} and Ly$\alpha$ in absorption. The latter spectral feature differs from what observed in known LyC emitters, which show strong Ly${\alpha}$ emission. An {\it HST} far-UV color map reveals that the LyC emission escapes from a region of the galaxy that is bluer than the rest. The F410M image shows that the centroid of the LyC emission is offset from the centroid of the non-ionizing UV emission by 0.12''$\pm$0.03'', corresponding to 0.85$\pm$0.21 pkpc, and that its morphology is likely moderately resolved. These morphological characteristics favor a scenario where the LyC photons produced by massive stars escape from low \ion{H}{1} column--density ``cavities'' in the ISM. We also collect the VIMOS U-band images of 107 Lyman-break galaxies at $3.40<z_{spec}<3.95$, i.e. sampling the LyC, and stack them with inverse-variance weights. No LyC emission is detected in the stacked image, resulting in a 32.5 m$_{\rm{AB}}$ flux limit (1$\sigma$) and an upper limit of absolute LyC escape fraction $f_{esc}^{abs} \le 0.63\%$.
\end{abstract}


\section{Introduction} \label{sec:intro}

The identification and characterization of the sources responsible for producing the bulk of the ionizing photons (or LyC, $\rm{\lambda < 912 \AA}$) that have reionized the Universe at redshift $\rm{z\approx 7}$ \citep{Planck2016}, and that have kept it ionized ever since, still elude us. AGNs and massive stars are among such sources, but we do not know if the abundance of the former \citep{Giallongo2015,Georgakakis2015} and the escape fraction of the latter and, the faint end luminosity distribution of star-forming galaxies \citep[e.g.][]{Bouwens2015,Atek2015,Castellano2016,Yue2018,Finkelstein2019} can provide enough LyC photons to explain the ionization conditions of the intergalactic medium (IGM) and its evolution. The rapid increase of the IGM neutral hydrogen (\ion{H}{1}) optical depth with redshift makes the direct detection of ionizing LyC at the Epoch of Reionization (EoR) virtually impossible \citep[e.g.][]{Madau1995, Vanzella2010, Vanzella2012, Steidel2018}. The physical properties of the LyC emitters at the EoR, therefore, can hopefully be inferred by studying their analogs at lower redshifts, although one has always to keep in mind the possibility that a rapid evolution in the properties of early galaxies can make such inferences fraught with systematic errors.

In recent years, large efforts have been made to identify sources of LyC radiation in the local universe and at intermediate-redshifts \citep[$\rm{1<z<4}$, e.g. ][]{Malkan2003, Shapley2006,Siana2010, Vanzella2010, Nestor2013, Siana2015, Steidel2018}. In the local Universe, searches for star-forming galaxies with escaping LyC have mostly yielded upper limits to the LyC escape fraction from individual sources \citep[e.g.][]{Grimes2007, Cowie2009}, although some (very rare) LyC emitters have been securely identified \citep{Leitet2011,Leitet2013,Borthakur2014, Izotov2016, Izotov2016b, Leitherer2016, Izotov2018b}. At the Cosmic Noon, \citet{Bian2017} reported on the {\it Hubble Space Telescope} ({\it HST}) WFC/F275W direct detection of the LyC emission from A2218-Flanking, a lensed compact dwarf galaxy at redshift z $\approx$ 2.5. Also, \citet{RiveraThorsen2019} recently reported a {\it HST} detection of the leaking LyC emission of the Sunburst Arc, a bright gravitationally lensed galaxy at z $=$ 2.4, where the LyC emission arises from a compact star-forming region, possibly consistent with a young massive star cluster \citet{Vanzella2019}. 
  
At higher redshifts and in particular at $\rm{3<z<4}$, the rest-frame $\lambda\lambda$ 912\AA\ Lyman Limit is redshifted to the U-band ($\approx$ 4000\AA), where the sensitivity of ground-based imaging and spectrograph is high, the sky background is relatively low, and the IGM is still relatively transparent (e.g. \citealt{Vanzella2010}), which makes this redshift range a ``sweet spot'' for the identification and study of individual LyC emitters \citep[e.g.][]{Steidel2018}. A complication in these searches arises from the contamination by faint interlopers, i.e. lower redshift galaxies located between the observer and the candidate LyC emitters, whose isophots at least partially overlap at the angular resolution of the ground-based instrumentation. These interlopers can very effectively mimic the presence of LyC emission from the real candidate, resulting in false positive detections (e.g. \citealt{Vanzella2010MNRAS}). The secure identifications of LyC emitters at these redshifts, therefore, require both spectroscopy to get accurate redshift measurements and deep high angular resolution images at different wavelengths to constrain the possibility of foreground contamination \citep{Siana2015}. These images need to be taken in the observed UV and optical window and currently only {\it HST} is capable to obtain them. To date, there are only three such robust LyC emitters identified at $\rm{z\approx 3\sim4}$, i.e. Ion2 \citep{Vanzella2016, debarros2016,Vanzella2019}, Ion3 \citep{Vanzella2018,Vanzella2019}, and Q1549-C25 \citep{Shapley2016}. For Q1549-C25, the LyC is detected in a Keck/Low Resolution Imaging Spectrometer (LRIS) spectrum while undetected in the {\it HST} image of LyC, i.e. the image that only samples the rest-frame UV light bluer than the Lyman Limit. For Ion3, the LyC is detected in the Very Large Telescope/FOcal Reducer and low dispersion Spectrograph (VLT/FORS) spectrum but without {\it HST} LyC imaging data available yet. For Ion2, the LyC is detected both in the {\it HST} LyC image and in a deep VLT/VIsible Multi Object Spectrograph (VIMOS) spectrum. Very recently, a sample of narrow-band selected Ly$\alpha$-emitting galaxies at z$\sim$ 3.1 have been imaged by {\it HST} WFCS/UVIS F336W (LyC, $\approx$ 30.2 AB mag flux limit, 3$\sigma$) and WFC/IR F160W (rest-frame optical) during the LymAn Continuum Escape Survey \citep[LACES,][]{Fletcher2019}, resulting in an increase in the number of individually detected LyC emitters at z$\sim$3 and a significant higher (20\%) successful LyC detection rate than seen in previously selected Lyman Break Galaxies (LBGs). It however remains to quantify the probability of being an interloper for individual LACES candidate, as requiring {\it HST} images observed in the rest-frame non-ionizing UV \citep{Siana2015}.
  
In addition to the mere detection of the LyC radiation, further insight is gained from the spectroscopic and morphological properties of LyC emitters, which help in constraining the physical conditions that allow the LyC photons to successfully escape from the host galaxy into IGM \citep[e.g.][]{Heckman2011, Vanzella2016, Jaskot2017,Steidel2018, Jaskot2019,Plat2019,Nakajima2019}.

In this work, we report on a distant galaxy, dubbed Ion1 \citep{Vanzella2012}, whose LyC emission is detected in deep images obtained both with the VLT/VIMOS in the U-band and {\it HST}/WFC3 in the F410M band.  A deep 20-hour spectrum obtained with the VLT/VIMOS is also available which allows us to study the far-UV properties of this galaxy, in addition to providing an accurate measure of its redshift. Throughout this paper, we adopt the AB magnitude system, and a $\Lambda$CDM cosmology with $\Omega_m = 0.3$, $\Omega_\Lambda=0.7$ and $\rm{h=H_0/(100\;kms^{-1}Mpc^{-1})=0.7}$.

\section{Observations and Data Sets} \label{sec:data}

Ion1 (J033216.64-274253.3) is a star-forming galaxy and a source of ionizing radiation with LyC relative escape fraction $f_{esc}^{rel}>35\%$ \citep{Vanzella2012}. It was originally selected as a LBG at redshift $z\approx 4$ (B-band dropout) based on the {\it HST}/ACS B$_{435}$, V$_{606}$, i$_{775}$ and z$_{850}$ photometry obtained as part of the Great Observatories Origins Deep Survey program in the Chandra Deep Field South (GOODS-South, \citealt{Giavalisco2004}). The galaxy was observed spectroscopically with the VLT/VIMOS by \citet{Popesso2009} and \citet{Balestra2010}, but no conclusive redshift could be obtained. The first measure of its redshift was obtained by \citet{Vanzella2010}, who reported the tentative value $z=3.795$ from a relatively low signal-to-noise ratio (SNR) Keck/DEep Imaging Multi-Object Spectrograph (DEIMOS) spectrum. \citet{Vanzella2010,Vanzella2015} also identified Ion1 as a LyC emitter based on an ultra--deep U-band image obtained by \citet{Nonino2009} with VLT/VIMOS (at $z=3.795$, the U-band is entirely to the blue of the Lyman limit), where the galaxy is detected with high SNR (see Section \ref{sec:IMAGE}). Finally, \citet{Vanzella2012} measured the integrated properties of the stellar populations of Ion1, including star-formation rate (SFR $\approx$ 50 $M_\sun/yr$) and stellar mass ($M_*\sim2.3\times10^9M_\sun$), by means of SED fitting. 

Here we present a very deep spectrum of Ion1, with on-target exposure time $T_{exp}=20$ hours obtained with the VLT/VIMOS as part of the VANDELS survey \citep{McLure2018,Pentericci2018}, and confirm the redshift of the galaxy to be $z=3.794$ (from the \ion{C}{3}] emission). While the spectral range of the VIMOS spectrum does not include the rest-frame LyC, the ionizing radiation of Ion1 is well detected (with SNR = 6.62) in the ultra-deep VLT/VIMOS U-band image at rest-frame $\lambda\lambda$ 700-830\AA. We also present a medium-deep image of Ion1 obtained with {\it HST}/WFC3 in the F410M bandpass, obtained as part of the {\it HST} Cycle 23 program of high-resolution imaging of the morphology of LyC emitters at $z\approx 3.2-3.8$ that we have identified in the GOODS-South field (Cycle 23 program GO 14088, PI: Eros Vanzella). Part of these data, namely WFC3 images in the F336W filter of another $z=3.2$ LyC-emitting galaxy (dubbed {\it Ion2}) totaling 47,600 sec of integration, has already been published by \citet{Vanzella2016}, where we have also discussed the data reduction procedures that we have adopted. The Ion1
image, which has total integration time $T_{exp}=$ 35,700 seconds on target and, at the redshift of Ion1, covers the rest-frame far-UV continuum at $\lambda\lambda$ 820-890\AA, was obtained to study the morphology of the ionizing radiation and also to investigate the possibility that a foreground interloper could mimic the LyC emission. To overcome the significant degradation of the charge transfer efficiency (CTE) of the WFC3/UVIS CCDs due to radiation bombardment from the space environment \citep{Biretta2013}, we have adjusted both our observational strategies and data reduction methods. In particular, we have used post-flashing to increase the background of the image by $\sim12e^-\,pix^{-1}$, thus mitigating the impact on photometric accuracy by the CTE losses \citep{Biretta2013}, and we have also placed the target near the read-out edge of the detector, which can greatly reduce the losses due to charge transfer inefficiency. During the data reduction, we paid particular attention to the dark current subtraction as it contributed nearly about half of the total background. The default data reduction pipeline dark frames consisted of a constant value, preventing us to capture dark current gradients and blotchy pattern \citep{Teplitz2013}. In addition, as already pointed out by \citet{Rafelski2015}, more than half of the hot and warm pixels were not properly masked because the pipeline dark current frames were not CTE-corrected. We therefore applied the same data reduction procedure as used in \citet{Vanzella2016}, which adopted the dark processing method suggested by \citet{Rafelski2015}. All raw dark images were first cosmic rays removed and CTE-corrected in the anneal cycle of our visits, after which we found and masked the hot pixels. We made a mean super dark using all of the darks in the anneal cycle and then dark subtracted the science images while kept the hot pixels masked. We processed all of our CTE-corrected raw data using STSDAS task CALWF3, including subtraction of our new super dark. These calibrated images were then background subtracted, cosmic-ray rejected, geometric distortion corrected and combined using Astrodrizzle \citep{Gonzaga2012}. The final combined F410M image has a pixel scale of 0.06 arcsec.

We also use the {\it HST}/ACS images in the $B_{435}$, $V_{606}$, $i_{775}$, $z_{850}$ and {\it HST}/WFC3 images in the F098M, $J_{125}$, $H_{160}$ obtained during the GOODS \citep{Giavalisco2004} and CANDELS \citep{Gorgin2011,Koekemoer2011} programs, to compare the morphology of Ion1's ionizing and non-ionizing emission and to look for low-redshift interlopers in close proximity with Ion1. We also take advantage of the $Y_{105}$ image observed during the large {\it HST}/WFC3 G102 grism survey CANDELS Lyman-alpha Emission At Reionization Experiment (CLEAR, PI: Casey Papovich), and the ultra deep $Ks$-band image taken by the HAWK-I imager on the VLT during the Hawk-I UDS and GOODS Survey (HUGS\footnote{Based on data products from observations made with ESO telescopes at the La Silla Paranal Observatory under programme IDs 181.A-0717(D), 181.A-0717(H), 186.A-0898(B), 186.A-0898(D), 186.A-0898(F), 186.A-0898(H)}, \citealt{Fontana2014}). Finally, we use the 7 Ms Chandra Deep Field-South (CDF-S) image to investigate the X-ray properties of Ion1 \citep{Luo2017}.

\section{Results}

\subsection{Spectral energy distribution}\label{sec:SED}

\begin{figure*}
\gridline{\fig{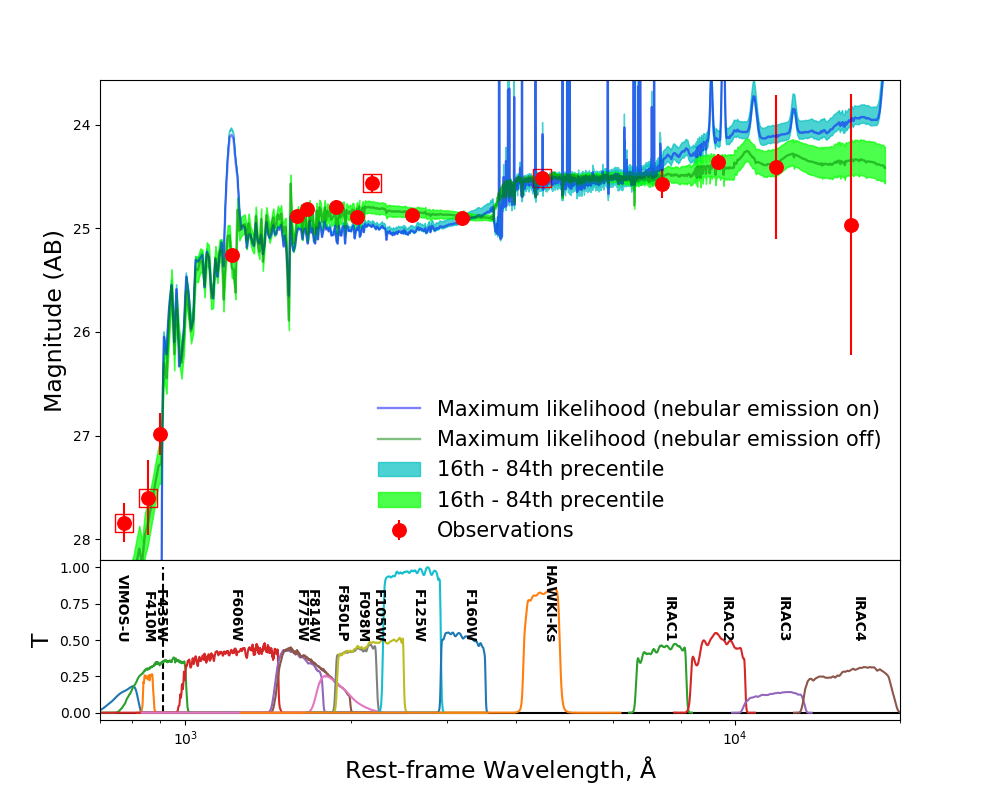}{0.77\textwidth}{}
}
\gridline{
	\fig{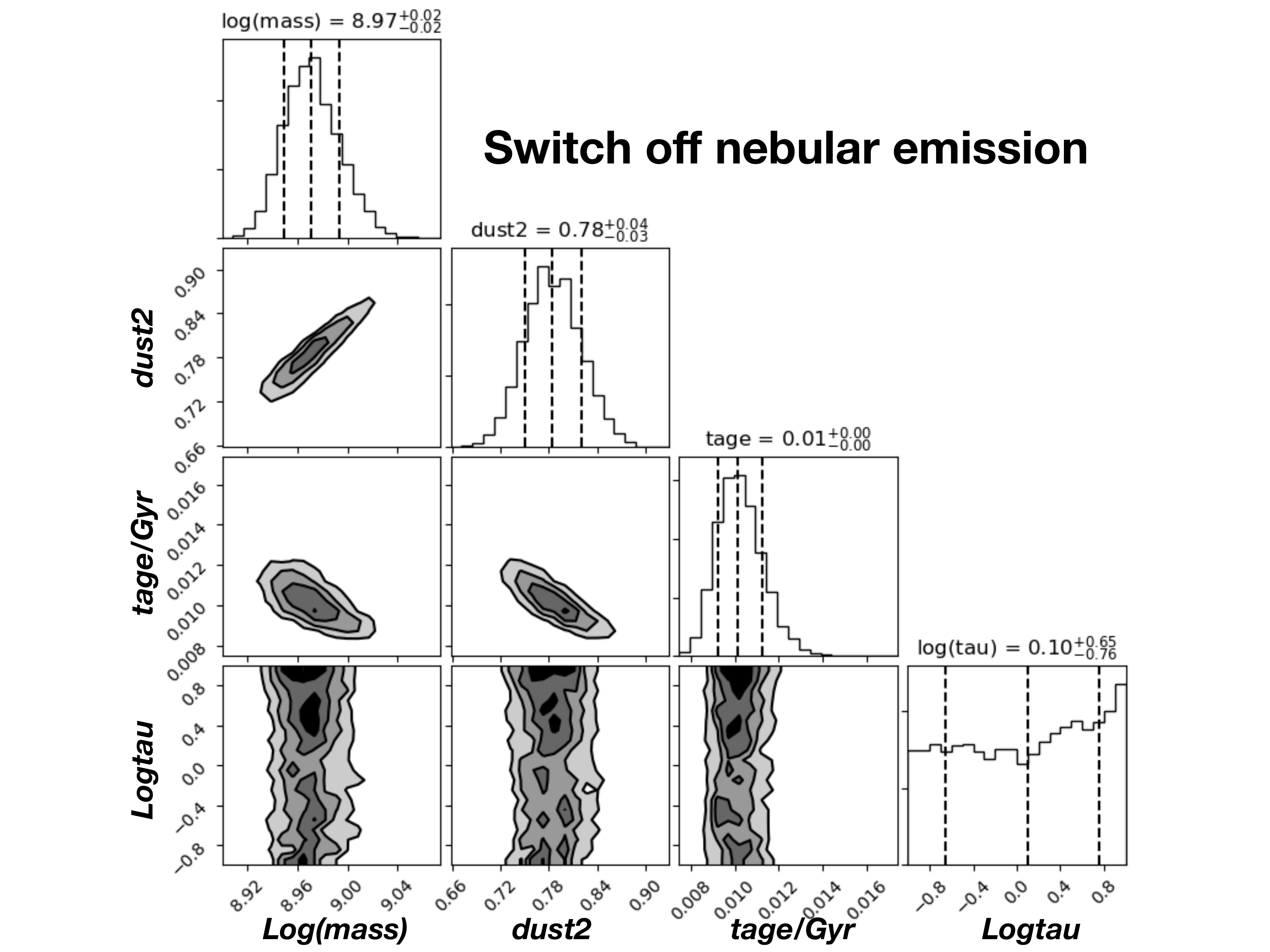}{0.5\textwidth}{}
	\fig{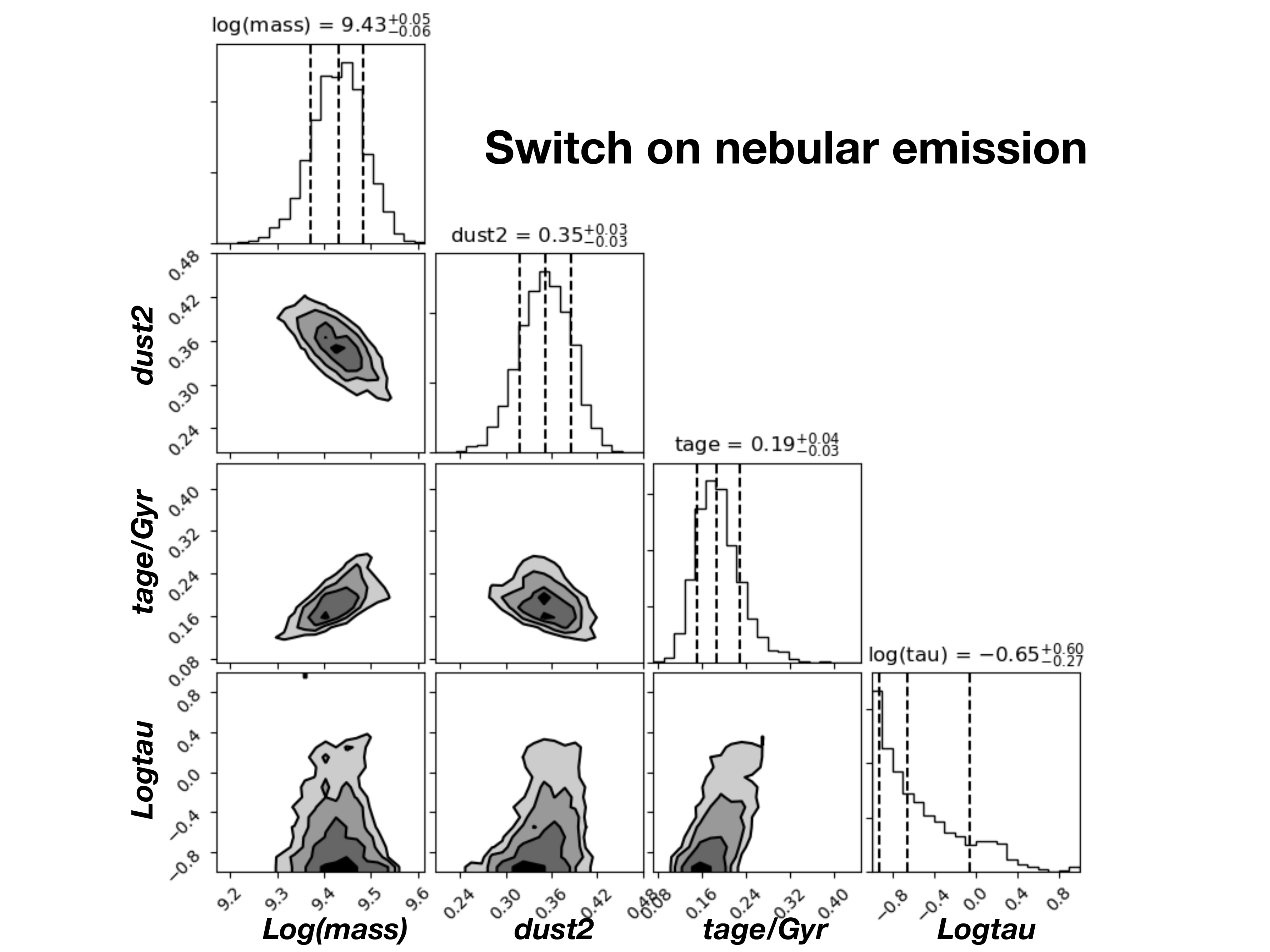}{0.5\textwidth}{}
       }
\caption{{\bf Upper:} The spectral energy distribution of Ion1. The majority of photometric data are taken from CANDELS catalog \citep{Guo2013}. The new photometry obtained in this work for Ion1 is marked with red open squares, including VLT/VIMOS U-band, {\it HST} WFC3/UVIS F410M, WFC3/IR F105W and VLT/HAWK-I $Ks$-band. The three photometric bands covering the LyC emission are not included during the SED fitting. The over-plotted blue (switch on the nebular emission model) and green (switch off the nebular emission model) spectra are the best-fit ones obtained by {\sc Prospector} \citep{Leja2017}, assuming a delay-tau ($SFR\propto te^{-t/\tau}$) star formation history. The corresponding color-coded shadows are 1$\sigma$ ranges of the fits. Transmission curves of the corresponding photometric bandpasses are also shown in the figure. {\bf Bottom:} Posteriors of the fitting parameters (switch off/on the nebular emission model), including stellar mass $log(mass)$, stellar age $tage$ and e-folding factor $log(tau)$ of the delay-tau model, and the V-band dust optical depth $dust2$ which can be converted to the color excess through $E(B-V) = (1.086\cdot dust2)/4.05$ for \citet{Calzetti2000} dust attenuation law. Dashed lines in each histogram mark the 16th, 50th and 84th percentiles.} \label{fig:SED}
\end{figure*}

We first investigate the general properties of Ion1 by looking at its spectral energy distribution (SED). Apart from the existing photometric data directly taken from the CANDELS catalog \citep{Guo2013}, we have conducted new photometry in other 4 bandpasses which are marked with red open squares in Figure \ref{fig:SED}. The newly added photometry includes the VLT/VIMOS U-band and {\it HST} WFC3/UVIS F410M, probing the rest-frame LyC of Ion1 (see Section \ref{sec:IMAGE} for details). We have also added the {\it HST} WFC3/IR F105W photometry using the imaging data observed during the CLEAR survey. Ion1 is barely detected in the median deep $Ks$-band image observed by \citet{Retzlaff2010} with the VLT/ISSAC, which was used to add the $Ks$-band photometry to the CANDELS catalog. We have replaced the low SNR ISSAC photometry with the new photometry (SNR $\approx$ 40) using the ultra-deep $Ks$-band image, which reaches a 27.8 M$\rm{_{AB}}$ 1$\sigma$-depth, obtained with the VLT/HAWK-I imager during the HUGS survey. The photometric bands covering the rest-frame LyC emission of Ion1, i.e. VLT/VIMOS, F410M and F435W, were excluded from the SED fitting due to our scant knowledge of the accurate IGM transmission (see Section \ref{sec:fesc} and e.g. \citealt{Vanzella2016,Steidel2018}). Nevertheless, we have still checked that the derived parameters cannot be significantly affected when including the LyC photometric bands to the SED fitting.

Physical parameters of Ion1 were derived by fitting its SED with {\sc Prospector} \citep{Leja2017}, an SED-fitting routine that is built on the Flexible Stellar Populations Synthesis (FSPS) stellar populations code \citep{Conroy2009,Conroy2010} and the Monte Carlo Markov Chain (MCMC) sampling framework, providing unbiased parameters and realistic error bars. During the fitting procedure, we adopted the affine-invariant ensemble sampler {\sc emcee} \citep{emcee} and assumed the priors of all fitting parameters to be uniform distributions. We fixed the redshift z = 3.794 and metallicity $Z = Z_\sun$, assumed the \citet{Kroupa2001} initial mass function and the \citet{Calzetti2000} dust attenuation law, and fit the SED with a delay-tau parametric star formation history (SFH), i.e. SFR $\propto te^{-t/\tau}$. Motivated by that a young stellar population is typically associated to strong nebular emission (even if Ly$\alpha$ is dumped), like in many local blue metal-poor dwarf galaxies such as IZ18 or SBS0335-052, we have tested the effect of including the nebular emission line model on the SED fitting, where {\sc Prospector} generated nebular continuum and line emission using the {\sc CLOUDY} \citep{Ferland1998, Ferland2013} implementation within FSPS \citep{Byler2017}. The upper panel of Figure \ref{fig:SED} shows the best-fit spectra, where the best-fit model with the nebular emission line model switched off is shown in green and the best-fit model with the nebular emission line model switched on is shown in blue. In the bottom panel of Figure \ref{fig:SED}, we show parameter posteriors for both cases. The fits suggest that Ion1's SED is consistent with a galaxy with stellar age $\approx(10, 200)$ Myr and stellar mass LogM/M$_\sun\approx(9.0, 9.4)$, where the former and latter values correspond to the models with nebular emission model switched off and on respectively. The derived stellar mass is similar to the stellar mass obtained by \citet{Vanzella2012}, also similar to the value (LogM/M$_\sun\approx9.15$) obtained using the SED fitting routine developed by \citet{Lee2018}, which adopts an advanced MCMC procedure that treats SFH as a free parameter. We do see the change of the derived stellar age by including or not including the nebular emission line model in the fitting. We, however, remind that many other assumptions can also significantly affect the derived stellar age, e.g. metallicity and SFH. Regardless of the exact value of Ion1's stellar age, the fitting suggests that Ion1's SED is consistent with a young star-forming galaxies with stellar mass $\sim10^9$ M$_\sun$.

\subsection{Spectrum}
\begin{figure*}
\gridline{\fig{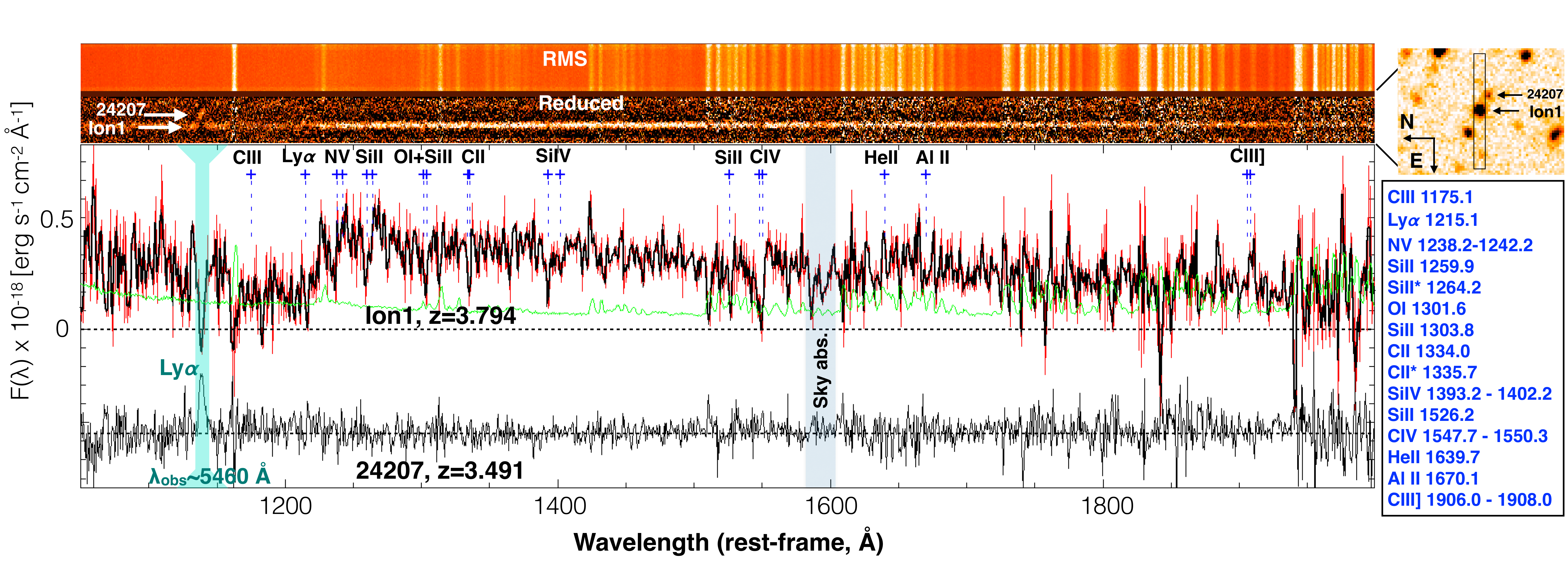}{0.97\textwidth}{}
       }
\caption{The 20 hours spectrum of Ion1 obtained during the VANDELS survey by the VLT/VIMOS spectrograph. Observed wavelengths have been converted to the rest-frame using the redshift (z=3.794) of Ion1. The blue ``+'' symbols mark the spectral features, the details of which are listed on the right blue table. The 1D spectrum of Ion1 is shown in red and the superimposed thick black line is the same after smoothing with a Gaussian kernel with sigma = 1 pixel. The 1$\sigma$ error spectrum is shown as the green line. The cyan band in the figure marks an absorption feature observed at $\lambda_{obs}\approx 5460$ \AA , which is due to an absorber with CANDELS ID=24207, z=3.491 and is $\approx$1.8'' away from Ion1 (also marked in the 2D spectrum with the white arrow and see Section \ref{sec:absorber} for details). The bottom spectrum (black) is extracted for the absorber. The spectrum has been shifted in the Y-axis for clarity. The horizontal dashed black line shows the zero level. The top-right shows the slit (of 1'' width) position and it is superimposed to the VLT/VIMOS R-band image.} \label{fig:SPEC}
\end{figure*}

The redshift of Ion1 has been initially reported to be z=3.795 by \citet{Vanzella2012} based on a relatively shallow spectrum obtained with Keck/DEIMOS, where the redshift was determined using a set of UV absorption features seen in the spectrum, including Ly$\alpha$, \ion{C}{2}$\lambda$1334+1336,  \ion{Si}{4}$\lambda$1393+1402 and \ion{C}{4}$\lambda$1550. With the new deeper spectrum, totaling an on-target exposure time $T_{exp}=20$ hours, obtained with VLT/VIMOS as part of the VANDELS program \citep{McLure2018} and shown in Figure \ref{fig:SPEC}, we have derived the redshift of Ion1 to be 3.794 using the \ion{C}{3}]$\lambda$1909 emission feature (SNR $\approx$ 6, EW = 3.5$\pm$0.5\AA). Considering the existence of other objects in the vicinity (see the top-right of Figure \ref{fig:SPEC}) which could affect the accuracy of the sky subtraction when extracting the spectrum, we adopted the customized data reduction method used by \citet{Vanzella2014} for the extraction of VLT/FORS2 multi-object spectra, which we adapted to the case of VLT/VIMOS. This includes algorithms that optimize the estimate of the sky emission and its removal. In particular, on top of the classical ``A-B'' dithering scheme, in order to take into account the possible local sky counts differences among the partial frames A-B and B-A, this method implements an ``A-B'' subtraction with a zero or first order polynomial fit of the sky along columns before combining the partial frames. The final 2D spectra are weight-averaged combined  and 2D sky-subtracted partial frames are also combined to produce the weighted RMS map. 

The spectrum of Ion1 shows the P-Cygni profile of \ion{C}{4}$\lambda$1550, indicating the presence of very young massive stars, consistent with the results of SED fitting. It also shows the tentative detection of \ion{He}{2}$\lambda$1640 in emission (SNR $\approx$ 2), although the line can be strongly affected by nearby sky emission lines (see Figure \ref{fig:SPEC}). The spectrum also reveals low-ionization (e.g. \ion{Si}{2}$\lambda$1260, \ion{O}{1}$\lambda$1302, \ion{Si}{2}$\lambda$1304, \ion{C}{2}$\lambda$1334+1336) and high-ionization (e.g. \ion{N}{5}$\lambda$1240, \ion{Si}{4}$\lambda$1393+1402) interstellar metal absorption lines. Some of the low-ionization absorption lines can be used to infer the \ion{H}{1} covering fractions and column densities. For instance, \citet{Heckman2011} proposed that, with high spectral resolution ($R=\lambda/\Delta\lambda\sim 10000$) and good SNR data, the relative residual intensity in the cores of interstellar absorption-lines such as \ion{Si}{2} series and \ion{C}{2} series can be used to measure the photoelectric opacity of the ISM. \citet{Gazagnes2018} showed that the neutral gas covering fraction can be reasonably derived by simultaneously fit the stellar continuum, dust attenuation, metal absorption, and \ion{H}{1} properties for spectral data with a wide range of SNR and resolutions (SNR per pixel $>$ 2 and $R>3000$, or SNR per pixel $>$ 5 and $R>600$). Similarly, \citet{Chisholm2018} found the significant correlation between the \ion{H}{1} covering fraction and \ion{Si}{2} inferred covering fraction, which can be measured by either the strong \ion{Si}{2}$\lambda$1260 or the \ion{Si}{2}$\lambda$1190 doublet. The detection of \ion{Si}{2} $\lambda$1260 in Ion1's spectrum is faint but we can clearly see the residual flux at the line center, which is also observed in other low-ionization interstellar absorptions such as \ion{C}{2}$\lambda$1334, \ion{O}{1}$\lambda$1302 and \ion{Si}{2}$\lambda$1304, suggesting the partial covering fraction of the neutral gas in Ion1. It is worth pointing out that the spectrum also seems to have a tentative detection of \ion{O}{3}]$\lambda$1661,1666 doublets emission, which, combined with the detection of \ion{C}{3}]$\lambda$1909 emission, can be used to derive the carbon-to-oxygen ratio C/O to infer galaxy metallicity \citep[e.g.][]{Amorin2017}. The detection, however, has SNR$\lesssim$2 and can be strongly affected by nearby sky emission lines, preventing us from making any conclusive statement at the moment.

The spectrum also reveals Ly$\alpha$ in absorption, with the line center consistent with zero flux and no clear evidence of any emission component. This spectroscopic feature is not observed among other known LyC emitters and its presence needs to be taken into account in future models of LyC emitters when guiding future searches (see the discussion in Section \ref{sec:outlook}), as well as for understanding the relationship between Ly$\alpha$ and escaping LyC emission since galaxies with detected LyC also are often expected to be relatively strong Ly$\alpha$ emitters \citep[e.g.][]{Verhamme2015,Dijkstra2016}. Given the current scant knowledge of the spectral properties of LyC emitters, the fact that Ion1 shows Ly$\alpha$ in absorption is, empirically speaking, interesting. The coexistence of LyC emission and Ly$\alpha$ absorption could be explained by the partial covering fraction of neutral gas \citep{McKinney2019}. The fact that the metal absorption lines don't reach zero flux is consistent with this picture. However, we note that our spectrum has relatively low resolution, $R\approx 580$ \citep{McLure2018}, and since it is integrated over the entire galaxy, it is possible that the region where the LyC of Ion1 escapes actually has Ly$\alpha$ in emission which is smeared out by the stronger absorption feature. Also, while the conditions for Ly$\alpha$ photons to escape are generally similar to those for LyC photons, the optical thickness of Ly$\alpha$ photons is $\approx 10^4$ times larger than that of LyC photons, and thus the \ion{H}{1} column density threshold for optically thick Ly$\alpha$ is correspondingly lower than for LyC. Another possibility for the Ly$\alpha$ to be intrinsically weak is if the channels through which LyC photons escape are ``clean'' without any neutral \ion{H}{1} gas, making nebular emission like Ly$\alpha$ no longer pumped \citep{Vanzella2012}. Under such a circumstance when all ionizing photons are escaping, nebular emission lines such as Ly$\alpha$ drop to intrinsically small equivalent widths \citep{Schaerer2003}. Meanwhile, if no Ly$\alpha$ emission from other sight-lines is scattered into observers' sight-line, we then expect to see escaping LyC while no Ly$\alpha$ emission. For example, this scenario could be the case of Haro 11, a nearby LyC-leaking galaxy at z$\sim$0.021 with stellar mass $\rm{M_*\sim2.4\times10^{10}M_\sun}$ \citep{Bergvall2006,Ostlin2015} and is known to have three photon-ionizing knots. \citet{Keenan2017} mapped the emission line ratio [\ion{O}{3}]$\lambda$5007/[\ion{O}{2}]$\lambda$3727 for Haro 11, with which they could infer where the LyC is escaping. They found evidence that the LyC emission of Haro 11 is escaping from a young star-forming knot with weak Ly$\alpha$, rather than another knot with strong Ly$\alpha$ emission, indicating the potential independency between LyC and Ly$\alpha$ emission. Although the Haro11-like galaxies might be more common at high-redshift since 50\% of the star-formation activity is embedded in star clusters \citep{Adamo2011}, it is unclear, however, the extent to which Haro 11 might be representative of Ion1.

Finally, an interesting question, which is relevant to the moderate \ion{C}{3}]$\lambda$1909 emission line observed in the spectrum of Ion1, is whether its relatively large EW $\sim$ 3.5\AA\  is consistent with the large escape fraction of ionizing radiation ($f_{esc}$, see Section \ref{sec:fesc} for details). After all, if a substantial amount of the ionizing photons escape the galaxy, one would expect relatively weak photo-ionized emission features including the \ion{C}{3}] emission line, as it has been predicted, for example, by \citet{Jaskot2016}. Empirically, however, galaxies with strong leaking LyC (relative escape fraction $f_{esc}^{rel}$ up to $\approx 90$\%) have been observed to have intense \ion{C}{3}] emission lines, with rest-frame equivalent widths comparable or larger than the one observed in Ion1 (e.g. \citealt{Schaerer2018,Vanzella2019}). Particularly interesting are the cases of Ion2, also discussed later in this paper (see Section \ref{sec:ind} for details), which is at $z\approx3.2$ with $f_{esc}^{rel}$ at least $\approx 50$\% and \ion{C}{3}] equivalent width $\approx 4$\AA\ \citep{Vanzella2019}, and of Sunburst, a star-forming super star cluster at $z\approx 2.4$ with $f_{esc}^{rel}\approx 90$\% and \ion{C}{3}] equivalent width $\approx 4$ \AA\ (Vanzella et al., in preparation). Evidently, whatever fraction of ionizing radiation remains in the galaxies seems to be sufficient to excite relatively intense \ion{C}{3}]\ emission lines. Alternatively, the \ion{C}{3}] emission line and the LyC could come from two independent regions of the galaxy even if at the spatial resolution of the images and spectra they are blended into one single source. If it is the case, the strength of \ion{C}{3}] and the $f_{esc}$ of LyC are independent from each other and when integrated over the whole galaxy their relationship characterized by large scatter. Moreover, one should also note that the EW of \ion{C}{3}] is highly dependent on age, metallicity, and ionization parameter \citep{Jaskot2016}. Therefore, a \ion{C}{3}] EW like Ion1’s could still be consistent with substantial ionizing photons escape depending on the values of these other parameters (See Figure 11 of \citealt{Jaskot2016}).

\subsection{Images} \label{sec:IMAGE}

\begin{figure*}
\gridline{\fig{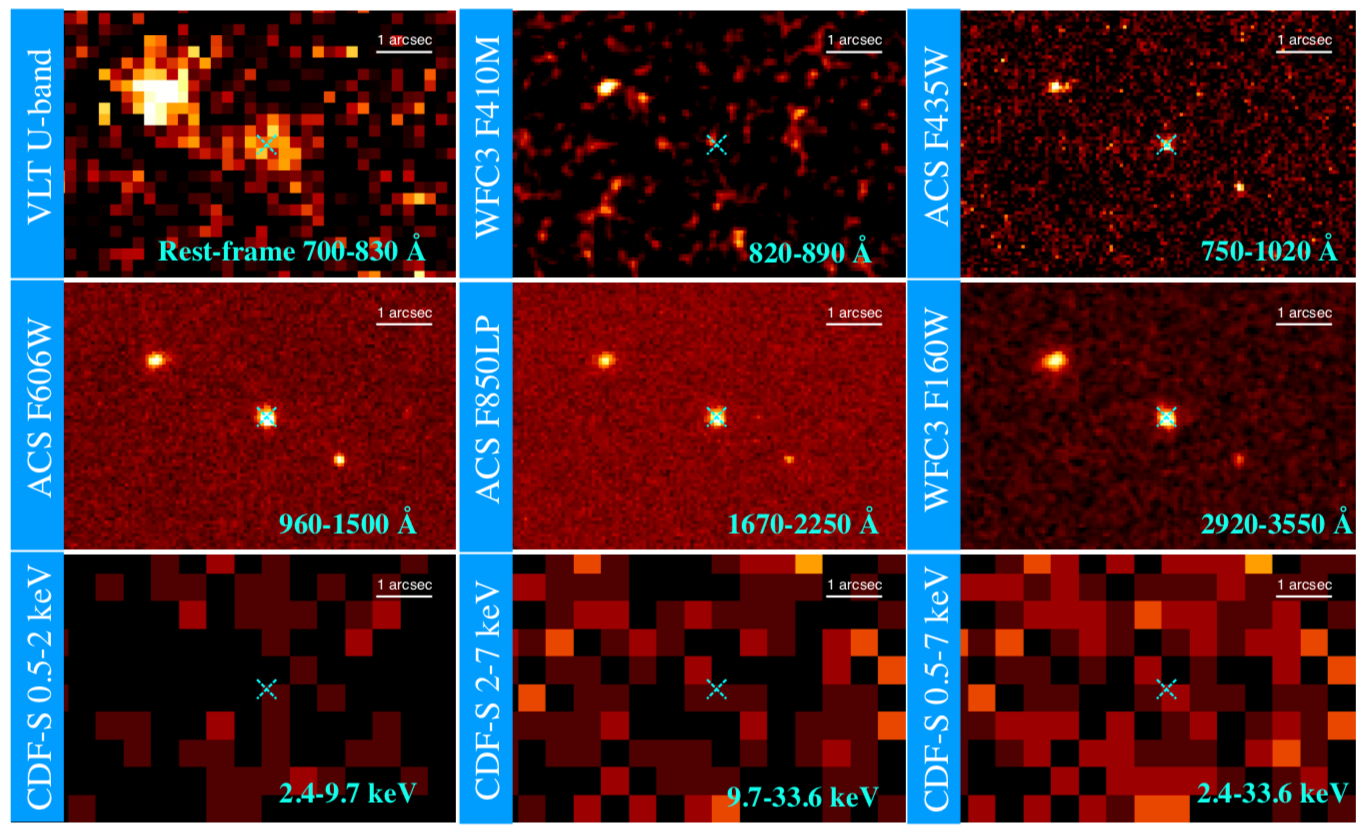}{0.9\textwidth}{(a)}
        }
\gridline{\fig{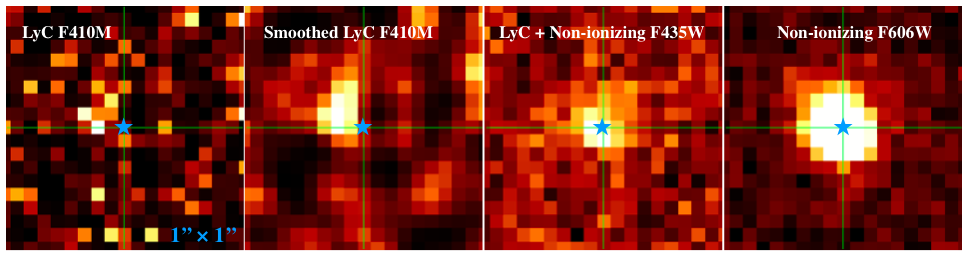}{0.9\textwidth}{(b)}
       }
\caption{{\bf (a)} Images of Ion1 in the VLT/VIMOS U-band, {\it HST} F410M, F435W, F606W, F850LP, F160W and Chandra CDF-S, where the corresponding rest-frame band coverage is also shown in each image. The cyan `X' marks the center of Ion1 in F435W image determined by CENTER in PyRAF with CALGORITHM = ``CENTROID''. {\bf (b)} Zoom-in 1'' $\times$ 1'' view of the {\it HST} postage stamps in the F410M (LyC), F435W (LyC+non-LyC) and F606W (non-LyC).} \label{fig:IMAGE}
\end{figure*}

Figure \ref{fig:IMAGE} shows images of Ion1 observed by the VLT and {\it HST}, including the VLT/VIMOS in the U-band and {\it HST}/WFC3 in the F410M, both probing the ionizing LyC. It also shows the {\it HST}/ACS image of the galaxy in the F435W band, which samples the rest-frame range $\lambda\lambda$ 750--1020\AA, namely a combination of LyC and non-ionizing far-UV radiation, and in the F606W band, which samples the rest-frame range $\lambda\lambda$ 960--1500\AA, i.e. non-ionizing UV radiation. Finally, Figure \ref{fig:IMAGE} shows the image of the galaxy in the 7 Ms {\it Chandra} map in the 0.5-2, 2-7 and 0.5-7 keV bands, where no flux is detected.

\begin{figure*}
\gridline{\fig{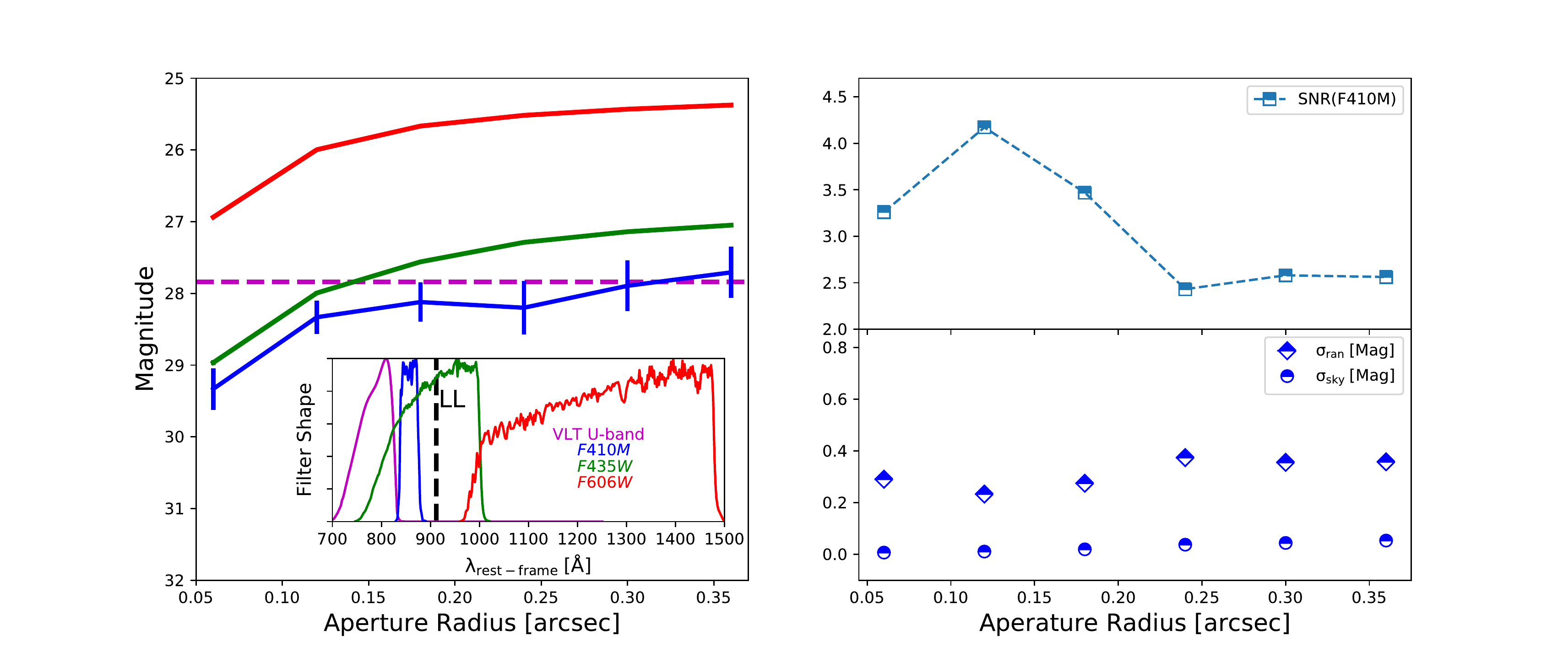}{0.97\textwidth}{}
        }
\caption{{\it HST} aperture photometry for Ion1. {\bf Left}: Photometric curves of growth in F410M (blue), F435W (green) and F606W (red). Error bars shown here only account for random errors (Poisson). The horizontal dashed magenta line marks the magnitude of Ion1 measured in the VLT/VIMOS U-band. We also show the corresponding rest-frame band coverages of different filters, where the vertical black dashed line is the Lyman limit, i.e. 912\AA. {\bf Right}: Signal-to-noise ratio (SNR), random errors $\sigma_{ran}$ and systematic errors $\sigma_{sky}$ introduced by sky subtraction with different aperture sizes in the F410M photometry.}\label{fig:HST_phot}
\end{figure*}

\begin{figure}
\gridline{\fig{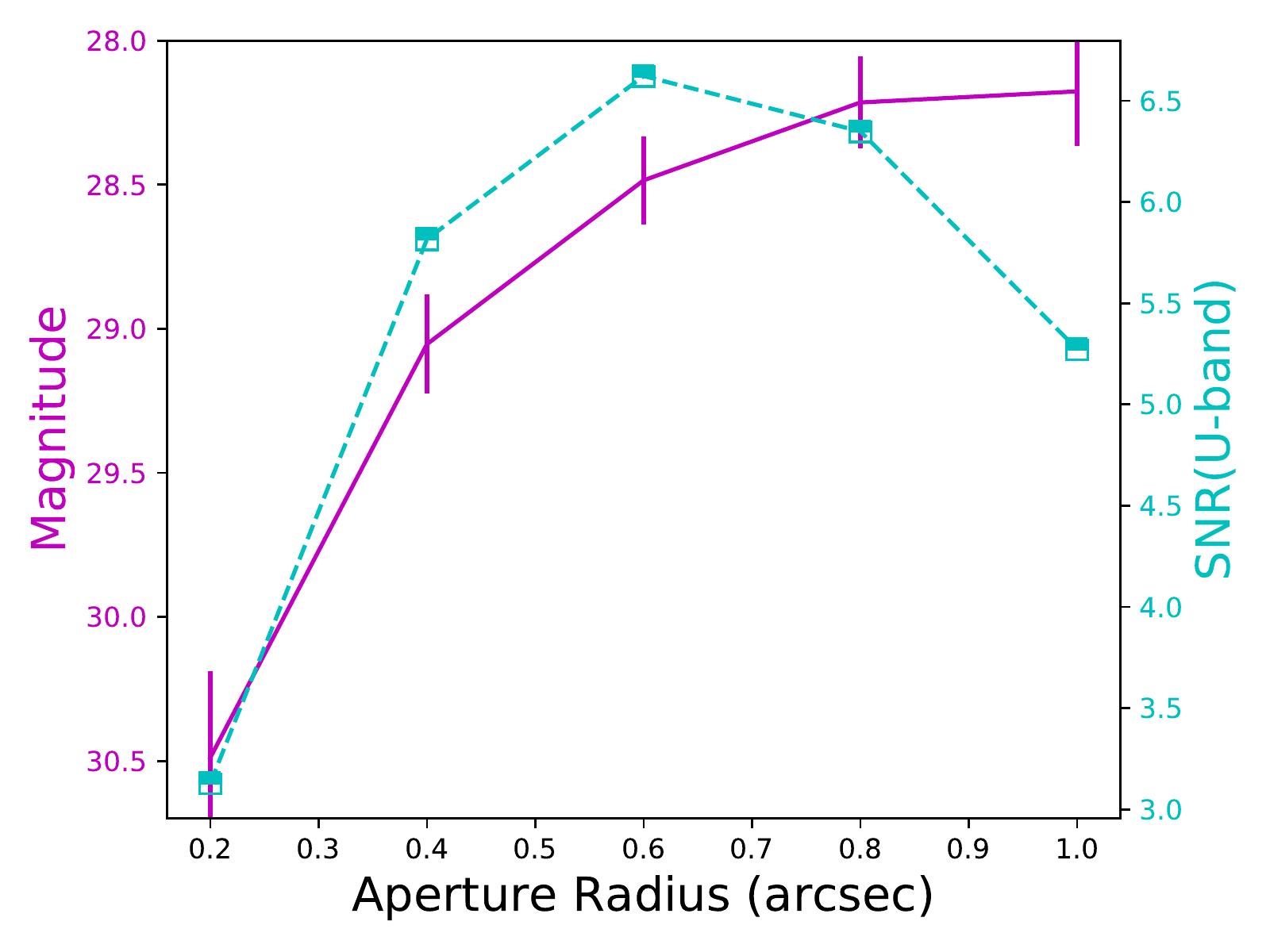}{0.47\textwidth}{}
        }
\caption{Photometric curve of growth (magenta) and signal-to-noise ratio (SNR, cyan) for Ion1 in the VLT/U-band.}\label{fig:VLT_phot}
\end{figure}

Figure \ref{fig:HST_phot} shows the aperture photometry of Ion1 in the F410M image, which probes the rest-frame range $\lambda\lambda$ 820-890 \AA. The galaxy is detected with marginal SNR in this image and we have carried out Monte Carlo simulations to quantify the significance of this detection (as we shall discuss later). We have carried out aperture photometry within a maximum radius of 0.36 arcsec to avoid contamination from nearby sources and we have checked that our measures are robust if we adopt a slightly larger or smaller ($\pm$ 0.06 arcsec) aperture. The photometry has been conducted by fixing the aperture centroid at the pixel with peak flux. We have applied the aperture correction for a point source derived using a bright star in the F410M image itself to estimate the ``missing'' flux outside the aperture \citep{Howell1989}. Note that the aperture correction is valid only if Ion1 is a point source in the F410M image. As will be shown in Section \ref{sec:410mor}, from the Monte Carlo simulations there is evidence that the galaxy is actually moderately resolved, but the effect on the measured flux is negligible. As Figure \ref{fig:HST_phot} shows, we find the F410M magnitude to be $27.71 \pm 0.36$  within a r = 0.36 arcsec aperture, corresponding to $m_{410}=27.60 \pm 0.36$  after aperture correction. The integrated SNR of the LyC detection in the F410M image, as shown in the right panel of Figure \ref{fig:HST_phot}, reaches its maximum value SNR $=$ 4.17 at r $=$ 0.12 arcsec and is SNR $=$ 2.56 at r $=$ 0.36 arcsec. The LyC is not only detected in the F410M image, as already discussed by \citet{Vanzella2010}, it is also detected in the VLT/VIMOS U-band image, which samples the rest-frame range $\lambda\lambda$ 700-830 \AA\ at z $=$ 3.794. We carried out new aperture photometry in the U-band, shown in Figure \ref{fig:VLT_phot}, where the measured flux within a r $=$ 1 arcsec aperture is 28.18 $\pm$ 0.19   and the asymptotic magnitude is $m_{U}= 27.84\pm 0.19$ , including the aperture correction for a point source. The figure also shows the SNR as a function of the aperture ratio and we find the peak to be SNR $=$ 6.62 at radius r $=$ 0.6 arcsec.

The centroids of the light in both the F435W and F606W images are cospatial, as we have verified by running the task CENTER in PyRAF\footnote{PyRAF is a product of the Space Telescope Science Institute, which is operated by AURA for NASA}. There is, however, an offset of $\approx$ 2 pixels, corresponding to 0.12 arcsec or 0.85 kpc physical separation at $z=3.794$, between the F410M and F435W centroids (and hence also F606W). We have cross-matched the centroids of 93 bright compact sources (stars and compact galaxies) in the F435W and F410M images and found the relative astrometric accuracy of the F410M and F435W images to coincide to within 0.05 $\pm$ 0.3 pixels and 0.09 $\pm$ 0.2 pixels in the horizontal and vertical array coordinates respectively, which is substantially smaller than the observed 2 pixel offset, indicating that this is not the result of astrometric uncertainty. Due to the faintness of Ion1 in F410M, we have further checked the offset by inserting 10,000 fake point sources with the same magnitude as Ion1's into the F410M image (see Section \ref{sec:410mor} for details), measuring their centroids with PyRAF/CENTER and calculating the distribution of centroid deviations. Our simulation shows the median centroid deviation is $\approx0.5$ pixels with the 16- to 84-percentile range (0.1, 0.8) pixels. Figure \ref{fig:IMAGE} also compares the F410M image to the F435W, which is contributed by both the LyC and non-ionizing radiation, showing that while the ionizing and non-ionizing radiation do spatially overlap in part, the peak of their emission do not coincide. 

Given the faintness of Ion1 in the F410M band, we have also quantified the robustness of the photometry against the adopted procedure for sky subtraction (see the right panel of Figure \ref{fig:HST_phot}), which can be affected by the selection of areas of the sky used to estimate the sky and the adopted algorithm. To test both, we have re-done the sky evaluation by taking annuli with different outer radii and adopting different sky fitting algorithms in PyRAF FITSKY. It turned out these can introduce an uncertainty of the order of $\rm{\approx 10^{-5}\;e^{-}/s/pixel}$. Moreover, although FITSKY can automatically reject the bright pixels which are likely from other sources when evaluating sky level, some undetected faint sources may still affect the sky fitting. To further check how this would affect FITSKY, we used the CANDELS F160W segmentation map to mask out all pixels with potential contribution from F160W detected sources. The reason for using F160W instead of bluer bands' segmentation maps is that the larger PSF of H-band enables us to minimize the residual flux from the outskirts of adjacent isophots. In any case, the results did not change. We therefore believe the uncertainty as introduced sky subtraction is $\rm{\approx\; 10^{-5}\;e^{-}/s/pixel}$, which is smaller than random errors. 

Finally, we note that Ion1 is not detected in the 7 Ms CDF-S image (both the soft and hard X-ray bands), which provides an upper limit to the X-ray luminosity of the galaxy equal to L$_X<$ 8.7$\times$10$^{41}$ ergs s$^{-1}$. The non-detection in the X-ray image and the centroid offset between the F410M image and that in the other bands strongly argue against an AGN origin of Ion1's LyC radiation.

\subsection{LyC escape fraction} \label{sec:fesc}
The relative escape fraction $f_{esc}^{rel}$ was introduced by \citet{Steidel2001} and it is defined as the fraction of escaping ionizing LyC photons divided by the fraction of escaping non-ionizing (nLyC) photons at 1500 \AA . Following the formulation proposed by \citet{Siana2007}, the $f_{esc}^{rel}$ of Ion1 can be derived from the flux densities ratio in the LyC and non-LyC (nLyC) UV bandpasses through
\begin{equation}
f_{esc}^{rel}= \frac{(L_{nLyC}/L_{LyC})_{int}}{(f_{nLyC}/f_{LyC})_{obs}}exp(\tau_{IGM}^{LyC})
\end{equation} 
$f_{esc}^{rel}$ is related to the absolute escape fraction $f_{esc}^{abs}$ through
\begin{equation}
f_{esc}^{abs} = f_{esc}^{rel}\times 10^{-0.4\times A_{nLyC}}
\end{equation}
where $(L_{nLyC}/L_{LyC})_{int}$ is the intrinsic luminosity density ratio between the non-LyC and LyC bands, $(f_{nLyC}/f_{LyC})_{obs}$ is the observed flux density ratio, $\tau_{IGM}^{LyC}$ is the IGM opacity for the LyC photons (IGM transmission is $T = e^{-\tau_{IGM}^{LyC}}$), and $A_{nLyC}$ is dust extinction at the non-LyC wavelength. 

Here we use the F410M ($\lambda\lambda$ 850\AA) and F775W ($\lambda\lambda$ 1600\AA) images to measure the observed LyC and non-LyC UV emission. As pointed out by \citet{Vanzella2012}, this measurement need to be conducted in the same spatial region where the ionizing and non-ionizing radiation arise. Otherwise, the measured $(f_{nLyC}/f_{LyC})_{obs}$ can be significantly overestimated (hence underestimate $f_{esc}$) because the LyC may arise from a sub-region of the galaxy while the nLyC from a much larger area, making the ``correct'' $(f_{nLyC})_{obs}$ value smaller. We did the photometry in the F775W image with the exactly same (centroid and radius $r=0.36$ arcsec) aperture as used in Section \ref{sec:IMAGE} for the F410M image, resulting in $(f_{850})_{obs} = 0.03\pm0.01\mu Jy$ and $(f_{1600})_{obs} = 0.377\pm0.008\mu Jy$, which translates to $(f_{1600}/f_{850})_{obs} = 12.6\pm4.2$. A more accurate measure for the flux density ratios should further take different PSF sizes into account, i.e. using the LyC morphology in the F410M as a prior to extract the corresponding nLyC flux from a PSF de-convolved F775W image. This requires a quantitative constrain on the LyC morphology in the F410M, which we do not have owing to the low SNR detection in the F410M. We point out that, however, when Ion1's LyC emission is not too far away from being a point source, which is likely the case (see Section \ref{sec:410mor}), the flux recovered by the aperture should not be sensitive to whether or not the two PSFs are matched because a $r=0.36$ arcsec aperture is already comparatively large relative to the PSFs (FWHM$_{F410M} \approx 0.12$ arcsec; FWHM$_{F775W} \approx 0.14$ arcsec).

Assuming the \citet{Calzetti2000} dust attenuation law and adopting the best-fit $E(B-V) = 0.21\pm0.01$ obtained by the SED fitting with the nebular emission model switched off (the bottom-left panel of Figure \ref{fig:SED}, see Section \ref{sec:SED} for details), we got $A_{1600} = k(1600\AA)E(B-V) = 2.09\pm0.10$. If we instead adopt the best-fit parameters derived by the SED fitting with the nebular emission model switched on, where $E(B-V)$ is $\approx2\times$ smaller (the bottom-right panel of Figure \ref{fig:SED}), then the derived $A_{1600}$ will decrease by a factor of 2, translating into a factor of 2.6 increase of $f_{esc}^{abs}$ according to Equation (2). In the following calculations of $f_{esc}^{abs}$, we will use $A_{1600} = 2.09\pm0.10$, i.e. the value derived from the SED fitting with the nebular emission model switched off, but one should keep in mind that there is a factor of 2.6 uncertainty on the inferred $f_{esc}^{abs}$, introduced by whether we include the nebular emission model to the SED fitting or not. 

We adopted the \citet{Inoue2014} IGM model at $z=3.8$ and carried out the Monte Carlo simulation over 10,000 line of sights to calculate the IGM transmission. We assumed the F410M filter probes the rest-frame wavelength range 835-876\AA\ with a box-like shape, and found the mean IGM transmission is 0.142 at $z=3.8$, with the minimum and maximum transmission being 0.0 and 0.548 among the 10,000 realizations. 

The precise determinations of escape fractions require detailed knowledge of the intrinsic luminosity density ratio $(L_{nLyC}/L_{LyC})_{int}$ and the accurate value of IGM transmission along a given line-of-sight. Both parameters, however, are quite uncertain for an individual galaxy. We therefore can only derive a range of plausible values of the escape fraction. As pointed out by \citet{Vanzella2015}, the method according to which Ion1 was originally selected as a LyC candidate favors a relative transparent IGM and/or ISM. We therefore expect a higher IGM transmission along Ion1’s line-of-sight than the averaged IGM transmission at the same redshift, and the the maximum IGM transmission provides with us a lower limit of escape fractions. If we assume an intrinsic value for $(L_{1600}/L_{850})_{int} = 3$ and the maximum IGM transmission $T = 0.548$ (thus lower limits of the escape fractions), the escape fractions are $f_{esc}^{rel} = 43\pm14\%$ and $f_{esc}^{abs} = 6\pm2\%$. If we assume $(L_{1600}/L_{850})_{int} = 5$, then the corresponding escape fractions are $f_{esc}^{rel} = 72\pm24\%$ and $f_{esc}^{abs} = 11\pm4\%$. The intrinsic luminosity density ratio can vary in the range of $1.5\lesssim(L_{1600}/L_{850})_{int}\lesssim7.0$ according to predictions of stellar population synthesis models \citep[e.g.,][]{Leitherer1999,Steidel2018}. If we use the intrinsic luminosity ratio predicted by the best-fit SED model obtained in Section \ref{sec:SED}, which is $(L_{1600}/L_{850})_{int} = 2.2$, the escape fractions change to $f_{esc}^{rel} = 32\pm11\%$ and $f_{esc}^{abs} = 5\pm2\%$. Many other factors can also introduce uncertainty of the escape fraction measurement, such as the choice of SFH \citep{Siana2007} and dust attenuation law, the contribution from binary stars etc., which are beyond the scope of this paper. 

\subsection{Monte Carlo simulations and the morphology of Ion 1 in the F410M image}\label{sec:410mor}

Ion1 is robustly identified as a LyC emitter in the VLT/U-band image but it is only marginally detected in the F410M image. To quantify the significance of this detection, as well as the combined effects of flux and morphology in the observed photometry, we have run a set of Monte Carlo simulations.

Firstly, we have tested the null hypothesis that the observed flux of Ion1 is actually due to a noise spike, including the observed growth curve measured for Ion1 within the r $=$ 0.36 arcsec aperture. To do so, we masked out objects in the F410M image with the aid of the CANDELS F160W segmentation map. Then we randomly placed a circular aperture with radius r $=$ 0.36 arcsec 20,000 times into the image in a 90 arcsec $\times$ 90 arcsec region centered around Ion1 and measured the growth curve of these realizations. Only in 12 realizations out of 20,000, corresponding to a probability $p=0.0006$, the flux inside the $r=0.36$ arcsec aperture is as bright or brighter than Ion1 (m $=$ 27.71 ) with SNR $\ge$ 2.56 and with a peak SNR $\ge$ 4.17 at r $=$ 0.12 arcsec. In other words, we reject the null hypothesis with a 99.94\% confidence level. 

\begin{figure}
\gridline{\fig{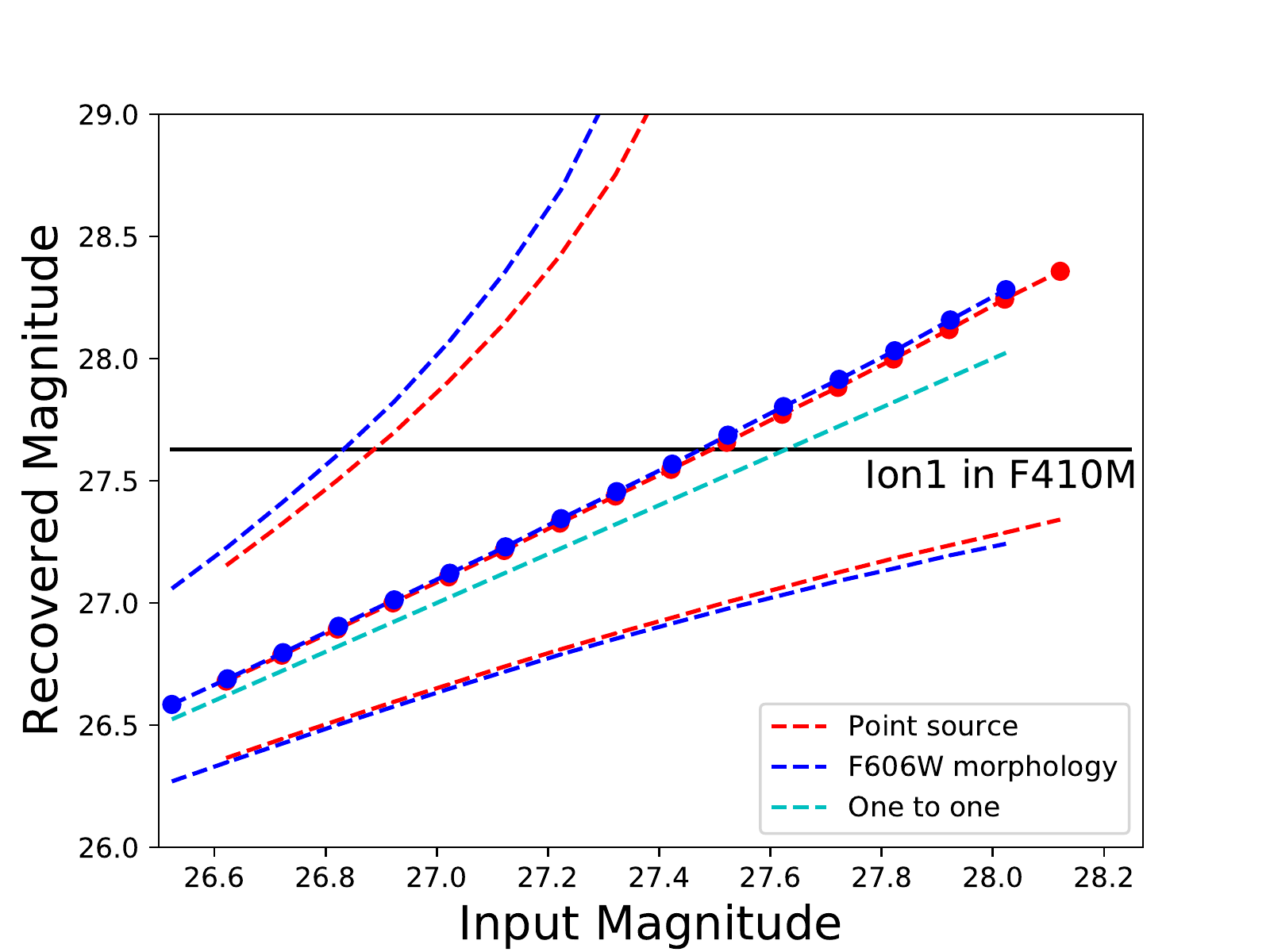}{0.5\textwidth}{}
        }
\caption{Simulations to test on the intrinsic magnitude of Ion1 in F410M (see Section \ref{sec:410mor} for details). The results for point sources and for sources with Ion1's F606W morphology are shown in red and blue respectively. The dashed curves mark the 16-th and 84-th percentiles of the simulations; the dashed lines with dots over-plotted mark the median. The cyan line shows the one-to-one relation. The measured magnitude ($m_{410}$, aperture correction included) of Ion1 is marked as the black horizontal line.}\label{fig:SIMULATION}
\end{figure}

We have subsequently run a set of simulations to estimate the magnitude of systematic errors that might be inherent in the photometry of marginally detected sources and infer the most likely ``intrinsic magnitude'' of Ion1 in the F410M. In these simulations we have injected artificial sources of assigned magnitude into empty regions of the image, again after masking detected sources in the F160W, and carried out aperture photometry of these sources with the same procedure used for Ion1. The values of the input magnitude cover the range from m $=$ 26.6  to m $=$ 28.2  in steps of $\Delta$m $=$ 0.1  and we have made 10,000 realizations at each step. Figure \ref{fig:SIMULATION} shows the run of the median of the recovered magnitude in the r $=$ 0.36 arcsec circular aperture (including the aperture correction) versus the input magnitude in two cases: 1- assuming a point source; 2- assuming that the source has the same light profile as the one observed in the F606W image. Also plotted are the curves marking the 16-th and 84-th percentiles of the realizations. The figure shows that the effect of a marginally resolved morphology is minimal, and that the most likely value of the ``intrinsic'' magnitude of Ion1 in the F410M band is $m_i=27.49^{+2.16}_{-0.50}$. The meaning of the asymmetric error bars is that it represents the 1-$\sigma$ dispersion (68\% of the realizations) from the simulations. The error on the estimate of the intrinsic magnitude, i.e. the difference between the input magnitude and the median of the recovered magnitude, is actually much smaller.

\begin{figure*}
\gridline{\fig{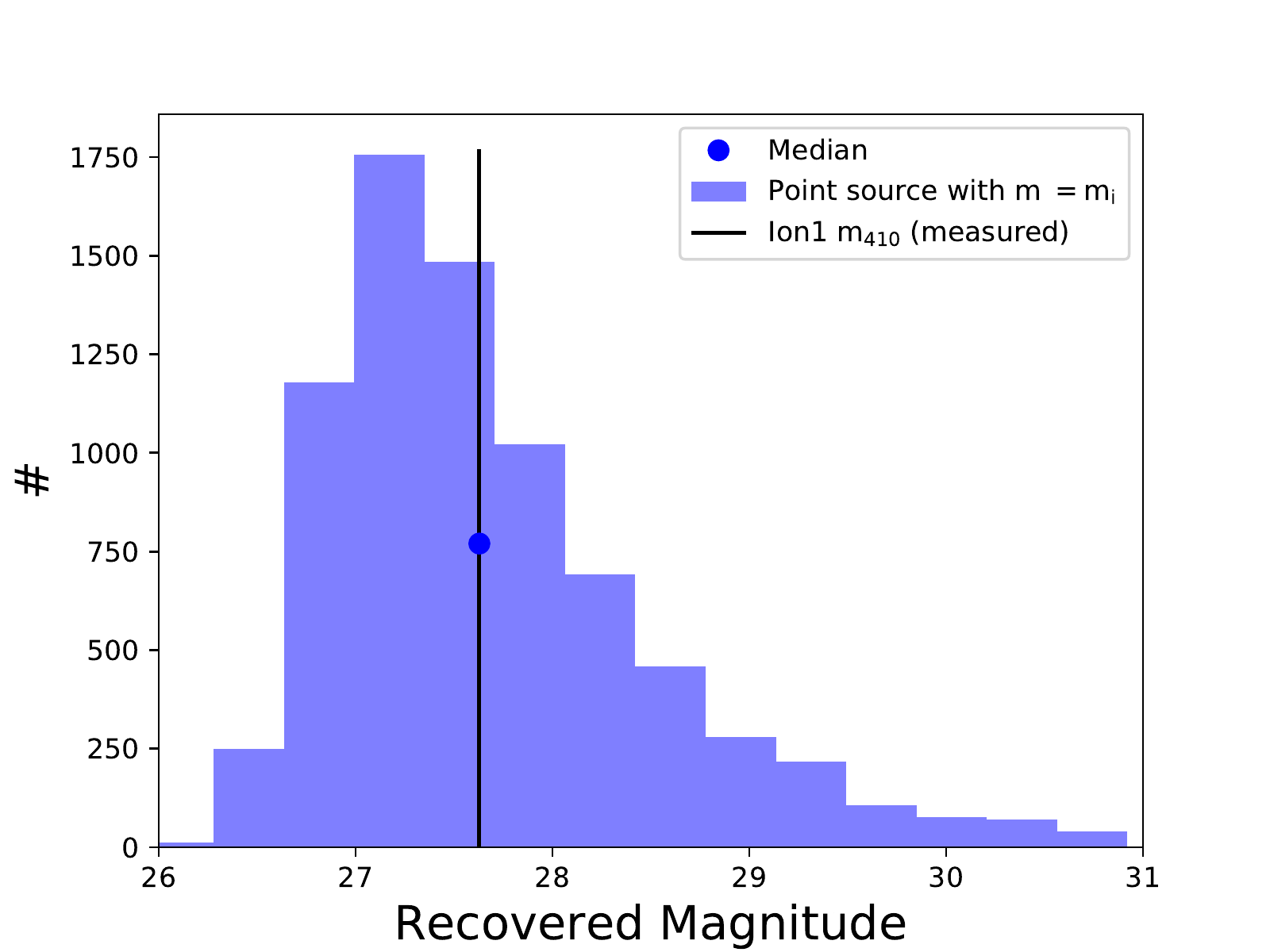}{0.5\textwidth}{}
	\fig{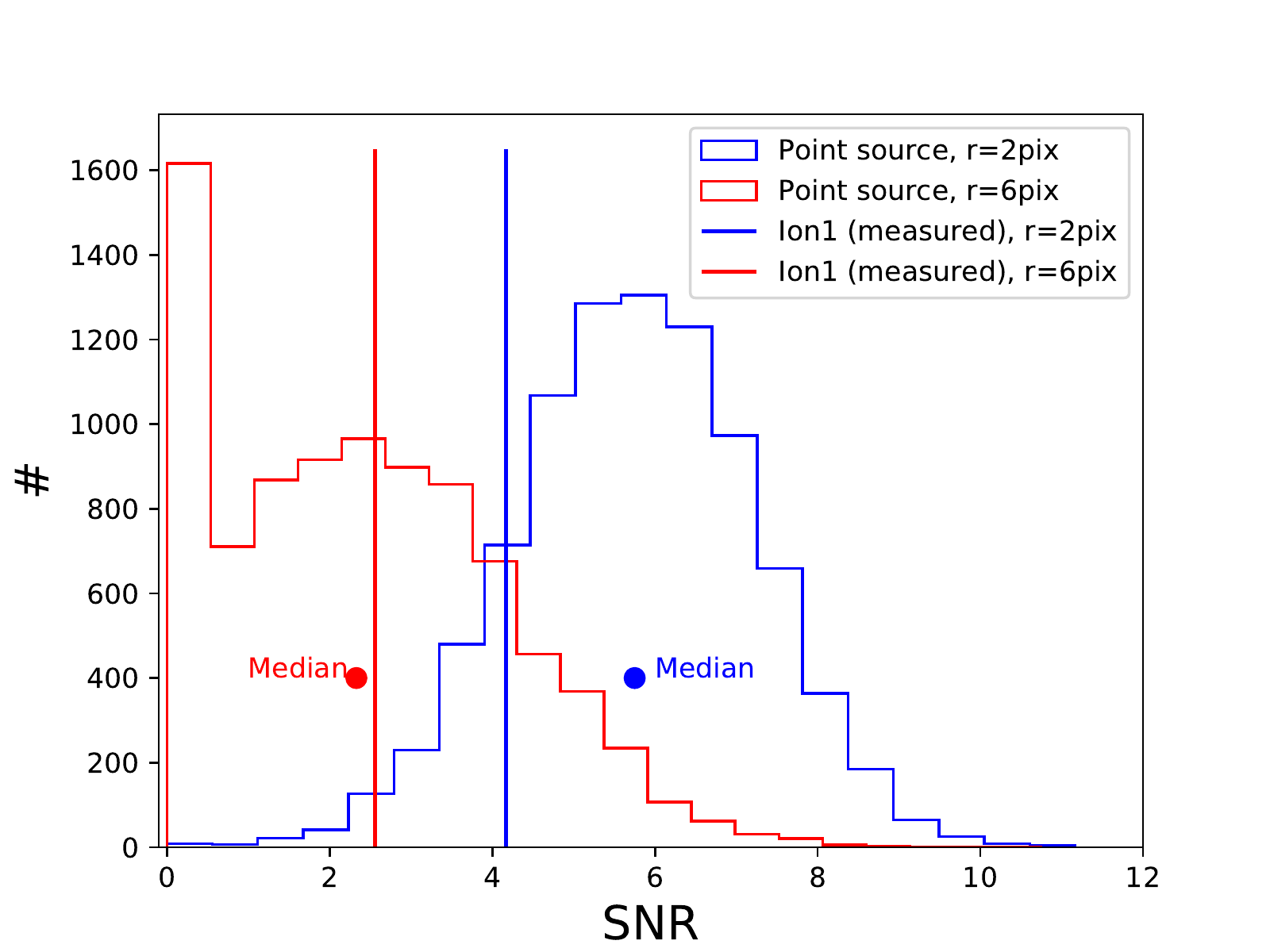}{0.5\textwidth}{}
        }
\caption{Simulations to test the plausibility of the intrinsic magnitude ($m_i$) estimation for Ion1 in F410M. {\bf Left:} The distribution of the recovered magnitude for point sources which are assumed to be as bright as $m_i$. The blue dot marks the median of the distribution. The black vertical line marks the measured magnitude of Ion1 in F410M ($m_{410}$, aperture correction included). {\bf Right:} The distributions of the recovered signal-to-noise ratio (SNR) at two aperture radii, with the case of r $=$ 2 pixels marked in blue and r $=$ 6 pixels marked in red. The dots with corresponding colors mark the median of the distributions; the vertical lines with corresponding colors mark the observed SNR of Ion1 in F410M at the two radii.}\label{fig:SIMULATION_PS}
\end{figure*}

\begin{figure*}
\gridline{\fig{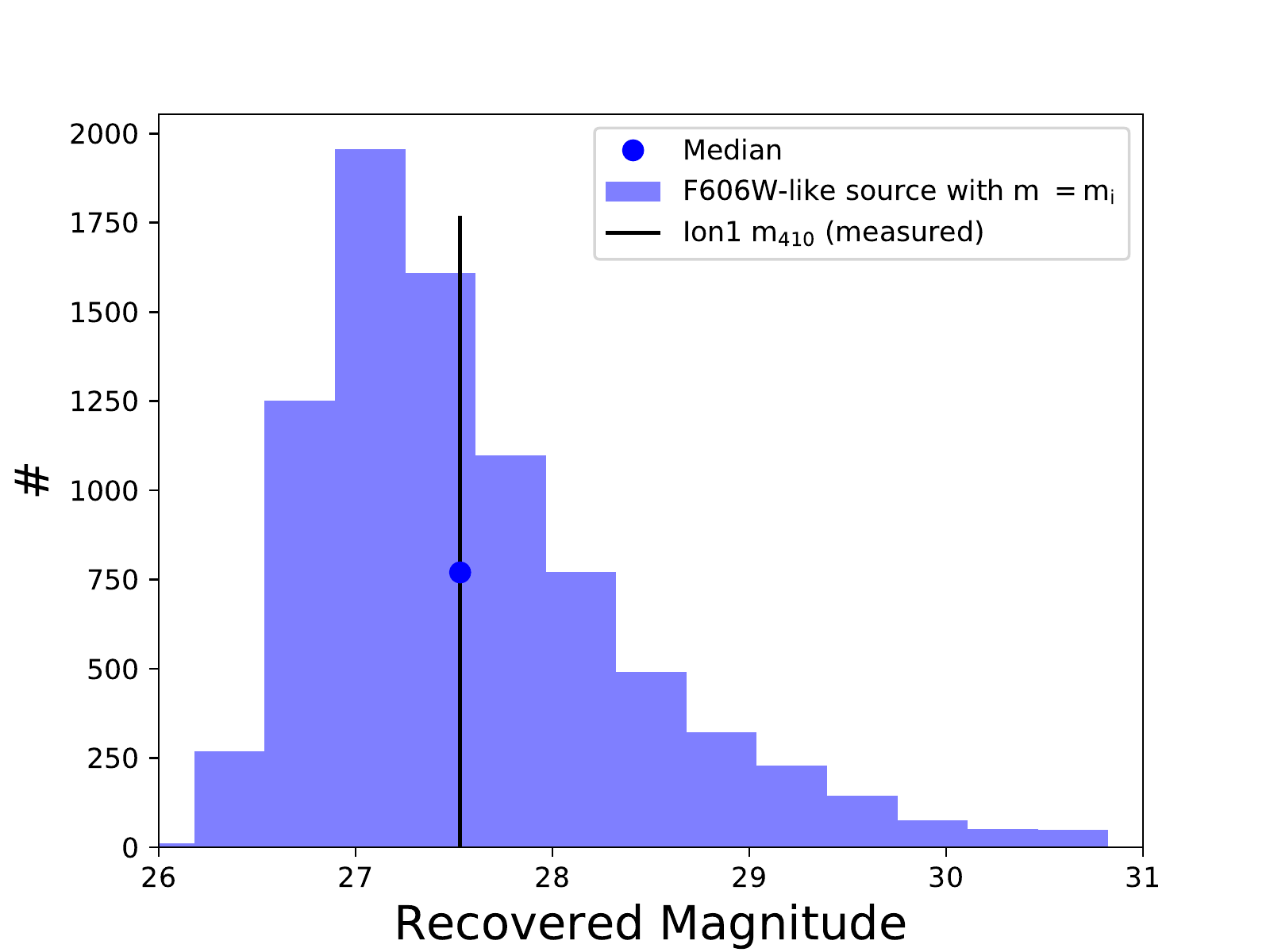}{0.5\textwidth}{}
	\fig{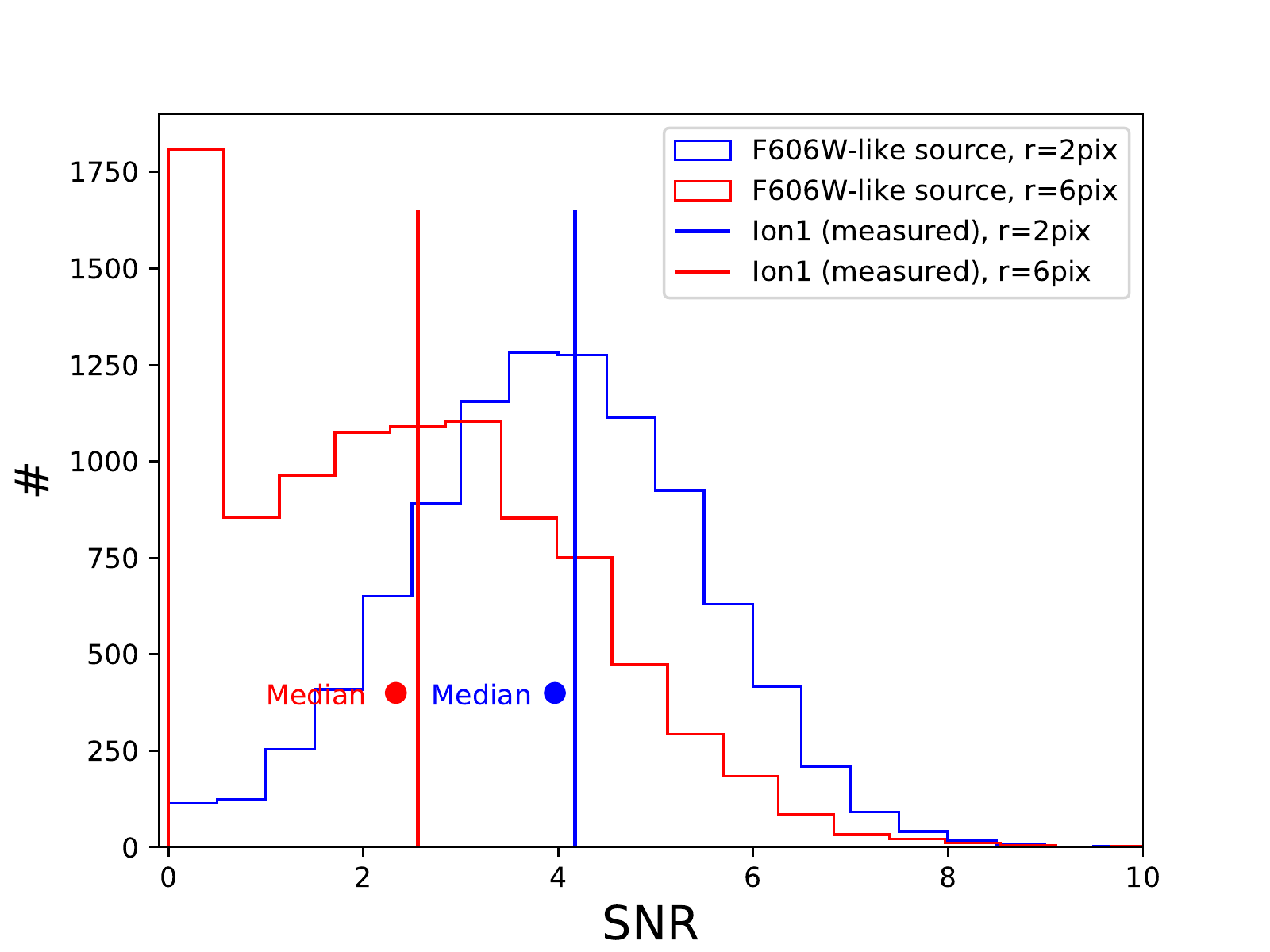}{0.5\textwidth}{}
        }
\caption{Similar as Figure \ref{fig:SIMULATION_PS}, but for the case of sources with Ion1's F606W morphology.}\label{fig:SIMULATION_606}
\end{figure*}

We have run a further set of simulations to test the plausibility of our estimate of the intrinsic magnitude. We have inserted 20,000 realizations of a source with magnitude $m=m_i$ and assigned morphology in the F410M image and measured its magnitude in the same $r=0.36$ arcsec aperture. We assumed the case of 1) a point source; 2) a source with the same morphology as the one Ion1 has in the F606W image. If our measure of $m_i$ is accurate and our assumption about the morphology realistic then we expect that the observed magnitude, growth curve and SNR of Ion1 to be close to the median or the mode of the distribution of the realizations. Such a comparison is illustrated in Figure \ref{fig:SIMULATION_PS} and \ref{fig:SIMULATION_606}, which plot the distributions of the recovered magnitude, as well as SNR in the $r=0.12$ arcsec aperture (blue) and in the $r=0.36$ arcsec one (red). The solid vertical lines mark the observed value for Ion1, the colored circles mark the median of the distributions. Figure \ref{fig:SIMULATION_PS}  is the case of the point source; Figure \ref{fig:SIMULATION_606} is the case of the F606W morphology. In both cases, the recovered magnitude is in a very good agreement with the observed value. There is also a good agreement between the observed SNR in the $r=0.36$ arcsec aperture, which is comparatively large relative to the PSF (FWHM $\approx$ 0.12 arcsec) of the image and thus fairly insensitive to small deviations from the a point source. In the case of the smaller aperture, however, which is quite sensitive to the peak value of the light profile, the observed SNR is substantially smaller than the median value of the simulations for the case of the point source, while it is similar to the median for the case of the F606W morphology. We should clarify that this does not indicate the similar morphology of the LyC and non-LyC emission, as already discussed in Section \ref{sec:IMAGE} that there is an centroid offset between the F410M and F606W images. This test only suggests that, by inserting fake sources with slightly extended morphology (such as F606W morphology) instead of point sources, the simulation can simultaneously recover the observed SNR both in the small and large aperture. This is evidence that the morphology of the F410M image of the LyC emission from Ion1 is moderately resolved. 

\subsection{GALFIT analysis of the morphologies of Ion1 in the F435W, F606W and F160W images}\label{sec:morp_other}

\begin{figure}
\gridline{\fig{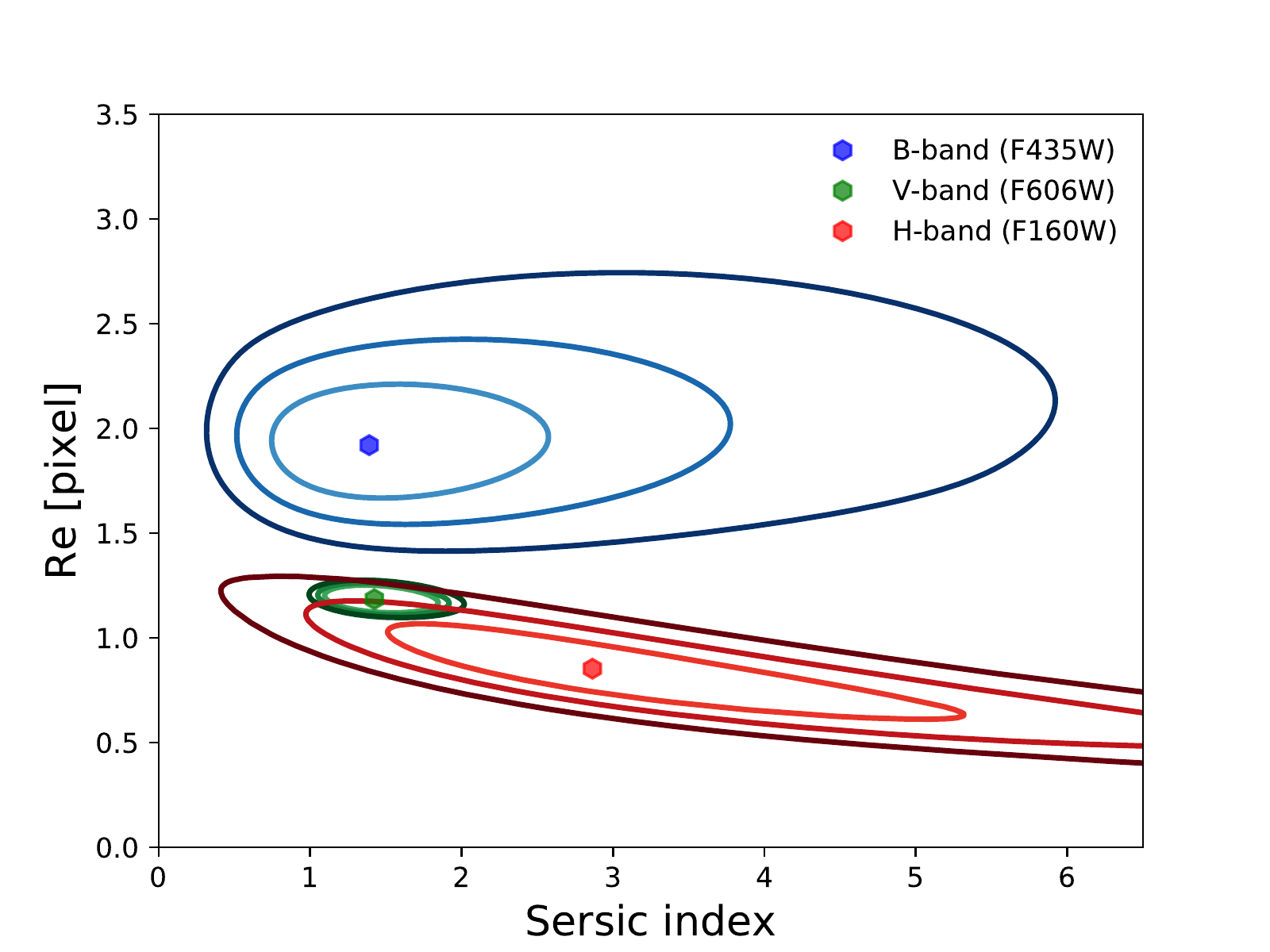}{0.5\textwidth}{}
        }        
\caption{Morphologies of Ion1 in the F435W, F606W and F160W. The analysis is conducted by using {\sc GALFIT} to fit each one of the images with a single 2D S\`{e}rsic profile. The best-fit values are shown as the dots. The covariance of n and $\rm{R_e}$ is measured, where the three contours show the 1$\rm{\sigma}$, 2$\rm{\sigma}$ and 3$\rm{\sigma}$ regions respectively. 1 pixel corresponds to $\rm{0.06''}$, which is $\rm{\approx}$ 0.43(2.04) kpc in physical(comoving) scale at z = 3.794.}\label{fig:sersic}
\end{figure}

In this section, we aim to jointly analyze the morphologies of Ion1 in the bandpasses of F435W, F606W and F160W with {\sc GALFIT} v3.0 \citep{Peng2010}. We have adopted the PSFs built by the CANDELS team \citep{vanderWel2012} and fit each one of the images with a single 2D S\'{e}rsic profile. In addition to the best-fit values, we have also measured the covariance of the S\'{e}rsic index (n) and effective radius ($R_e$). To do so, we first ran {\sc GALFIT} to get the best-fit values of all fitting parameters and then used {\sc GALFIT} to generate a number of models by changing n and $R_e$ while fixing any other parameters to the best-fit values. We have calculated the $\rm{\chi^2}$ of these models and measured the covariance of n and $R_e$ for the single S\'{e}rsic profile fitting of the three images, the results of which are shown in Figure \ref{fig:sersic}. Our measurements point out one issue which has not been carefully treated in S\'{e}rsic fitting, that is very often n and $\rm{R_e}$ cannot be individually well-determined even though the photometric SNR is high. The morphological analysis (by S\'{e}rsic fitting) of galaxies therefore should in principle take this into account. 

\begin{figure}
\gridline{\fig{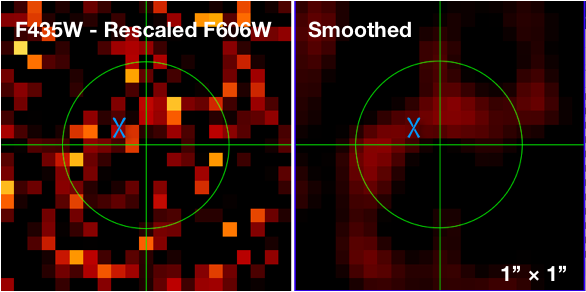}{0.47\textwidth}{}
        }        
\caption{{\bf Left}: The F435W image of Ion1 subtracted by its rescaled F606W image. The F606W image is rescaled according to the ratio between the central flux of F435W and F606W. The blue cross marks the pixel with Ion1's peak flux in F410M. {\bf Right}: The smoothed version of the left one. }\label{fig:f435-f606}
\end{figure}

While the S\'{e}rsic indexes in the F435W and F160W images are not well-constrained, the fits suggest n$\sim$1 (disc-like morphology) in the F606W, i.e. $\lambda\lambda$960-1500\AA. The fits also yield $R_e = 498\pm26$ pc in the F606W, suggesting the non-ionizing UV radiation emerges from the compact and resolved star-forming regions. It also suggests that the $R_e$ of Ion1 is significantly larger in the F435W ($\lambda\lambda$ 750-1020 \AA) than in the F606W. We have also obtained a B-V color map of Ion 1 by subtracting a rescaled F606W image from the F435W image, where the rescaled factor was chosen to be the ratio between (i) the flux of the central pixel; (ii) the total flux within the central 2 pixels $\times$ 2 pixels region; (iii) the total flux within the central 3 pixels $\times$ 3 pixels region. We have checked that the choice of the rescaled factors does not qualitatively affect any results. As Figure \ref{fig:f435-f606} shows, we see the evidence that the photons in F435W prefer (relative to the photons in F606W) to escape from the upper-left of Ion1. The causes can be either the non-uniform dust obscuration across the galaxy which results in the ``excess'' of bluer light (including the LyC) from the low dust obscuration parts of the galaxy, or different sight-line intervening absorbers, or both. The effects of dust on UV photons are monotonically increasing from near-UV to far-UV \citep[see, e.g.][]{Calzetti1994,Battisti2016}. If dust distributes non-uniformly, the dust obscuration will vary across the galaxy and should be stronger in the regions with higher dust column density, which can result in the different morphologies in F435W and F606W. As for the case of intervening absorbers, we will discuss in detail in Section \ref{sec:absorber}, there is an intervening Ly$\alpha$ absorber (1.8 arcsec from Ion1), as well as Ly$\alpha$ emitter, at z=3.491 along the sight line to Ion1 unveiled by the deep VIMOS 2D spectrum. Depending on the distribution and covering factor of gas, this intervening galaxy in principle can also generate such the subtracted image's pattern as seen in Figure \ref{fig:f435-f606}. Regardless of the exact causes, the F435W-F606W image subtraction indicates the lower absorption of far-UV photons in the upper-left part of Ion1. This is coincidently overlapped with the location where the offset LyC is detected, which is consistent with the picture of LyC photons escaping from the regions with low \ion{H}{1} column density.

\section{Discussions}

\subsection{Could the LyC emission of Ion1 actually be an interloper?}

We now discuss the possibility that the source that we have so far identified as the LyC emission from Ion1 is actually a lower redshift interloper accidentally located within Ion1's light profile. We have firstly checked the VIMOS spectrum and noticed that, except an absorption feature seen at $\lambda\approx 5460$ \AA\ (will be discussed in Section \ref{sec:absorber}), no other obvious unidentified emission and absorption lines in the spectrum, i.e. no clear evidence of a spectroscopic ``contamination'' from a lower-redshift object.

We remind that the light centroid of the putative LyC source is displaced relative to that of the non-ionizing radiation of the galaxy by $\approx$ 0.12 arcsec. \citet{Vanzella2012} calculate the probability that, given a galaxy at redshift $z_g$ with apparent magnitude $m_g$ (and thus absolute magnitude $M_g$ in a given band), a lower-redshift interloper located at $z_i<z_g$ with $m_i$ ($M_i$) is observed at an angular distance $\theta_{gi}$. For our calculation, we adopt the same assumptions as \citet{Vanzella2012}, who  adopts the U-band number counts of \citet{Nonino2009}. We estimate that the probability that the LyC source of Ion1 is an interloper with redshift $z_i<3.794$ located at $\theta_{gi}=0.12$ arcsec is $p_i\approx 1\times 10^{-5}$ (the value of the redshift has been chosen to that at least half of the F410M passband contains non-ionizing radiation from the interloper). The physical meaning of this probability is that if we toss at random Ion1 on the celestial sphere $p_i^{-1}$ times, we expect that in one case its light centroid will land at 0.12 arcsec or less from that of a galaxy with redshift $z_i<$3.794. This probability is sufficiently small that it provides reassurance that the detection of LyC emission is real. Nonetheless we have performed additional checks to investigate if there is any evidence of an interloper in the data.

\begin{figure}
\gridline{\fig{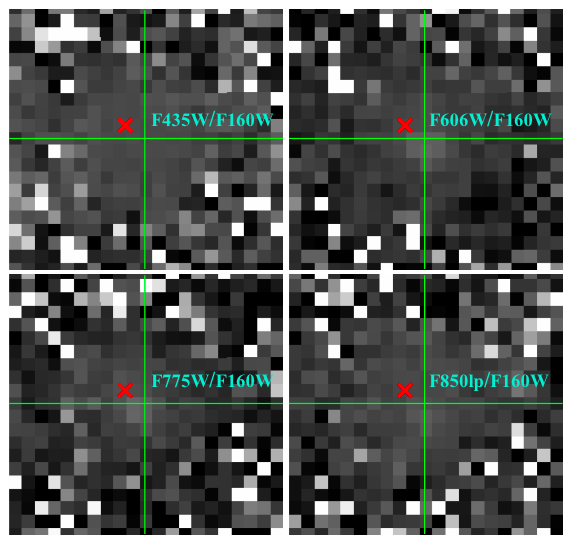}{0.47\textwidth}{}
        }        
\caption{The ratio of {\it HST} images of Ion1 in the F435W, F606W, F775W and F850LP to its F160W image. The red cross marks the position of LyC peak in the F410M.}\label{fig:ratio}
\end{figure}

In the first test, we have produced color maps of Ion1, i.e. the ratio of images of the galaxy taken in two different filters, for all the available {\it HST} passbands. For example, Figure \ref{fig:ratio} shows the ratio of {\it HST} images of Ion1 in the F435W, F606W, F775W and F850LP bandpass to its image taken with the F160W filter. There is no clear evidence that a source is detected at the position of the LyC emission, which indicates that either no additional source of light exists at the position of the LyC emission or, if it exists, it has exactly the same colors as those of Ion1.

\begin{figure*}
\gridline{\fig{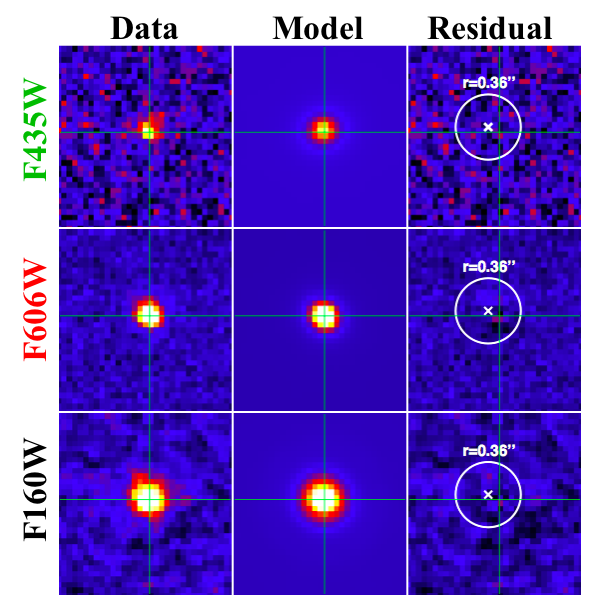}{0.43\textwidth}{(a)}
	\fig{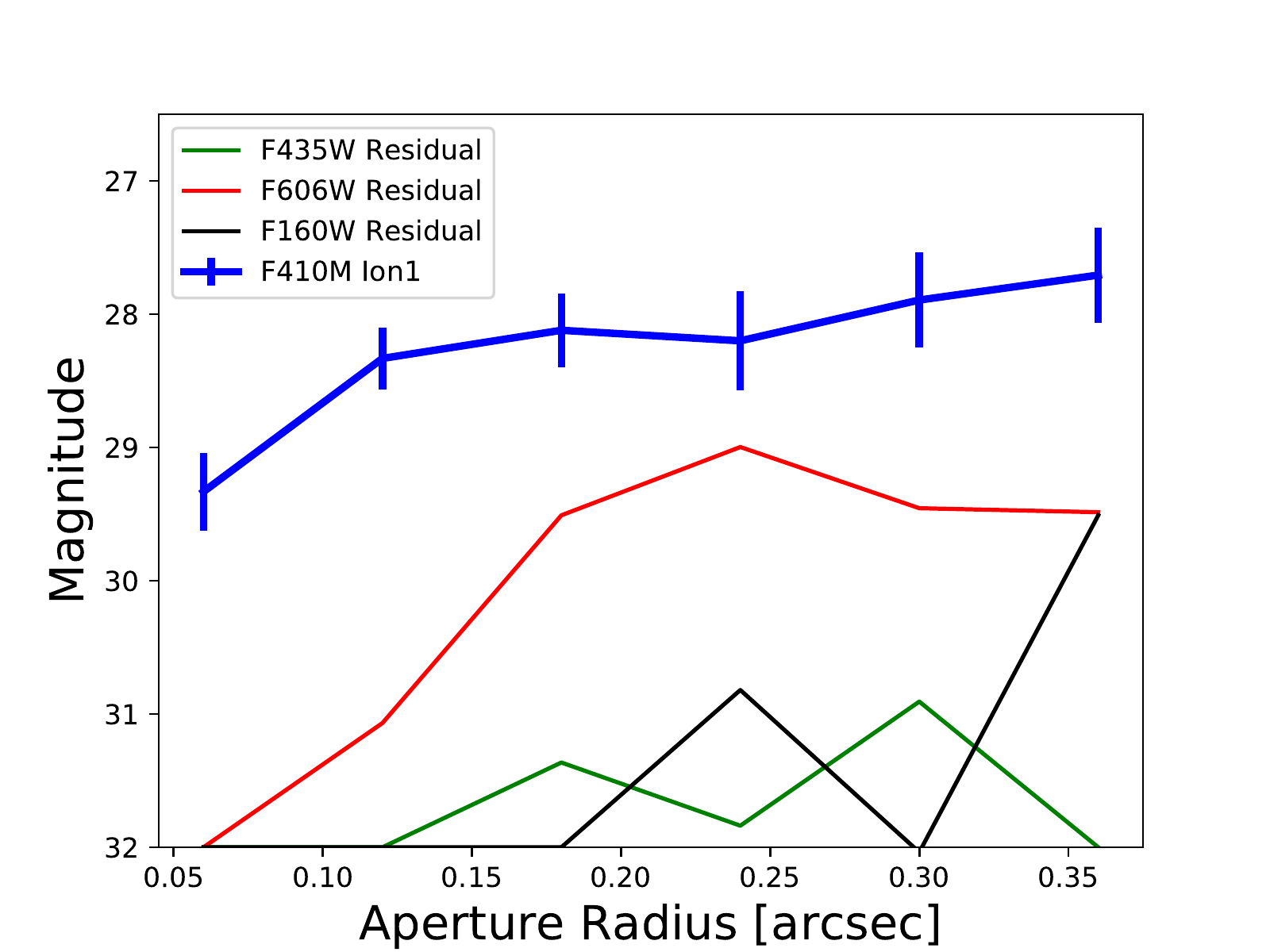}{0.57\textwidth}{(b)}
        }   
\caption{{\bf (a)} Single S\`{e}rsic fitting for Ion1 in F435W, F606W and F160W. The observed images are shown in the first column and the best-fit 2D S\'{e}rsic models and corresponding residuals are shown in the second and third columns. The white circle and cross as marked in the residual are the aperture and its center with which the photometry of Ion1 is conducted in F410M. {\bf (b)} Photometric curves of growth measured in the residual images of F435W (green), F606W (red) and F160W (black). The photometry is conducted with the same aperture as used for Ion1 F410M. The curve of growth of Ion1 F410M is shown as the blue solid line with error bars.}\label{fig:ion1_gal_res}
\end{figure*}

In the second test we have modeled the emission of Ion1 in the F435W, F606W and F160W filters with a S\`{e}rsic light profile (see Section \ref{sec:morp_other} for details) and subtracted the model galaxy from the data to search for a galaxy at the position of the LyC source. As shown in Figure \ref{fig:ion1_gal_res} (a), none was found. We have further conducted aperture photometry in the residual images by using the same aperture as used for the F410M photometry. As Figure \ref{fig:ion1_gal_res} (b) shows, the photometric curves of growth as measured in the residual images verify that no interloper exists. The fact that the total residual flux within the aperture is much fainter than that measured in F410M indicates that we would have detected an interloper as bright as the LyC source if it were there (we have assumed the interloper to be a flat spectrum source, $f_{\nu}\propto \nu^0$ in the UV). This is a conservative assumption, since in practice all faint galaxies have a spectrum which is redder than the flat one. In all cases we would have identified the source if it were there.

We should also consider the possibility that the source detected in the F410M passband is actually a line-emitting interloper with a continuum too faint to be detected in the other {\it HST} images (after subtracting the image of Ion1). Given the observed magnitude, this line would have flux $\approx 3\times 10^{-17}$ erg s$^{-1}$ cm$^{-2}$. While, unfortunately, there is no spectral coverage in any of the spectra of Ion1 in our possession, the fact that flux from this source is detected both in the VLT/U-band and in the F410M passbands, the wavelength coverages of which are adjacent but not overlapping (see Figure \ref{fig:HST_phot}) rules out this possibility.

Finally, if the light detected in the U-band and F410M were coming from an interloper, the most likely possibility given the faintness of the source in the F410M would be that the U band filter probes a bright emission line and the F410M one probes the continuum. In this case, the only possibility would be that the line in the U-band is Ly$\alpha$ at $z\approx 2.1$. If this were the case, then the H-band filter should include the [\ion{O}{3}]$\lambda$5007 line. As we already discussed in the early of this Section, we have modeled the H-band emission of Ion1 and subtracted it from the image (see Figure \ref{fig:ion1_gal_res}). Photometry on the residual image finds no light, which, in addition to the extremely small value of the probability of an interloper, strongly argues against the interloper interpretation of the detected U-band and F410M flux. 

We conclude noting that, ultimately, any flux detection at wavelengths of the LyC of a high-redshift galaxy, either an image or a spectrum, can always be interpreted as coming from a co-spatial, or quasi co-spatial, suitably faint, lower-z interloper. While it is extremely difficult to completely rule out such a possibility with current instrumentation and sensitivity, all our tests and analysis consistently argues that the probability of such a possibility for Ion1 is vanishingly small (for example, our analysis has shown that it is much more likely that the flux detection in the F410M passband is due to a statistical fluke ($p=0.0006$, see Section \ref{sec:410mor}) than it is due to the faint continuum of an interloper).

\subsection{An intervening absorption system at $z=3.491$}\label{sec:absorber}

As the spectrum of Ion1 shows (Figure \ref{fig:SPEC}), there is an absorption feature at $\lambda\approx 5460$ \AA\ and, at the same wavelength but offset upward by $\approx 1.8$ arcsec from Ion1's spectrum, an emission line with no obvious detected continuum (see the extracted 1D spectrum in the bottom panel of Figure \ref{fig:SPEC}). A visual inspection of the {\it HST}/ACS F606W and VLT/VIMOS R-band images with the spectrograph slit over-imposed (see the top-right panel in Figure \ref{fig:SPEC}),  reveals that the source of the emission very likely is a blue galaxy identified in the CANDELS catalog as ID=24207, with apparent magnitude $H_{160}=$ 26.54 , $z_{850}=$ 26.68  and $i_{775}-z_{850}=-0.2$  \citep{Guo2013}. The galaxy line centroid is $\approx 0.9$ arcsec away from the center of the slit, but a fraction of its light must have entered the slit due to the atmospheric seeing, causing the observed offset emission feature. If the emission line is Ly$\alpha$, as strongly suggested by the simultaneous detection of the offset emission and the absorption in Ion1's spectrum at the source wavelength, as further supported by the asymmetric 1D line profile of the object (bottom panel of Figure \ref{fig:SPEC}), then the redshift of the intervening galaxy is $z=3.491$, which is also very consistent with the photometric redshift $z_{phot}=3.35$ reported by the CANDELS team \citep{Dahlen2013}. Thus, neutral gas associated with this galaxy is very likely causing intervening Ly$\alpha$ absorption in the spectrum of Ion1. A crude estimate of the column density $N_{HI}$ can be obtained by assuming the optical depth at the Ly$\alpha$ line center is larger than 1, i.e. 
\begin{equation}
\tau(Ly\alpha) = N_{\rm{HI}} \sigma_0 \ge 1
\end{equation}
where $\sigma_0$ is the Ly$\alpha$ cross section at the line center and it equals to 
\begin{equation}
\sigma_0 = 5.9\times10^{-14}(\frac{T}{10^4 \rm{K}})^{-0.5}\ \rm{cm}^2
\end{equation}
this yields $N_{\rm{HI}}\ge 1.7\times 10^{13}\ \rm{cm}^{-2}$ if $T = 10^4$ K. 

The difference in redshift between Ion1 and the galaxy that gives rise to the Ly$\alpha$ absorption is sufficiently small that the LyC photons that we observe in the U-band image at rest-frame $\lambda\lambda$ 790 \AA\ reach the intervening galaxy at $\lambda\lambda$ 844 \AA, namely they still are ionizing radiation. If the sight-line to the absorbing gas is the same as that to the Ion1's LyC emission, this sets a crude upper limit to the column density $N_{HI}\le 10^{17.5}$ cm$^{-2}$ or else the LyC photons from Ion1 would not be observed. If the sight-line to the absorption system is not cospatial with the LyC emission, nothing can be said. We remind, however, that the faint detection of LyC in the F410M filter from Ion1, corresponding to rest-frame $\lambda\lambda$ 855 \AA, is offset by about 0.12 arcsec from the centroid of the non-ionizing far-UV continuum of the galaxy, or about 1 kpc, suggesting that the two sight-lines are at least not exactly co-spatial. Regardless of the exact alignment of the two sight-lines and neglecting the differences in the covering factor of the gas, the large difference in optical depth between Ly$\alpha$ photons and LyC photons, approximately a factor $10^4$, is very likely a major contributing factor to the simultaneous presence of absorption of Ly$\alpha$ photons and transmission of LyC ones.

Finally, we note that the mean free path of ionizing photons with energy around Rydberg at $z\approx 3.8$ is of the order of 50-60 Mpc \citep{Worseck2014}, which is approximately the radial distance between Ion1 and the intervening $z=3.491$ galaxy. LyC photons with wavelength $\lambda=790$ \AA\ emitted at $z=3.794$ are redshifted into non-ionizing wavelengths at redshift $z\approx 3.153$. Since we detect LyC from Ion1, this implies that the ionizing photons have actually traveled at least $\approx 120$ Mpc, about twice as large as the mean free path by \citet[][see their Figure 10]{Worseck2014}, suggesting that either their mean value has been underestimated or that the scatter of their relationship is large (or both), or also an unusually large under-density of neutral gas at these redshifts towards Ion1.

\subsection{Physical mechanisms of the escaping LyC of Ion1}\label{sec:ind}
A number of scenarios have been proposed for the mechanisms by which LyC photons can escape from host galaxies. Whilst a big fraction (or the entirety) of ionizing photons generated by massive stars are absorbed by neutral hydrogen in the surrounding ISM, stellar wind and/or supernova can carve the low \ion{H}{1} column density cavities through which ionizing photons can escape \citep[e.g.,] []{Kimm2014,Wise2014,Calura2015}. If our line of sight is coincident with the opening angles of these cavities, and the opacity of intervening \ion{H}{1} along the line of sight is not too large, we can directly detect the ionizing photons. In such case, LyC emission should be compact and could be highly anisotropic \citep{Paardekooper2015}. 

Other scenarios predict that the morphology of escaping LyC emission can also be extended. The interaction between the Cosmic Microwave Background (CMB) photons and galaxies is much more important in the early Universe since  the energy density of the CMB is proportional to $(1+z)^4$. \citet{Oh2001} showed that the ``hot'' CMB photons at high redshifts can be cooled through thermal emission and the inverse Compton scattering with relativistic electrons, which were presumably accelerated in star-forming regions during supernovae explosions and then moved into the Circum-Galactic Medium (CGM). The inverse Compton scattering can produce X-ray emission, the soft tail of which can effectively ionized the IGM \citep{Oh2001}. Therefore, if the observed ionizing radiation is originated from the CMB-CGM interaction, we should expect the extended LyC. Another scenario that can also produce the extended LyC is through hydrogen bound-free process in ``matter-bounded" nebulae \citep{Inoue2010}. Multiple supernovae and stellar winds can make the ISM diffuse and highly ionized such that the opacity for LyC is negligible. \citet{Inoue2010} showed that LyC emission can be produced in such diffuse ISM through the hydrogen bound-free process. We can then imagine a picture of galactic-wide outflow of ionized gas, within which the bound-free process can produce the extended LyC. The morphology of LyC, therefore, is key to constrain the physical mechanisms of escaping LyC photons.

The results that we have obtained in the previous Sections suggest that Ion1's non-ionizing UV radiation probed by F606W emerges from a compact ($\sim$ 500 pc) resolved region and the detected LyC emission is (i) offset from the center of non-ionizing UV emission; (ii) coincidently overlapped with the part of the galaxy where we see the evidence of relatively low FUV absorption. The marginal SNR of Ion1 in F410M prevents us from quantitatively analyzing its LyC morphology, although our simulations indicate the emission is moderately resolved (see Section \ref{sec:410mor}), suggesting the effective radius of the LyC emission is $R_e\gtrsim240$ pc, i.e. the resolution of the F410M image. These morphological properties favor the scenario that the UV radiation of Ion1 originates from star-forming regions, presumably young star-forming complexes (i.e. clustered young star clusters and associations) considering the comparable size as observed for local galaxies \citep[e.g.][]{Grasha2017}. While the non-ionizing UV photons can escape from most of the star-forming regions, the ionizing photons can only escape from the ones with low \ion{H}{1} column density ``cavities'' in the surrounding ISM. The detected LyC photons are those escaped from such sight-lines with low intervening absorption.

\subsubsection{Comparison of Ion1 with other LyC emitters at intermediate redshifts -- morphological and spectroscopical properties} \label{sec:comp-mor-spec}

We now compare the morphological and spectroscopical properties of Ion1 with
those of other four $1<z<4$ LyC emitters mentioned in Section \ref{sec:intro}, i.e. the Sunburst Arc at z $=$ 2.37 \citep{RiveraThorsen2017,RiveraThorsen2019}, Q1549-C25  at z $=$ 3.15 \citep{Shapley2016}, Ion2 at z $=$ 3.21 \citep{debarros2016,Vanzella2016,Vanzella2019} and Ion3 at z $=$ 4.00 \citep{Vanzella2018,Vanzella2019}.

The Sunburst Arc is gravitationally lensed by a z $=$ 0.44 massive galaxy cluster into multiple images from which the LyC has been recently detected by {\it HST}  WFC3/UVIS observations in the F275W bandpass \citep{RiveraThorsen2019}. The observations reveal the LyC source to be a $3.0\pm0.1$ Myr old, bright and compact object \citep{Chisholm2019}, compatible with being a young massive star cluster with a physical effective radius $R_e < 20$ pc and a stellar mass of a few $10^6$ M$_\sun$ \citep{Vanzella2019}. The LyC radiation from Ion2 is detected in the {\it HST}/WFC3 F336W bandpass with apparent magnitude $m_{336}=27.57$ and SNR $\approx$ 10) in an image with total exposure time $T_{exp} = 47.6$ ks, \citealt{Vanzella2016}). Morphological analysis with GALFIT shows that the non-ionizing UV radiation from Ion2 is compact and spatially resolved with $R_e=340\pm25$ pc, while the LyC is spatially unresolved with $R_e\le200$ pc. The absence of {\it HST} imaging of Ion3's LyC prevents us from conclusively constraining its morphology \citep{Vanzella2018}. Other LyC  emitters at similar redshift that have been imaged with {\it HST} show a degree of morphological similarity with Ion1, primarily their being compact in the rest-frame UV. Since there is no evidence suggesting that the LyC emission might powered by an AGN in any of the observed cases, a common interpretation is that the ionizing radiation powered by young star forming regions presumably escapes from ``cavities'' in the ISM, namely regions with low \ion{H}{1} column density. The two cases of spatially-unresolved LyC morphology in the Sunburst and Ion2 suggest that the regions can be either a single gravitationally bounded super star cluster or an association of  unresolved star clusters \citep{Vanzella2019}. The case of Ion1, whose LyC is also very compact but appears to be moderately spatially resolved, would still remain consistent with the case of an association of young star clusters. 

High-ionization interstellar lines are commonly observed in the rest-frame UV spectra of the LyC emitters, including the P-Cygni profile of \ion{N}{5}$\lambda$ 1240 of Ion3 \citep{Vanzella2018}, P-Cygni profile of \ion{C}{4} $\lambda$1550 and broad \ion{He}{2}$\lambda$ 1640 of the Sunburst and Ion2 \citep{Vanzella2019}. Despite of the relatively low spectral resolution, similar features are also seen in Ion1's spectrum, such as the P-Cygni profile of \ion{C}{4}$\lambda$1550. These spectral features are evidence of the presence of stellar winds from massive stars, again consistent with the picture of LyC escaping from star-forming regions. 

Ly$\alpha$ is among the most studied properties of LyC emitters and is particularly important since its brightness makes it observable for large samples of galaxies throughout the cosmic time. The four LyC emitters are all observed to have strong Ly$\alpha$ emission. High resolution spectra (R $\sim$ 5000) has further revealed that the Ly$\alpha$ profiles of the Sunburst \citep{RiveraThorsen2017}, Ion2 and Ion3 \citep{Vanzella2019} are triple-peaked, consistent with the theoretical predictions of Ly$\alpha$ photons escaping through a porous neutral medium (see Figure 1 in \citealt{RiveraThorsen2017}, also see Figure 4 in \citealt{Vanzella2019} for a comparison among Ly$\alpha$ profiles of LyC emitters with the $f_{esc}^{rel} >50\%$). The Ly$\alpha$ line of Ion1, however, is observed in absorption, but the spectral resolution of the VLT/VIMOS spectrum, R$\approx$580, prevents us from studying the Ly$\alpha$ profile in details, e.g. if there is any faint emission component on top of the absorption feature. Regardless of the exact shape of the Ly$\alpha$ profile, the overall absorption feature makes Ion1 an empirically interesting case in the sense that the sources is observed to simultaneously have escaping LyC photons and Ly$\alpha$ in absorption, suggesting that we have not yet fully characterized the dispersion of properties of LyC emitters at these redshifts.
 
The properties of low-ionization interstellar lines vary among the LyC emitters. While absorption features such as \ion{Si}{2}$\lambda$1260, \ion{O}{1}$\lambda$1303 and \ion{C}{2}$\lambda$1334 are very weak or absent in the observed spectrum of Ion2 \citep{debarros2016, Vanzella2016} and Ion3 \citep{Vanzella2018}, they are clearly observed in that of Sunburst (\ion{Si}{2}$\lambda$1260, \citealt{RiveraThorsen2017}), Q1549-C25 (\ion{Si}{2}$\lambda$1260, \ion{O}{1}$\lambda$1303, \ion{C}{2}$\lambda$1334, \citealt{Shapley2016}), and of Ion1 (see Figure \ref{fig:SPEC}). The residual intensity in the cores of these saturated absorption features has been proposed as an indirect diagnostic of the escape of ionizing radiation \citep[e.g.][]{Alexandroff2015}. The scattering of the low-ionization absorption properties among the emitters, therefore, indicates a corresponding scattering in the coverage of low-ionization gas along the line of sight towards the sources of UV continuum, namely young stars. But, in turns, this also suggests that incomplete covering by that gas can happen in other regions of the galaxies, too. In other words, the gas covering can be patchy. In such a scenario strong interstellar absorption can still be observed in the UV continuum while LyC is escaping from some other region of the galaxy, not cospatial with that whose UV continuum is used to observed the absorption lines (recall that the 1D spectrum of a galaxy is basically a luminosity-weighted average), as seems to be the case with Ion1.  Thus, the apparent diversity of properties among the observed LyC leakers at high redshift is likely in part the result of the very small number of galaxies studied so far and, in part, reflects the intrinsic scatter of the conditions under which ionizing radiation can escape from host galaxies. These conditions appear to be {\it local}, namely LyC photons can escape from some localized region of the galaxy if there exists low neutral gas surrounding stars from
where the ionizing emission arise. Other portions of the same galaxy could not have any emerging LyC radiation, i.e. the LyC emission is likely anisotropic. In the case of Ion2/Ion3-like galaxies, the escaping LyC and the weak (or absent) low-ionization absorption can be explained by an overall low covering fraction of neutral gas around the brightest (i.e. dominant in the 1D spectra) regions where the LyC also originates from (the fact that Ion2's LyC is nucleated is consistent with this, see \citealt{Vanzella2016}). As for galaxies like the Sunburst, Q1549-C25 and Ion1, the simultaneous observing of LyC and low-ionization metal absorptions (although they don't fully reach zero flux) can be explained by a configuration such that LyC photons can still escape from the low neutral gas covering regions even though the luminosity-weighted neutral gas covering fraction is relatively high. This configuration is also consistent with the fact that the observed LyC of Ion1 is offset from its non-ionizing UV emission (Section \ref{sec:IMAGE}). 

\subsubsection{Comparison of Ion1 with other LyC emitters at intermediate redshifts -- absolute escape fractions}

We caution that, however, all the comparisons done in Section \ref{sec:comp-mor-spec} ignore the fact that the inferred absolute escape fractions ($f_{esc}^{abs}$, see Equation 1 and 2) among individual LyC leakers are different. Since the determination of $f_{esc}^{abs}$ is essentially the ensemble of galaxy intrinsic properties ($E(B-V)$ and $(L_{nLyC}/L_{nLyC})_{int}$) and IGM transmission along a given line-of-sight, any uncertainties on these parameters will affect $f_{esc}^{abs}$ and of course the comparisons between different measurements. Before further discussing indications that can be made according to different $f_{esc}^{abs}$ values, we therefore first remind readers of the important systematics which can be introduced by different assumptions made by different authors when measuring $f_{esc}^{abs}$. Uncertainties on the derived galaxy intrinsic properties are mainly introduced by different assumptions of SED fittings, such as adopted SFHs, stellar population synthesis codes, dust attenuation laws etc \citep[e.g.][]{Siana2007,Plat2019}. Also, as already mentioned in Section \ref{sec:SED} and Section \ref{sec:fesc}, whether to include the nebular emission modeling during the fittings or not can easily introduce an uncertainty on Ion1's $f_{esc}^{abs}$ by a factor of few. The IGM transmission depends on sight-lines, and it is impossible to precisely estimate for a single sight-line \citep[e.g.][]{Vanzella2016,Steidel2018}, which is directly translated into the uncertainty of $f_{esc}^{abs}$. In principle, the IGM uncertainties can be significantly reduced by including many sight-lines (i.e. find many LyC emitters at similar redshifts). However, even the averaged IGM transmission at a given redshift can vary depending on assumed IGM models. For example, \citet{Inoue2014} IGM model is more transparent than the model used by \citet{Steidel2018}, which, apart from the IGM absorption, also takes the CGM absorption into account. Although it is still challenging to precisely determine $f_{esc}^{abs}$ of ionizing radiation, it seems, under reasonable assumptions, that the $f_{esc}^{abs}$ of Ion1 ($5\sim11 \%$, see Section \ref{sec:fesc}) is smaller that that of Ion2, Ion3 and Q1549-C25 ($\gtrsim 50\%$, \citealt{Vanzella2016,Vanzella2019,Shapley2016}). In the following, we outline the possible scenarios which can be used to interpret the different $f_{esc}^{abs}$ among different LyC leakers at intermediate redshifts. 

The smaller $f_{esc}^{abs}$ measured for Ion1 than for Ion2, Ion3 and Q1549-C25 could be (partly) due to the spatially varying IGM transmission. It is possible that the ``true'' IGM transmission along Ion1's line-of-sight is lower than the value predicted by \citet{Inoue2014}, i.e. the IGM model we adopted, hence resulting in an under-estimated $f_{esc}^{abs}$. We remind that \citealt{Inoue2014} model only takes the IGM absorption into account, which can under-estimate the IGM transmission if there are foreground absorbers located within small impact parameters from the observer's line-of-sight. In such case, we would expect that the absorption from CGM becomes non-negligible, making the ``true'' IGM transmission smaller than predictions from the IGM-absorption-only models. In fact, as discussed in Section \ref{sec:absorber}, we know there is at least one promising absorber at $z=3.491$ with a $\approx1.8$ arcsec impact parameter from Ion1. We, however, cannot quantitatively estimate how the CGM absorption would affect the inferred Ion1's $f_{esc}^{abs}$ since it depends on sight-lines and requires observational constrains on the \ion{H}{1} column density distribution of CGM out to $z\sim4$. The IGM model built by \citet{Steidel2018} takes both the IGM and CGM absorptions into account, adopting the \ion{H}{1} column density distributions of IGM and CGM obtained by the Keck Baryonic Structure Survey \citep{Rudie2013}. We caution, however, the \ion{H}{1} column density distribution of CGM from \citet{Rudie2013} was built on a sample of galaxies within the redshift range of $2.0\lesssim z\lesssim 2.8$, meaning some extrapolations are necessary to get the CGM absorption at higher redshifts. Regardless of the potential uncertainties of the CGM absorption, we empirically and roughly estimate the effect of the CGM absorption on the IGM transmission by using the Table 12 of \citealt{Steidel2018} to calculate the ratio of $\langle exp(-\tau_{IGM}^{LyC})\rangle/\langle exp(-\tau_{IGM+CGM}^{LyC}) \rangle\approx$ 1.15 at $z=4$. This translates into a factor of 1.15 increase of $f_{esc}^{abs}$ (Equation 1 and 2), which seems too small and indicates the spatially varying IGM transmission itself might not be enough to explain the different $f_{esc}^{abs}$. However, no matter what IGM transmission model is used, the model needs to be ``calibrated'' using the \ion{H}{1} column density distribution derived based on high-redshift quasar absorption lines observations, which usually is the ensemble of several tens of quasar sight-lines and hence by no means enough to sample the all-sky environments. For example, \citet{Mawatari2017} used a sample of galaxies at $3.3\lesssim z \lesssim 3.5$ as background light sources to probe the intervening Ly$\alpha$ absorption, from which they showed that the \ion{H}{1} absorption in the $z = 3.1$ SSA22 proto-cluster region is systematically larger than that in the other two control fields. The LACES survey found that, among their Ly$\alpha$ emitting galaxies selected in the SSA22 field, because the LyC detections and non-detections appear clustered on small scales, the spatially varying \ion{H}{1} might be an important factor that affects the detection of ionizing radiation \citep{Fletcher2019}. It is finally worth pointing out the Sunburst Arc, where {\it one} compact and bright star-forming region is multiply imaged because of strong gravitational lensing and its ionizing radiation is detected at least in 12 images \citep{RiveraThorsen2019}. Interestingly, the derived $f_{esc}^{abs}$ for the 12 images are different by a factor of few, ranging from a few percent to $\sim$ 20 percent (see Figure 2 of \citealt{RiveraThorsen2019}). Because these 12 images have different light travel paths, it indicates that the absorption of ionizing radiation can be very different en route from the emitting region, even on a very small transverse scale ($\lesssim 10$ kpc). It is therefore essentially important to get better constrains on the \ion{H}{1} column density distribution with a much larger sky sampling so that we will be able do the IGM transmission correction for $f_{esc}^{abs}$ on a slight-line-dependent basis, rather than using a ``standard'' averaged IGM model.

The intrinsically different origins of ionizing photons can also result in different $f_{esc}^{abs}$. Although there is no clear evidence of bright AGNs in Ion1, Ion2, Ion3 and Q1549-C25, it is hard to rule out the possibility of faint AGNs, which can affect $f_{esc}^{abs}$ via either contributing to entire/fractional ionizing photons, or making ISM more transparent through AGN feedback, or both \citep[e.g.][]{Trebitsch2018}. It is commonly assumed that $f_{esc}^{abs}$ of an AGN is 100\% \citep[e.g.][]{Giallongo2015}, the higher $f_{esc}^{abs}$ seen in Ion2, Ion3 and Q1549-C25 than Ion1 therefore might indicate the faint AGN or, a mixture of stellar and faint AGN nature of the LyC emission from Ion2, Ion3 and Q1549-C25 while a more ``pure'' stellar origin of Ion1's LyC. It is also possible that {\it previous} AGN activities in Ion2, Ion3 and Q1549-C25 make their ISM more transparent (relative to Ion1) with lower \ion{H}{1} column densities along the LyC leaking channels, resulting in higher $f_{esc}^{abs}$. We caution that, although a 100\% $f_{esc}^{abs}(AGN)$ is consistent with observations of bright quasars \citep[e.g.][]{Cristiani2016}, our current knowledge on $f_{esc}^{abs}$ of faint and highly-obscured AGN populations is still very scant. In order to quantitatively answer whether or not, or to what extend, the faint AGN nature can explain different $f_{esc}^{abs}$ among the LyC emitters, we need better observational constrains on $f_{esc}^{abs}$ of faint AGN populations and, better theoretical knowledge on how the faint AGN activities affect ISM transparency. 

We finally mention the possibility of time varying escaping ionizing radiation to explain the different $f_{esc}^{abs}$ among the LyC emitters. The LyC leakage is stochastic with a number of physical processes internal and external to galaxies, such as accretion from cosmic webs, intensity of star formation activities and feedback effects on ISM. It has been widely shown by simulations that $f_{esc}^{abs}$ is not a fixed value, instead, it is fluctuating with time and correlated with the energy and momentum input from star formation, in particular supernovae explosions \citep[e.g.][]{Kimm2014,Ma2015,Paardekooper2015,Trebitsch2017}. We therefore would expect time-varying $f_{esc}^{abs}$ and, galaxies like Ion1 are observed during the relative quiescent phase of leaking ionizing radiation while Ion2, Ion3 and Q1549-C25-like galaxies are observed during the relatively active phase with higher $f_{esc}^{abs}$.

\subsection{The LyC Emission of the Stacked LBGs} \label{sec:stack}

\begin{figure*}
\gridline{\fig{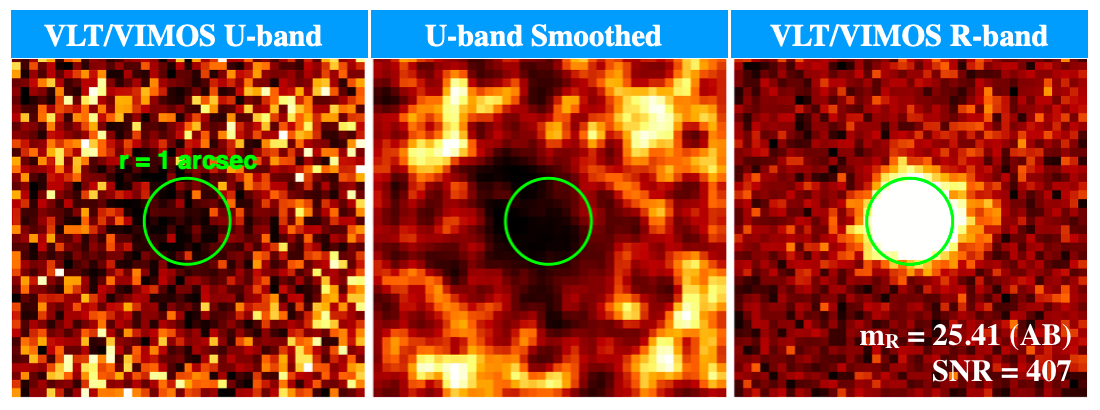}{0.87\textwidth}{}
        }
\caption{Inverse-variance stacked VLT VIMOS/U-band and R-band images of a sample of carefully selected 107 star-forming galaxies (see Section \ref{sec:stack} for details) spectroscopically confirmed at redshift z $\ge$ 3.4, i.e. such that U-band images probe the rest-frame LyC emission.}\label{fig:stack}
\end{figure*}

Apart from individual detections, it is important to constrain the LyC emission property of larger samples of galaxies. The deep {VLT/VIMOS} U-band image in the GOODS-S, which reaches 1$\sigma$ flux limit of the $\approx$ 29.8 m$\rm{_{AB}}$ for point sources \citep{Nonino2009}, offers us an opportunity to investigate the ionizing radiation from star-forming galaxies at z $>$ 3.4, i.e. such that U-band probes the rest-frame ionizing LyC. We have collected galaxy samples at $3.4\le z \le 4.0$ from spectroscopical surveys conducted in the GOODS-S field, including the VLT/FORS2 spectroscopy \citep{Vanzella2005,Vanzella2006,Vanzella2008}, VLT/VIMOS spectroscopy \citep{Popesso2009,Balestra2010} and VANDELS \citep{McLure2018,Pentericci2018}. We started with 190 LBGs whose redshifts have been securely derived based on multiple absorption/emission features, and the redshift solutions are $> 95\%$ reliable (see details in the survey papers as mentioned above). We visually checked the {\it HST} z$_{850}$ and H$_{160}$ images and excluded the galaxies which have low redshift interlopers within their U-band 1 arcsec vicinities. We also checked the VLT/U-band image and excluded the galaxies which are at the positions where we can clearly see contamination from adjacent objects' isophots. We further checked the 7 Ms {\it Chandra} X-ray data \citep{Luo2017} and excluded the galaxies which are identified as AGNs. The final culled sample contains 107 star-forming galaxies with redshift in the range $3.403\le z_{spec}\le 3.951$. Before stacking, we first ran the {\sc SExtractor} \citep{Bertin1996} to generate the U-band segmentation map with a 1.5$\sigma$ detection threshold, and verified that none of the 107 galaxies were detected. We cut 10 arcsec $\times$ 10 arcsec U-band images centered on each galaxy, masked out all the nearby detected objects and then stacked the images with inverse-variance weights. We noticed that, however, even we have used {\sc SExtractor} to mask out all detected sources, there remains some sources which were not successfully detected by {\sc SExtractor} due to either they are below the 1.5 $\sigma$ detection limit or very extended (see the stacked U-band image in Figure \ref{fig:stack}). We used the same procedure to stack the VLT/VIMOS R-band images. 

Figure \ref{fig:stack} shows the stacked images, where the U- and R-band probe the rest-frame $\lambda\lambda$ 730--860\AA\ and $\lambda\lambda$ 1260--1540\AA\ respectively at $\bar{z} = 3.6$, the mean redshift of the stacked galaxies. The stacked source in the R-band has $m_R = 25.41$ (r = 1 arcsec aperture) with SNR $=$ 407. No LyC emission is detected at the position of the stacked U-band image, resulting in a 1$\sigma$ upper limit of $\approx$ 32.5 m$\rm{_{AB}}$ LyC emission (r = 1 arcsec aperture), or a 31.3 m$\rm{_{AB}}$ 3$\sigma$ upper limit. This flux level is too faint for current surveys, revealing the likely reason of the large rate of unsuccessful identifications of ionizing emission from star--forming galaxies at the similar redshift. While IGM transmission of a single sight-line is uncertain by a factor of few \citep[e.g.][]{Vanzella2015,Steidel2018}, ensemble of many sight-lines can significantly reduce the uncertainty. For example, \citet{Steidel2018} showed that, by using their IGM model, uncertainty of the IGM transmission at rest-frame 900\AA\ of ensemble of 120 sight-lines can reduce to only $\approx 5\%$ at $z\sim3$. Similar to \citet{Vanzella2012}, we adopted the \citet{Inoue2014} IGM transmission model and convolved the model with the VIMOS/U-band filter to get the average IGM transmission for our stacked sample, which is $\langle T\rangle = 0.157$ in the redshift interval $3.4 - 3.9$ for the U-band. We then corrected the derived upper limit of the apparent (uncorrected for IGM) stacked LyC emission with this mean IGM transmission, which gave us a 30.5 m$\rm{_{AB}}$ (1$\sigma$, or 29.3 m$\rm{_{AB}}$ for 3$\sigma$) upper limit of emergent ionizing radiation. The emergent ionizing emission is fainter than the apparent LyC emission of Ion1 and the other known LyC emitters at the similar redshift, indicating the successful detections of LyC emission from these galaxies are not due to the variation of the foreground transmission. Instead, the detected LyC emitters are likely the sources which are at the bright-end of the LyC luminosity function.

The stacked U- and R-band images gives a 1$\sigma$ lower limit of flux ratio $(F_R/F_U)_{obs} = 660.7$, consistent with the limits derived from the similar deep ($\approx$ 30.2 magnitude) UV imaging for $z>3$ LBGs obtained by the Large Binocular Camera (LBC) at the Large Binocular Telescope \citep{Boutsia2011,Grazian2016,Grazian2017}. In particular, \citet[][hereafter G17]{Grazian2017} stacked 69 star-forming galaxies at z $\sim$ 3.3 in the COSMOS, CANDELS/GOODS-N and EGS fields. The G17 sample is about 0.4 magnitude brighter as ours, with the R-band absolute magnitude being $M_R =  -22.1$ and $M_R^{G17} = -22.5$ respectively. No LyC was detected in the stacked G17 sample, resulting in a 1$\sigma$ upper limit of $(F_R/F_U)_{obs} = 640.2$. 

A further step is to convert $(F_R/F_U)_{obs}$ into the escape fraction ($f_{esc}$), which, however, is non-trivial because it requires both the intrinsic SED of a galaxy and knowledge of the opacity of the foreground IGM \citep[e.g.][]{Madau1995} and CGM \citep[e.g.][]{Rudie2013}. Since the intrinsic far-UV spectrum is not well constrained due to the dust attenuation, the relative escape fraction ($f_{esc}^{rel}$) instead is usually derived from observations \citep{Steidel2001}. The correction for the IGM opacity is even more difficult in the sense that the transmission depends on the sight-line and can vary by a factor of few for a single sight-line \citep[e.g.][]{Inoue2008,Inoue2014,Steidel2018}. Stacking, on one side, can help to reduce the variance of mean transmission as it is the ensemble of many sight-lines. On the other side, since the UV imaging probes different rest-frame wavelengths for each individual stacked galaxy and again, their intrinsic far-UV spectra is not a priori, stacking introduces the extra systematics of the $f_{esc}^{rel}$ measurement. G17 adopted the intrinsic $(L_{1500}/L_{900})_{int} = 3$ and the mean IGM transmission $\langle T\rangle = 0.28$ at $z\sim3.3$ to get the upper limit of $f_{esc}^{rel}|_{z\sim3.3} \le 1.7\%$. Assuming the same $(L_{1500}/L_{900})_{int} = 3$ and $\langle T\rangle\approx 0.157$ for the stacked sample, we derived the upper limit of relative escape fraction is $f_{esc}^{rel}|_{z\sim3.6} \le 2.9\%$. If we assume dust attenuation of \citet{Calzetti2000} and adopt the mean $\langle E(B-V)\rangle\approx0.16$ of our stacked galaxies, we can convert the $f_{esc}^{rel}$ to $f_{esc}^{abs}$ through
\begin{equation}
f_{esc}^{abs} = f_{esc}^{rel}/10^{0.4A_{1500}}= f_{esc}^{rel}/4.59 \le 0.63\%.
\end{equation}

Another approach to empirically constrain the mean $f_{esc}^{abs}$ over a broad redshift range, first adopted by \citet{Chen2007}, is through the far-UV spectroscopic observations of long duration gamma-ray burst (GRB) afterglows, namely fit their Ly$\alpha$ absorption features to directly measure the line-of-sight $N_{\rm{H I}}$ and convert it to $f_{esc}^{abs}$. An individual detection of the escaping LyC has been reported from a GRB host galaxy at z = 3.35 by \citet{Fynbo2009}. Very recently, \citet{Tanvir2019} compiled 140 GRBs in the redshift range $1.6<z<6.7$ to derive a mean escape fraction $\langle f_{esc}^{abs}\rangle \approx 0.5\%$, without clear redshift evolution between $z\sim2$ to $z\sim5$, consistent with the $f_{esc}^{abs}$ upper limit derived from our stacked LBGs. Despite this method is very promising in deriving $N_{\rm{HI}}$ at high redshift, the systematics related to the GRB progenitors and the regions where they explode still need to be carefully investigated with larger samples of high-redshift GRBs \citep{Cen2014}.

\section{Outlook for future surveys}\label{sec:outlook}

The identification and characterization of the sources responsible for reionizing most of the hydrogen at the EoR stand as one of the primary scientific goals of future astronomical surveys, e.g. JWST, TMT, ELT. Because of the increasing opacity of IGM absorption, it will be impossible to direct detect the ionizing LyC at EoR. Particular attention and efforts therefore have been put on investigating the indirect diagnostics of LyC emitters, such as large Ly$\alpha$ equivalent wavelength and its profile (e.g. \citealt{Verhamme2015,Verhamme2017,Dijkstra2016,Marchi2018,Steidel2018}), the large line ratio of [O III]/[O II] (e.g. \citealt{Jaskot2013,Nakajima2014,Schaerer2016,Rutkowski2017,Izotov2018,Naidu2018}), weak low-ionization absorption lines (e.g. \citealt{Chisholm2018}). These indirect indicators, however, remain poorly quantified. The direct detection of LyC from Ion1, a galaxy with strong interstellar low-ionization absorption and the absence of Ly$\alpha$, immediately suggests that we have not yet fully characterized the dispersion of properties of LyC emitters. Moreover, the spectroscopic confirmation of Ion1-like galaxies without strong emission lines at EoR will be very difficult as requiring the measurements of the continuum break. These will make the identification of such sources and the quantification of their relative contribution to the EoR challenging, making the detections of more Ion1-like LyC emitters and obtain statistical constrains on their escaping ionizing LyC photons important.

\section{Summary}

We report on the direct detection of the LyC emission from Ion1 by the {\it HST} WFC3 in the F410M bandpass with peak SNR $=$ 4.17 in a r $=$ 0.12 arcsec aperture. The LyC is also detected by the VLT/VIMOS in the U-band with peak SNR $=$ 6.7 in a r $=$ 0.6 arcsec aperture.  The LyC emission is $m_{410}=27.60\pm0.36$ (AB) in the F410M and $m_U = 27.84\pm0.19$ (AB) in the U-band. Ion1 is not detected in the 7 Ms {\it Chandra} image and the high angular resolution {\it HST}/F410M observation shows the peak of the LyC emission is offset from the center of its non-ionizing UV radiation (probed by {\it HST}/F606W) by 0.12$\pm$0.03 arcsec, corresponding to 0.85$\pm$0.21 kpc physical scale. These suggest the origin of the LyC emission is not AGN. 

The redshift of the galaxy is measured to be $\rm{z = 3.794}$ using the \ion{C}{3}]$\lambda$1909 emission feature from a deep 20-hr spectrum with R$\sim$580 obtained during the VANDELS survey. The spectrum reveals low- and high- ionization interstellar metal absorption lines (\ion{C}{2}, \ion{O}{1}, \ion{N}{5} \ion{Si}{2}, \ion{Si}{4}), the P-Cyngi profile of \ion{C}{4} and Ly$\alpha$ in absorption. This is different from the typical far-UV spectrum of the known LyC emitters at the similar redshift, which shows strong Ly$\alpha$ in emission, suggesting that we have not yet fully characterized the dispersion of properties of LyC emitters. An absorption feature also has been revealed in the 2D spectrum at $\lambda \approx$ 5460 \AA\ and, at the same wavelength but offset upward by $\approx$ 1.8 arcsec from Ion1's spectrum, there is an emission line with no obvious detected continuum. A visual inspection of the {\it HST}/ACS F606W image suggests that the source of the emission very likely is an intervening blue galaxy which has already been identified in the CANDELS catalog. If the emission is Ly$\alpha$, as strongly suggested by the simultaneous absorption and emission at the source wavelength, then the intervening galaxy is at z $=$ 3.491, consistent with its z$_{phot}$ as reported by the CANDELS team. This makes Ion1 a very interesting case such that the escaping LyC is still detectable even though there exists an intervening absorber. 

A GALFIT analysis of Ion1's morphologies in the F435W ($\lambda\lambda$ 750-1020\AA), F606W ($\lambda\lambda$ 960-1500\AA) and F160W ($\lambda\lambda$ 2920-3550\AA) bandpasses suggests that 1) the disc-like morphology in the F606W image; 2) the non-ionizing UV radiation probed by F606W is compact with $R_e = 498\pm26$ pc, suggesting the radiation emerges from compact but moderately resolved star-forming regions; 3) $R_e$ is significantly larger in the F435W than in the F606W. By doing the image subtraction between the F435W and F606W, we find evidence that the detected LyC emission escapes from parts of the galaxy with low \ion{H}{1} absorption. Although the marginal detection (SNR $=$ 2.56 in a r $=$ 0.36 arcsec aperture) in the F410M prevents us from quantitatively studying Ion1's LyC morphology, Monte Carlo simulations based on the shape of photometric curve of growth suggests that Ion1's LyC is likely moderately resolved. The corresponding effective radius, $R_e\ge240$ pc, suggests the LyC emitting regions are young star-forming complexes, i.e. spatially resolved star clusters and associations. The ionizing photons can only escape from the ones with low \ion{H}{1} column density ``cavities'' in the surrounding ISM. The detected LyC photons are those escaped from the sight-lines with low intervening absorption.

We finally investigate the LyC emission properties of star-forming galaxies at z$\sim$3.6 by inverse-variance stacking a sample of 107 galaxies at $3.40\lesssim z_{spec}\lesssim3.95$ using a deep VLT/VIMOUS U-band image. No LyC emission is detected in the stacked image, resulting in a 1$\sigma$ flux limit of 32.5 magnitude and an upper limit of absolute escape fraction $f_{esc}^{abs} \le  0.63\%$. This flux level is too faint for current surveys, revealing the likely reason of the large rate of unsuccessful identifications of ionizing emission from star--forming galaxies at the similar redshift.The detected LyC emitters like Ion1 are likely at the bright-end of the LyC luminosity function.

\section{Acknowledgement}
We thank the anonymous referee for the careful review of our work. AJ acknowledges support by NASA through Hubble Fellowship grant HST-HF2-51392. The Cosmic Dawn center is funded by the DNRF.

\bibliography{ji_2018_Ion1}

\begin{thebibliography}{}
\expandafter\ifx\csname natexlab\endcsname\relax\def\natexlab#1{#1}\fi
\providecommand{\url}[1]{\href{#1}{#1}}

\bibitem[{{Adamo} {et~al.}(2011){Adamo}, {{\"O}stlin}, \&
  {Zackrisson}}]{Adamo2011}
{Adamo}, A., {{\"O}stlin}, G., \& {Zackrisson}, E. 2011, \mnras, 417, 1904

\bibitem[{{Alexandroff} {et~al.}(2015){Alexandroff}, {Heckman}, {Borthakur},
  {Overzier}, \& {Leitherer}}]{Alexandroff2015}
{Alexandroff}, R.~M., {Heckman}, T.~M., {Borthakur}, S., {Overzier}, R., \&
  {Leitherer}, C. 2015, \apj, 810, 104

\bibitem[{{Amor{\'\i}n} {et~al.}(2017){Amor{\'\i}n}, {Fontana},
  {P{\'e}rez-Montero}, {Castellano}, {Guaita}, {Grazian}, {Le F{\`e}vre},
  {Ribeiro}, {Schaerer}, {Tasca}, {Thomas}, {Bardelli}, {Cassar{\`a}},
  {Cassata}, {Cimatti}, {Contini}, {de Barros}, {Garilli}, {Giavalisco},
  {Hathi}, {Koekemoer}, {Le Brun}, {Lemaux}, {Maccagni}, {Pentericci}, {Pforr},
  {Talia}, {Tresse}, {Vanzella}, {Vergani}, {Zamorani}, {Zucca}, \&
  {Merlin}}]{Amorin2017}
{Amor{\'\i}n}, R., {Fontana}, A., {P{\'e}rez-Montero}, E., {et~al.} 2017,
  Nature Astronomy, 1, 0052

\bibitem[{{Atek} {et~al.}(2015){Atek}, {Richard}, {Jauzac}, {Kneib},
  {Natarajan}, {Limousin}, {Schaerer}, {Jullo}, {Ebeling}, {Egami}, \&
  {Clement}}]{Atek2015}
{Atek}, H., {Richard}, J., {Jauzac}, M., {et~al.} 2015, \apj, 814, 69

\bibitem[{{Balestra} {et~al.}(2010){Balestra}, {Mainieri}, {Popesso},
  {Dickinson}, {Nonino}, {Rosati}, {Teimoorinia}, {Vanzella}, {Cristiani},
  {Cesarsky}, {Fosbury}, {Kuntschner}, \& {Rettura}}]{Balestra2010}
{Balestra}, I., {Mainieri}, V., {Popesso}, P., {et~al.} 2010, \aap, 512, A12

\bibitem[{{Battisti} {et~al.}(2016){Battisti}, {Calzetti}, \&
  {Chary}}]{Battisti2016}
{Battisti}, A.~J., {Calzetti}, D., \& {Chary}, R.-R. 2016, \apj, 818, 13

\bibitem[{{Bergvall} {et~al.}(2006){Bergvall}, {Zackrisson}, {Andersson},
  {Arnberg}, {Masegosa}, \& {{\"O}stlin}}]{Bergvall2006}
{Bergvall}, N., {Zackrisson}, E., {Andersson}, B.-G., {et~al.} 2006, \aap, 448,
  513

\bibitem[{{Bertin} \& {Arnouts}(1996)}]{Bertin1996}
{Bertin}, E., \& {Arnouts}, S. 1996, \aaps, 117, 393

\bibitem[{{Bian} {et~al.}(2017){Bian}, {Fan}, {McGreer}, {Cai}, \&
  {Jiang}}]{Bian2017}
{Bian}, F., {Fan}, X., {McGreer}, I., {Cai}, Z., \& {Jiang}, L. 2017, \apjl,
  837, L12

\bibitem[{{Biretta} \& {Baggett}(2013)}]{Biretta2013}
{Biretta}, J., \& {Baggett}, S. 2013, {WFC3 Post-Flash Calibration}, Tech. rep.

\bibitem[{{Borthakur} {et~al.}(2014){Borthakur}, {Heckman}, {Leitherer}, \&
  {Overzier}}]{Borthakur2014}
{Borthakur}, S., {Heckman}, T.~M., {Leitherer}, C., \& {Overzier}, R.~A. 2014,
  Science, 346, 216

\bibitem[{{Boutsia} {et~al.}(2011){Boutsia}, {Grazian}, {Giallongo}, {Fontana},
  {Pentericci}, {Castellano}, {Zamorani}, {Mignoli}, {Vanzella}, \&
  {Fiore}}]{Boutsia2011}
{Boutsia}, K., {Grazian}, A., {Giallongo}, E., {et~al.} 2011, \apj, 736, 41

\bibitem[{{Bouwens} {et~al.}(2015){Bouwens}, {Illingworth}, {Oesch}, {Caruana},
  {Holwerda}, {Smit}, \& {Wilkins}}]{Bouwens2015}
{Bouwens}, R.~J., {Illingworth}, G.~D., {Oesch}, P.~A., {et~al.} 2015, \apj,
  811, 140

\bibitem[{{Byler} {et~al.}(2017){Byler}, {Dalcanton}, {Conroy}, \&
  {Johnson}}]{Byler2017}
{Byler}, N., {Dalcanton}, J.~J., {Conroy}, C., \& {Johnson}, B.~D. 2017, \apj,
  840, 44

\bibitem[{{Calura} {et~al.}(2015){Calura}, {Few}, {Romano}, \&
  {D'Ercole}}]{Calura2015}
{Calura}, F., {Few}, C.~G., {Romano}, D., \& {D'Ercole}, A. 2015, \apjl, 814,
  L14

\bibitem[{{Calzetti} {et~al.}(2000){Calzetti}, {Armus}, {Bohlin}, {Kinney},
  {Koornneef}, \& {Storchi-Bergmann}}]{Calzetti2000}
{Calzetti}, D., {Armus}, L., {Bohlin}, R.~C., {et~al.} 2000, \apj, 533, 682

\bibitem[{{Calzetti} {et~al.}(1994){Calzetti}, {Kinney}, \&
  {Storchi-Bergmann}}]{Calzetti1994}
{Calzetti}, D., {Kinney}, A.~L., \& {Storchi-Bergmann}, T. 1994, \apj, 429, 582

\bibitem[{{Castellano} {et~al.}(2016){Castellano}, {Yue}, {Ferrara}, {Merlin},
  {Fontana}, {Amor{\'\i}n}, {Grazian}, {M{\'a}rmol-Queralto}, {Micha{\l}owski},
  {Mortlock}, {Paris}, {Parsa}, {Pilo}, \& {Santini}}]{Castellano2016}
{Castellano}, M., {Yue}, B., {Ferrara}, A., {et~al.} 2016, \apjl, 823, L40

\bibitem[{{Cen} \& {Kimm}(2014)}]{Cen2014}
{Cen}, R., \& {Kimm}, T. 2014, \apj, 794, 50

\bibitem[{{Chen} {et~al.}(2007){Chen}, {Prochaska}, \& {Gnedin}}]{Chen2007}
{Chen}, H.-W., {Prochaska}, J.~X., \& {Gnedin}, N.~Y. 2007, \apj, 667, L125

\bibitem[{{Chisholm} {et~al.}(2019){Chisholm}, {Rigby}, {Bayliss}, {Berg},
  {Dahle}, {Gladders}, \& {Sharon}}]{Chisholm2019}
{Chisholm}, J., {Rigby}, J.~R., {Bayliss}, M., {et~al.} 2019, arXiv e-prints,
  arXiv:1905.04314

\bibitem[{{Chisholm} {et~al.}(2018){Chisholm}, {Gazagnes}, {Schaerer},
  {Verhamme}, {Rigby}, {Bayliss}, {Sharon}, {Gladders}, \&
  {Dahle}}]{Chisholm2018}
{Chisholm}, J., {Gazagnes}, S., {Schaerer}, D., {et~al.} 2018, \aap, 616, A30

\bibitem[{{Conroy} \& {Gunn}(2010)}]{Conroy2010}
{Conroy}, C., \& {Gunn}, J.~E. 2010, {FSPS: Flexible Stellar Population
  Synthesis}, , , ascl:1010.043

\bibitem[{{Conroy} {et~al.}(2009){Conroy}, {Gunn}, \& {White}}]{Conroy2009}
{Conroy}, C., {Gunn}, J.~E., \& {White}, M. 2009, \apj, 699, 486

\bibitem[{{Cowie} {et~al.}(2009){Cowie}, {Barger}, \& {Trouille}}]{Cowie2009}
{Cowie}, L.~L., {Barger}, A.~J., \& {Trouille}, L. 2009, \apj, 692, 1476

\bibitem[{{Cristiani} {et~al.}(2016){Cristiani}, {Serrano}, {Fontanot},
  {Vanzella}, \& {Monaco}}]{Cristiani2016}
{Cristiani}, S., {Serrano}, L.~M., {Fontanot}, F., {Vanzella}, E., \& {Monaco},
  P. 2016, \mnras, 462, 2478

\bibitem[{{Dahlen} {et~al.}(2013){Dahlen}, {Mobasher}, {Faber}, {Ferguson},
  {Barro}, {Finkelstein}, {Finlator}, {Fontana}, {Gruetzbauch}, {Johnson},
  {Pforr}, {Salvato}, {Wiklind}, {Wuyts}, {Acquaviva}, {Dickinson}, {Guo},
  {Huang}, {Huang}, {Newman}, {Bell}, {Conselice}, {Galametz}, {Gawiser},
  {Giavalisco}, {Grogin}, {Hathi}, {Kocevski}, {Koekemoer}, {Koo}, {Lee},
  {McGrath}, {Papovich}, {Peth}, {Ryan}, {Somerville}, {Weiner}, \&
  {Wilson}}]{Dahlen2013}
{Dahlen}, T., {Mobasher}, B., {Faber}, S.~M., {et~al.} 2013, \apj, 775, 93

\bibitem[{{de Barros} {et~al.}(2016){de Barros}, {Vanzella}, {Amor{\'{\i}}n},
  {Castellano}, {Siana}, {Grazian}, {Suh}, {Balestra}, {Vignali}, {Verhamme},
  {Zamorani}, {Mignoli}, {Hasinger}, {Comastri}, {Pentericci},
  {P{\'e}rez-Montero}, {Fontana}, {Giavalisco}, \& {Gilli}}]{debarros2016}
{de Barros}, S., {Vanzella}, E., {Amor{\'{\i}}n}, R., {et~al.} 2016, \aap, 585,
  A51

\bibitem[{{Dijkstra} {et~al.}(2016){Dijkstra}, {Gronke}, \&
  {Venkatesan}}]{Dijkstra2016}
{Dijkstra}, M., {Gronke}, M., \& {Venkatesan}, A. 2016, \apj, 828, 71

\bibitem[{{Ferland} {et~al.}(1998){Ferland}, {Korista}, {Verner}, {Ferguson},
  {Kingdon}, \& {Verner}}]{Ferland1998}
{Ferland}, G.~J., {Korista}, K.~T., {Verner}, D.~A., {et~al.} 1998, \pasp, 110,
  761

\bibitem[{{Ferland} {et~al.}(2013){Ferland}, {Porter}, {van Hoof}, {Williams},
  {Abel}, {Lykins}, {Shaw}, {Henney}, \& {Stancil}}]{Ferland2013}
{Ferland}, G.~J., {Porter}, R.~L., {van Hoof}, P.~A.~M., {et~al.} 2013, \rmxaa,
  49, 137

\bibitem[{{Finkelstein} {et~al.}(2019){Finkelstein},
  {D{\textquoteright}Aloisio}, {Paardekooper}, {Ryan}, {Behroozi}, {Finlator},
  {Livermore}, {Upton Sanderbeck}, {Dalla Vecchia}, \&
  {Khochfar}}]{Finkelstein2019}
{Finkelstein}, S.~L., {D{\textquoteright}Aloisio}, A., {Paardekooper}, J.-P.,
  {et~al.} 2019, \apj, 879, 36

\bibitem[{{Fletcher} {et~al.}(2019){Fletcher}, {Tang}, {Robertson}, {Nakajima},
  {Ellis}, {Stark}, \& {Inoue}}]{Fletcher2019}
{Fletcher}, T.~J., {Tang}, M., {Robertson}, B.~E., {et~al.} 2019, \apj, 878, 87

\bibitem[{{Fontana} {et~al.}(2014){Fontana}, {Dunlop}, {Paris}, {Targett},
  {Boutsia}, {Castellano}, {Galametz}, {Grazian}, {McLure}, \&
  {Merlin}}]{Fontana2014}
{Fontana}, A., {Dunlop}, J.~S., {Paris}, D., {et~al.} 2014, \aap, 570, A11

\bibitem[{{Foreman-Mackey} {et~al.}(2013){Foreman-Mackey}, {Hogg}, {Lang}, \&
  {Goodman}}]{emcee}
{Foreman-Mackey}, D., {Hogg}, D.~W., {Lang}, D., \& {Goodman}, J. 2013, PASP,
  125, 306

\bibitem[{{Fynbo} {et~al.}(2009){Fynbo}, {Jakobsson}, {Prochaska}, {Malesani},
  {Ledoux}, {de Ugarte Postigo}, {Nardini}, {Vreeswijk}, {Wiersema}, \&
  {Hjorth}}]{Fynbo2009}
{Fynbo}, J.~P.~U., {Jakobsson}, P., {Prochaska}, J.~X., {et~al.} 2009, \apjs,
  185, 526

\bibitem[{{Gazagnes} {et~al.}(2018){Gazagnes}, {Chisholm}, {Schaerer},
  {Verhamme}, {Rigby}, \& {Bayliss}}]{Gazagnes2018}
{Gazagnes}, S., {Chisholm}, J., {Schaerer}, D., {et~al.} 2018, \aap, 616, A29

\bibitem[{{Georgakakis} {et~al.}(2015){Georgakakis}, {Aird}, {Buchner},
  {Salvato}, {Menzel}, {Brandt}, {McGreer}, {Dwelly}, {Mountrichas}, {Koki},
  {Georgantopoulos}, {Hsu}, {Merloni}, {Liu}, {Nandra}, \&
  {Ross}}]{Georgakakis2015}
{Georgakakis}, A., {Aird}, J., {Buchner}, J., {et~al.} 2015, \mnras, 453, 1946

\bibitem[{{Giallongo} {et~al.}(2015){Giallongo}, {Grazian}, {Fiore}, {Fontana},
  {Pentericci}, {Vanzella}, {Dickinson}, {Kocevski}, {Castellano}, {Cristiani},
  {Ferguson}, {Finkelstein}, {Grogin}, {Hathi}, {Koekemoer}, {Newman}, \&
  {Salvato}}]{Giallongo2015}
{Giallongo}, E., {Grazian}, A., {Fiore}, F., {et~al.} 2015, \aap, 578, A83

\bibitem[{{Giavalisco} {et~al.}(2004){Giavalisco}, {Ferguson}, {Koekemoer},
  {Dickinson}, {Alexander}, {Bauer}, {Bergeron}, {Biagetti}, {Brandt},
  {Casertano}, {Cesarsky}, {Chatzichristou}, {Conselice}, {Cristiani}, {Da
  Costa}, {Dahlen}, {de Mello}, {Eisenhardt}, {Erben}, {Fall}, {Fassnacht},
  {Fosbury}, {Fruchter}, {Gardner}, {Grogin}, {Hook}, {Hornschemeier}, {Idzi},
  {Jogee}, {Kretchmer}, {Laidler}, {Lee}, {Livio}, {Lucas}, {Madau},
  {Mobasher}, {Moustakas}, {Nonino}, {Padovani}, {Papovich}, {Park},
  {Ravindranath}, {Renzini}, {Richardson}, {Riess}, {Rosati}, {Schirmer},
  {Schreier}, {Somerville}, {Spinrad}, {Stern}, {Stiavelli}, {Strolger},
  {Urry}, {Vandame}, {Williams}, \& {Wolf}}]{Giavalisco2004}
{Giavalisco}, M., {Ferguson}, H.~C., {Koekemoer}, A.~M., {et~al.} 2004, \apjl,
  600, L93

\bibitem[{{Gonzaga} \& {et al.}(2012)}]{Gonzaga2012}
{Gonzaga}, S., \& {et al.} 2012, {The DrizzlePac Handbook}

\bibitem[{{Grasha} {et~al.}(2017){Grasha}, {Calzetti}, {Adamo}, {Kim},
  {Elmegreen}, {Gouliermis}, {Dale}, {Fumagalli}, {Grebel}, {Johnson}, {Kahre},
  {Kennicutt}, {Messa}, {Pellerin}, {Ryon}, {Smith}, {Shabani}, {Thilker}, \&
  {Ubeda}}]{Grasha2017}
{Grasha}, K., {Calzetti}, D., {Adamo}, A., {et~al.} 2017, \apj, 840, 113

\bibitem[{{Grazian} {et~al.}(2016){Grazian}, {Giallongo}, {Gerbasi}, {Fiore},
  {Fontana}, {Le F{\`e}vre}, {Pentericci}, {Vanzella}, {Zamorani}, \&
  {Cassata}}]{Grazian2016}
{Grazian}, A., {Giallongo}, E., {Gerbasi}, R., {et~al.} 2016, \aap, 585, A48

\bibitem[{{Grazian} {et~al.}(2017){Grazian}, {Giallongo}, {Paris}, {Boutsia},
  {Dickinson}, {Santini}, {Windhorst}, {Jansen}, {Cohen}, \&
  {Ashcraft}}]{Grazian2017}
{Grazian}, A., {Giallongo}, E., {Paris}, D., {et~al.} 2017, \aap, 602, A18

\bibitem[{{Grimes} {et~al.}(2007){Grimes}, {Heckman}, {Strickland}, {Dixon},
  {Sembach}, {Overzier}, {Hoopes}, {Aloisi}, \& {Ptak}}]{Grimes2007}
{Grimes}, J.~P., {Heckman}, T., {Strickland}, D., {et~al.} 2007, \apj, 668, 891

\bibitem[{{Grogin} {et~al.}(2011){Grogin}, {Kocevski}, {Faber}, {Ferguson},
  {Koekemoer}, {Riess}, {Acquaviva}, {Alexander}, {Almaini}, {Ashby}, {Barden},
  {Bell}, {Bournaud}, {Brown}, {Caputi}, {Casertano}, {Cassata}, {Castellano},
  {Challis}, {Chary}, {Cheung}, {Cirasuolo}, {Conselice}, {Roshan Cooray},
  {Croton}, {Daddi}, {Dahlen}, {Dav{\'e}}, {de Mello}, {Dekel}, {Dickinson},
  {Dolch}, {Donley}, {Dunlop}, {Dutton}, {Elbaz}, {Fazio}, {Filippenko},
  {Finkelstein}, {Fontana}, {Gardner}, {Garnavich}, {Gawiser}, {Giavalisco},
  {Grazian}, {Guo}, {Hathi}, {H{\"a}ussler}, {Hopkins}, {Huang}, {Huang},
  {Jha}, {Kartaltepe}, {Kirshner}, {Koo}, {Lai}, {Lee}, {Li}, {Lotz}, {Lucas},
  {Madau}, {McCarthy}, {McGrath}, {McIntosh}, {McLure}, {Mobasher},
  {Moustakas}, {Mozena}, {Nandra}, {Newman}, {Niemi}, {Noeske}, {Papovich},
  {Pentericci}, {Pope}, {Primack}, {Rajan}, {Ravindranath}, {Reddy}, {Renzini},
  {Rix}, {Robaina}, {Rodney}, {Rosario}, {Rosati}, {Salimbeni}, {Scarlata},
  {Siana}, {Simard}, {Smidt}, {Somerville}, {Spinrad}, {Straughn}, {Strolger},
  {Telford}, {Teplitz}, {Trump}, {van der Wel}, {Villforth}, {Wechsler},
  {Weiner}, {Wiklind}, {Wild}, {Wilson}, {Wuyts}, {Yan}, \& {Yun}}]{Gorgin2011}
{Grogin}, N.~A., {Kocevski}, D.~D., {Faber}, S.~M., {et~al.} 2011, \apjs, 197,
  35

\bibitem[{{Guo} {et~al.}(2013){Guo}, {Ferguson}, {Giavalisco}, {Barro},
  {Willner}, {Ashby}, {Dahlen}, {Donley}, {Faber}, {Fontana}, {Galametz},
  {Grazian}, {Huang}, {Kocevski}, {Koekemoer}, {Koo}, {McGrath}, {Peth},
  {Salvato}, {Wuyts}, {Castellano}, {Cooray}, {Dickinson}, {Dunlop}, {Fazio},
  {Gardner}, {Gawiser}, {Grogin}, {Hathi}, {Hsu}, {Lee}, {Lucas}, {Mobasher},
  {Nandra}, {Newman}, \& {van der Wel}}]{Guo2013}
{Guo}, Y., {Ferguson}, H.~C., {Giavalisco}, M., {et~al.} 2013, \apjs, 207, 24

\bibitem[{{Heckman} {et~al.}(2011){Heckman}, {Borthakur}, {Overzier},
  {Kauffmann}, {Basu-Zych}, {Leitherer}, {Sembach}, {Martin}, {Rich},
  {Schiminovich}, \& {Seibert}}]{Heckman2011}
{Heckman}, T.~M., {Borthakur}, S., {Overzier}, R., {et~al.} 2011, \apj, 730, 5

\bibitem[{{Howell}(1989)}]{Howell1989}
{Howell}, S.~B. 1989, \pasp, 101, 616

\bibitem[{{Inoue}(2010)}]{Inoue2010}
{Inoue}, A.~K. 2010, \mnras, 401, 1325

\bibitem[{{Inoue} \& {Iwata}(2008)}]{Inoue2008}
{Inoue}, A.~K., \& {Iwata}, I. 2008, \mnras, 387, 1681

\bibitem[{{Inoue} {et~al.}(2014){Inoue}, {Shimizu}, {Iwata}, \&
  {Tanaka}}]{Inoue2014}
{Inoue}, A.~K., {Shimizu}, I., {Iwata}, I., \& {Tanaka}, M. 2014, \mnras, 442,
  1805

\bibitem[{{Izotov} {et~al.}(2016{\natexlab{a}}){Izotov}, {Orlitov{\'a}},
  {Schaerer}, {Thuan}, {Verhamme}, {Guseva}, \& {Worseck}}]{Izotov2016}
{Izotov}, Y.~I., {Orlitov{\'a}}, I., {Schaerer}, D., {et~al.}
  2016{\natexlab{a}}, \nat, 529, 178

\bibitem[{{Izotov} {et~al.}(2016{\natexlab{b}}){Izotov}, {Schaerer}, {Thuan},
  {Worseck}, {Guseva}, {Orlitov{\'a}}, \& {Verhamme}}]{Izotov2016b}
{Izotov}, Y.~I., {Schaerer}, D., {Thuan}, T.~X., {et~al.} 2016{\natexlab{b}},
  \mnras, 461, 3683

\bibitem[{{Izotov} {et~al.}(2018{\natexlab{a}}){Izotov}, {Schaerer}, {Worseck},
  {Guseva}, {Thuan}, {Verhamme}, {Orlitov{\'a}}, \& {Fricke}}]{Izotov2018b}
{Izotov}, Y.~I., {Schaerer}, D., {Worseck}, G., {et~al.} 2018{\natexlab{a}},
  \mnras, 474, 4514

\bibitem[{{Izotov} {et~al.}(2018{\natexlab{b}}){Izotov}, {Worseck}, {Schaerer},
  {Guseva}, {Thuan}, {Fricke}, \& {Orlitov{\'a}}}]{Izotov2018}
{Izotov}, Y.~I., {Worseck}, G., {Schaerer}, D., {et~al.} 2018{\natexlab{b}},
  \mnras, 478, 4851

\bibitem[{{Jaskot} {et~al.}(2019){Jaskot}, {Dowd}, {Oey}, {Scarlata}, \&
  {McKinney}}]{Jaskot2019}
{Jaskot}, A.~E., {Dowd}, T., {Oey}, M.~S., {Scarlata}, C., \& {McKinney}, J.
  2019, arXiv e-prints, arXiv:1908.09763

\bibitem[{{Jaskot} \& {Oey}(2013)}]{Jaskot2013}
{Jaskot}, A.~E., \& {Oey}, M.~S. 2013, \apj, 766, 91

\bibitem[{{Jaskot} {et~al.}(2017){Jaskot}, {Oey}, {Scarlata}, \&
  {Dowd}}]{Jaskot2017}
{Jaskot}, A.~E., {Oey}, M.~S., {Scarlata}, C., \& {Dowd}, T. 2017, \apjl, 851,
  L9

\bibitem[{{Jaskot} \& {Ravindranath}(2016)}]{Jaskot2016}
{Jaskot}, A.~E., \& {Ravindranath}, S. 2016, \apj, 833, 136

\bibitem[{{Keenan} {et~al.}(2017){Keenan}, {Oey}, {Jaskot}, \&
  {James}}]{Keenan2017}
{Keenan}, R.~P., {Oey}, M.~S., {Jaskot}, A.~E., \& {James}, B.~L. 2017, \apj,
  848, 12

\bibitem[{{Kimm} \& {Cen}(2014)}]{Kimm2014}
{Kimm}, T., \& {Cen}, R. 2014, \apj, 788, 121

\bibitem[{{Koekemoer} {et~al.}(2011){Koekemoer}, {Faber}, {Ferguson}, {Grogin},
  {Kocevski}, {Koo}, {Lai}, {Lotz}, {Lucas}, {McGrath}, {Ogaz}, {Rajan},
  {Riess}, {Rodney}, {Strolger}, {Casertano}, {Castellano}, {Dahlen},
  {Dickinson}, {Dolch}, {Fontana}, {Giavalisco}, {Grazian}, {Guo}, {Hathi},
  {Huang}, {van der Wel}, {Yan}, {Acquaviva}, {Alexander}, {Almaini}, {Ashby},
  {Barden}, {Bell}, {Bournaud}, {Brown}, {Caputi}, {Cassata}, {Challis},
  {Chary}, {Cheung}, {Cirasuolo}, {Conselice}, {Roshan Cooray}, {Croton},
  {Daddi}, {Dav{\'e}}, {de Mello}, {de Ravel}, {Dekel}, {Donley}, {Dunlop},
  {Dutton}, {Elbaz}, {Fazio}, {Filippenko}, {Finkelstein}, {Frazer}, {Gardner},
  {Garnavich}, {Gawiser}, {Gruetzbauch}, {Hartley}, {H{\"a}ussler},
  {Herrington}, {Hopkins}, {Huang}, {Jha}, {Johnson}, {Kartaltepe},
  {Khostovan}, {Kirshner}, {Lani}, {Lee}, {Li}, {Madau}, {McCarthy},
  {McIntosh}, {McLure}, {McPartland}, {Mobasher}, {Moreira}, {Mortlock},
  {Moustakas}, {Mozena}, {Nandra}, {Newman}, {Nielsen}, {Niemi}, {Noeske},
  {Papovich}, {Pentericci}, {Pope}, {Primack}, {Ravindranath}, {Reddy},
  {Renzini}, {Rix}, {Robaina}, {Rosario}, {Rosati}, {Salimbeni}, {Scarlata},
  {Siana}, {Simard}, {Smidt}, {Snyder}, {Somerville}, {Spinrad}, {Straughn},
  {Telford}, {Teplitz}, {Trump}, {Vargas}, {Villforth}, {Wagner}, {Wandro},
  {Wechsler}, {Weiner}, {Wiklind}, {Wild}, {Wilson}, {Wuyts}, \&
  {Yun}}]{Koekemoer2011}
{Koekemoer}, A.~M., {Faber}, S.~M., {Ferguson}, H.~C., {et~al.} 2011, \apjs,
  197, 36

\bibitem[{{Kroupa}(2001)}]{Kroupa2001}
{Kroupa}, P. 2001, \mnras, 322, 231

\bibitem[{{Lee} {et~al.}(2018){Lee}, {Giavalisco}, {Whitaker}, {Williams},
  {Ferguson}, {Acquaviva}, {Koekemoer}, {Straughn}, {Guo}, {Kartaltepe},
  {Lotz}, {Pacifici}, {Croton}, {Somerville}, \& {Lu}}]{Lee2018}
{Lee}, B., {Giavalisco}, M., {Whitaker}, K., {et~al.} 2018, \apj, 853, 131

\bibitem[{{Leitet} {et~al.}(2013){Leitet}, {Bergvall}, {Hayes}, {Linn{\'e}}, \&
  {Zackrisson}}]{Leitet2013}
{Leitet}, E., {Bergvall}, N., {Hayes}, M., {Linn{\'e}}, S., \& {Zackrisson}, E.
  2013, \aap, 553, A106

\bibitem[{{Leitet} {et~al.}(2011){Leitet}, {Bergvall}, {Piskunov}, \&
  {Andersson}}]{Leitet2011}
{Leitet}, E., {Bergvall}, N., {Piskunov}, N., \& {Andersson}, B.-G. 2011, \aap,
  532, A107

\bibitem[{{Leitherer} {et~al.}(2016){Leitherer}, {Hernandez}, {Lee}, \&
  {Oey}}]{Leitherer2016}
{Leitherer}, C., {Hernandez}, S., {Lee}, J.~C., \& {Oey}, M.~S. 2016, \apj,
  823, 64

\bibitem[{{Leitherer} {et~al.}(1999){Leitherer}, {Schaerer}, {Goldader},
  {Delgado}, {Robert}, {Kune}, {de Mello}, {Devost}, \&
  {Heckman}}]{Leitherer1999}
{Leitherer}, C., {Schaerer}, D., {Goldader}, J.~D., {et~al.} 1999, \apjs, 123,
  3

\bibitem[{{Leja} {et~al.}(2017){Leja}, {Johnson}, {Conroy}, {van Dokkum}, \&
  {Byler}}]{Leja2017}
{Leja}, J., {Johnson}, B.~D., {Conroy}, C., {van Dokkum}, P.~G., \& {Byler}, N.
  2017, \apj, 837, 170

\bibitem[{{Luo} {et~al.}(2017){Luo}, {Brandt}, {Xue}, {Lehmer}, {Alexander},
  {Bauer}, {Vito}, {Yang}, {Basu-Zych}, {Comastri}, {Gilli}, {Gu},
  {Hornschemeier}, {Koekemoer}, {Liu}, {Mainieri}, {Paolillo}, {Ranalli},
  {Rosati}, {Schneider}, {Shemmer}, {Smail}, {Sun}, {Tozzi}, {Vignali}, \&
  {Wang}}]{Luo2017}
{Luo}, B., {Brandt}, W.~N., {Xue}, Y.~Q., {et~al.} 2017, \apjs, 228, 2

\bibitem[{{Ma} {et~al.}(2015){Ma}, {Kasen}, {Hopkins}, {Faucher-Gigu{\`e}re},
  {Quataert}, {Kere{\v{s}}}, \& {Murray}}]{Ma2015}
{Ma}, X., {Kasen}, D., {Hopkins}, P.~F., {et~al.} 2015, \mnras, 453, 960

\bibitem[{{Madau}(1995)}]{Madau1995}
{Madau}, P. 1995, \apj, 441, 18

\bibitem[{{Malkan} {et~al.}(2003){Malkan}, {Webb}, \& {Konopacky}}]{Malkan2003}
{Malkan}, M., {Webb}, W., \& {Konopacky}, Q. 2003, \apj, 598, 878

\bibitem[{{Marchi} {et~al.}(2018){Marchi}, {Pentericci}, {Guaita}, {Schaerer},
  {Verhamme}, {Castellano}, {Ribeiro}, {Garilli}, {Le F{\`e}vre}, {Amorin},
  {Bardelli}, {Cassata}, {Durkalec}, {Grazian}, {Hathi}, {Lemaux}, {Maccagni},
  {Vanzella}, \& {Zucca}}]{Marchi2018}
{Marchi}, F., {Pentericci}, L., {Guaita}, L., {et~al.} 2018, \aap, 614, A11

\bibitem[{{Mawatari} {et~al.}(2017){Mawatari}, {Inoue}, {Yamada}, {Hayashino},
  {Otsuka}, {Matsuda}, {Umehata}, {Ouchi}, \& {Mukae}}]{Mawatari2017}
{Mawatari}, K., {Inoue}, A.~K., {Yamada}, T., {et~al.} 2017, \mnras, 467, 3951

\bibitem[{{McKinney} {et~al.}(2019){McKinney}, {Jaskot}, {Oey}, {Yun}, {Dowd},
  \& {Lowenthal}}]{McKinney2019}
{McKinney}, J.~H., {Jaskot}, A.~E., {Oey}, M.~S., {et~al.} 2019, \apj, 874, 52

\bibitem[{{McLure} {et~al.}(2018){McLure}, {Pentericci}, {Cimatti}, {Dunlop},
  {Elbaz}, {Fontana}, {Nandra}, {Amorin}, {Bolzonella}, {Bongiorno}, {Carnall},
  {Castellano}, {Cirasuolo}, {Cucciati}, {Cullen}, {De Barros}, {Finkelstein},
  {Fontanot}, {Franzetti}, {Fumana}, {Gargiulo}, {Garilli}, {Guaita},
  {Hartley}, {Iovino}, {Jarvis}, {Juneau}, {Karman}, {Maccagni}, {Marchi},
  {M{\'a}rmol-Queralt{\'o}}, {Pompei}, {Pozzetti}, {Scodeggio}, {Sommariva},
  {Talia}, {Almaini}, {Balestra}, {Bardelli}, {Bell}, {Bourne}, {Bowler},
  {Brusa}, {Buitrago}, {Caputi}, {Cassata}, {Charlot}, {Citro}, {Cresci},
  {Cristiani}, {Curtis-Lake}, {Dickinson}, {Fazio}, {Ferguson}, {Fiore},
  {Franco}, {Fynbo}, {Galametz}, {Georgakakis}, {Giavalisco}, {Grazian},
  {Hathi}, {Jung}, {Kim}, {Koekemoer}, {Khusanova}, {F{\`e}vre}, {Lotz},
  {Mannucci}, {Maltby}, {Matsuoka}, {McLeod}, {Mendez-Hernandez},
  {Mendez-Abreu}, {Mignoli}, {Moresco}, {Mortlock}, {Nonino}, {Pannella},
  {Papovich}, {Popesso}, {Rosario}, {Salvato}, {Santini}, {Schaerer},
  {Schreiber}, {Stark}, {Tasca}, {Thomas}, {Treu}, {Vanzella}, {Wild},
  {Williams}, {Zamorani}, \& {Zucca}}]{McLure2018}
{McLure}, R.~J., {Pentericci}, L., {Cimatti}, A., {et~al.} 2018, \mnras,
  doi:10.1093/mnras/sty1213

\bibitem[{{Naidu} {et~al.}(2018){Naidu}, {Forrest}, {Oesch}, {Tran}, \&
  {Holden}}]{Naidu2018}
{Naidu}, R.~P., {Forrest}, B., {Oesch}, P.~A., {Tran}, K.-V.~H., \& {Holden},
  B.~P. 2018, \mnras, 478, 791

\bibitem[{{Nakajima} {et~al.}(2019){Nakajima}, {Ellis}, {Robertson}, {Tang}, \&
  {Stark}}]{Nakajima2019}
{Nakajima}, K., {Ellis}, R.~S., {Robertson}, B.~E., {Tang}, M., \& {Stark},
  D.~P. 2019, arXiv e-prints, arXiv:1909.07396

\bibitem[{{Nakajima} \& {Ouchi}(2014)}]{Nakajima2014}
{Nakajima}, K., \& {Ouchi}, M. 2014, \mnras, 442, 900

\bibitem[{{Nestor} {et~al.}(2013){Nestor}, {Shapley}, {Kornei}, {Steidel}, \&
  {Siana}}]{Nestor2013}
{Nestor}, D.~B., {Shapley}, A.~E., {Kornei}, K.~A., {Steidel}, C.~C., \&
  {Siana}, B. 2013, \apj, 765, 47

\bibitem[{{Nonino} {et~al.}(2009){Nonino}, {Dickinson}, {Rosati}, {Grazian},
  {Reddy}, {Cristiani}, {Giavalisco}, {Kuntschner}, {Vanzella}, {Daddi},
  {Fosbury}, \& {Cesarsky}}]{Nonino2009}
{Nonino}, M., {Dickinson}, M., {Rosati}, P., {et~al.} 2009, \apjs, 183, 244

\bibitem[{{Oh}(2001)}]{Oh2001}
{Oh}, S.~P. 2001, \apj, 553, 499

\bibitem[{{{\"O}stlin} {et~al.}(2015){{\"O}stlin}, {Marquart}, {Cumming},
  {Fathi}, {Bergvall}, {Adamo}, {Amram}, \& {Hayes}}]{Ostlin2015}
{{\"O}stlin}, G., {Marquart}, T., {Cumming}, R.~J., {et~al.} 2015, \aap, 583,
  A55

\bibitem[{{Paardekooper} {et~al.}(2015){Paardekooper}, {Khochfar}, \& {Dalla
  Vecchia}}]{Paardekooper2015}
{Paardekooper}, J.-P., {Khochfar}, S., \& {Dalla Vecchia}, C. 2015, \mnras,
  451, 2544

\bibitem[{{Peng} {et~al.}(2010){Peng}, {Ho}, {Impey}, \& {Rix}}]{Peng2010}
{Peng}, C.~Y., {Ho}, L.~C., {Impey}, C.~D., \& {Rix}, H.-W. 2010, \aj, 139,
  2097

\bibitem[{{Pentericci} {et~al.}(2018){Pentericci}, {McLure}, {Garilli},
  {Cucciati}, {Franzetti}, {Iovino}, {Amorin}, {Bolzonella}, {Bongiorno}, \&
  {Carnall}}]{Pentericci2018}
{Pentericci}, L., {McLure}, R.~J., {Garilli}, B., {et~al.} 2018, \aap, 616,
  A174

\bibitem[{{Planck Collaboration} {et~al.}(2016){Planck Collaboration}, {Adam},
  {Aghanim}, {Ashdown}, {Aumont}, {Baccigalupi}, {Ballardini}, {Banday},
  {Barreiro}, {Bartolo}, {Basak}, {Battye}, {Benabed}, {Bernard}, {Bersanelli},
  {Bielewicz}, {Bock}, {Bonaldi}, {Bonavera}, {Bond}, {Borrill}, {Bouchet},
  {Boulanger}, {Bucher}, {Burigana}, {Calabrese}, {Cardoso}, {Carron},
  {Chiang}, {Colombo}, {Combet}, {Comis}, {Couchot}, {Coulais}, {Crill},
  {Curto}, {Cuttaia}, {Davis}, {de Bernardis}, {de Rosa}, {de Zotti},
  {Delabrouille}, {Di Valentino}, {Dickinson}, {Diego}, {Dor{\'e}}, {Douspis},
  {Ducout}, {Dupac}, {Elsner}, {En{\ss}lin}, {Eriksen}, {Falgarone}, {Fantaye},
  {Finelli}, {Forastieri}, {Frailis}, {Fraisse}, {Franceschi}, {Frolov},
  {Galeotta}, {Galli}, {Ganga}, {G{\'e}nova-Santos}, {Gerbino}, {Ghosh},
  {Gonz{\'a}lez-Nuevo}, {G{\'o}rski}, {Gruppuso}, {Gudmundsson}, {Hansen},
  {Helou}, {Henrot-Versill{\'e}}, {Herranz}, {Hivon}, {Huang}, {Ili{\'c}},
  {Jaffe}, {Jones}, {Keih{\"a}nen}, {Keskitalo}, {Kisner}, {Knox},
  {Krachmalnicoff}, {Kunz}, {Kurki-Suonio}, {Lagache}, {L{\"a}hteenm{\"a}ki},
  {Lamarre}, {Langer}, {Lasenby}, {Lattanzi}, {Lawrence}, {Le Jeune},
  {Levrier}, {Lewis}, {Liguori}, {Lilje}, {L{\'o}pez-Caniego}, {Ma},
  {Mac{\'{\i}}as-P{\'e}rez}, {Maggio}, {Mangilli}, {Maris}, {Martin},
  {Mart{\'{\i}}nez-Gonz{\'a}lez}, {Matarrese}, {Mauri}, {McEwen}, {Meinhold},
  {Melchiorri}, {Mennella}, {Migliaccio}, {Miville-Desch{\^e}nes}, {Molinari},
  {Moneti}, {Montier}, {Morgante}, {Moss}, {Naselsky}, {Natoli}, {Oxborrow},
  {Pagano}, {Paoletti}, {Partridge}, {Patanchon}, {Patrizii}, {Perdereau},
  {Perotto}, {Pettorino}, {Piacentini}, {Plaszczynski}, {Polastri}, {Polenta},
  {Puget}, {Rachen}, {Racine}, {Reinecke}, {Remazeilles}, {Renzi}, {Rocha},
  {Rossetti}, {Roudier}, {Rubi{\~n}o-Mart{\'{\i}}n}, {Ruiz-Granados},
  {Salvati}, {Sandri}, {Savelainen}, {Scott}, {Sirri}, {Sunyaev}, {Suur-Uski},
  {Tauber}, {Tenti}, {Toffolatti}, {Tomasi}, {Tristram}, {Trombetti},
  {Valiviita}, {Van Tent}, {Vielva}, {Villa}, {Vittorio}, {Wandelt}, {Wehus},
  {White}, {Zacchei}, \& {Zonca}}]{Planck2016}
{Planck Collaboration}, {Adam}, R., {Aghanim}, N., {et~al.} 2016, \aap, 596,
  A108

\bibitem[{{Plat} {et~al.}(2019){Plat}, {Charlot}, {Bruzual}, {Feltre},
  {Vidal-Garc{\'\i}a}, {Morisset}, {Chevallard}, \& {Todt}}]{Plat2019}
{Plat}, A., {Charlot}, S., {Bruzual}, G., {et~al.} 2019, \mnras, 490, 978

\bibitem[{{Popesso} {et~al.}(2009){Popesso}, {Dickinson}, {Nonino}, {Vanzella},
  {Daddi}, {Fosbury}, {Kuntschner}, {Mainieri}, {Cristiani}, {Cesarsky},
  {Giavalisco}, {Renzini}, \& {GOODS Team}}]{Popesso2009}
{Popesso}, P., {Dickinson}, M., {Nonino}, M., {et~al.} 2009, \aap, 494, 443

\bibitem[{{Rafelski} {et~al.}(2015){Rafelski}, {Teplitz}, {Gardner}, {Coe},
  {Bond}, {Koekemoer}, {Grogin}, {Kurczynski}, {McGrath}, {Bourque}, {Atek},
  {Brown}, {Colbert}, {Codoreanu}, {Ferguson}, {Finkelstein}, {Gawiser},
  {Giavalisco}, {Gronwall}, {Hanish}, {Lee}, {Mehta}, {de Mello},
  {Ravindranath}, {Ryan}, {Scarlata}, {Siana}, {Soto}, \&
  {Voyer}}]{Rafelski2015}
{Rafelski}, M., {Teplitz}, H.~I., {Gardner}, J.~P., {et~al.} 2015, \aj, 150, 31

\bibitem[{{Retzlaff} {et~al.}(2010){Retzlaff}, {Rosati}, {Dickinson},
  {Vandame}, {Rit{\'e}}, {Nonino}, {Cesarsky}, \& {GOODS Team}}]{Retzlaff2010}
{Retzlaff}, J., {Rosati}, P., {Dickinson}, M., {et~al.} 2010, \aap, 511, A50

\bibitem[{{Rivera-Thorsen} {et~al.}(2017){Rivera-Thorsen}, {Dahle}, {Gronke},
  {Bayliss}, {Rigby}, {Simcoe}, {Bordoloi}, {Turner}, \&
  {Furesz}}]{RiveraThorsen2017}
{Rivera-Thorsen}, T.~E., {Dahle}, H., {Gronke}, M., {et~al.} 2017, \aap, 608,
  L4

\bibitem[{{Rivera-Thorsen} {et~al.}(2019){Rivera-Thorsen}, {Dahle}, {Chisholm},
  {Florian}, {Gronke}, {Rigby}, {Gladders}, {Mahler}, {Sharon}, \&
  {Bayliss}}]{RiveraThorsen2019}
{Rivera-Thorsen}, T.~E., {Dahle}, H., {Chisholm}, J., {et~al.} 2019, Science,
  366, 738

\bibitem[{{Rudie} {et~al.}(2013){Rudie}, {Steidel}, {Shapley}, \&
  {Pettini}}]{Rudie2013}
{Rudie}, G.~C., {Steidel}, C.~C., {Shapley}, A.~E., \& {Pettini}, M. 2013,
  \apj, 769, 146

\bibitem[{{Rutkowski} {et~al.}(2017){Rutkowski}, {Scarlata}, {Henry}, {Hayes},
  {Mehta}, {Hathi}, {Cohen}, {Windhorst}, {Koekemoer}, {Teplitz}, {Haardt}, \&
  {Siana}}]{Rutkowski2017}
{Rutkowski}, M.~J., {Scarlata}, C., {Henry}, A., {et~al.} 2017, \apjl, 841, L27

\bibitem[{{Schaerer}(2003)}]{Schaerer2003}
{Schaerer}, D. 2003, \aap, 397, 527

\bibitem[{{Schaerer} {et~al.}(2018){Schaerer}, {Izotov}, {Nakajima}, {Worseck},
  {Chisholm}, {Verhamme}, {Thuan}, \& {de Barros}}]{Schaerer2018}
{Schaerer}, D., {Izotov}, Y.~I., {Nakajima}, K., {et~al.} 2018, \aap, 616, L14

\bibitem[{{Schaerer} {et~al.}(2016){Schaerer}, {Izotov}, {Verhamme},
  {Orlitov{\'a}}, {Thuan}, {Worseck}, \& {Guseva}}]{Schaerer2016}
{Schaerer}, D., {Izotov}, Y.~I., {Verhamme}, A., {et~al.} 2016, \aap, 591, L8

\bibitem[{{Shapley} {et~al.}(2006){Shapley}, {Steidel}, {Pettini},
  {Adelberger}, \& {Erb}}]{Shapley2006}
{Shapley}, A.~E., {Steidel}, C.~C., {Pettini}, M., {Adelberger}, K.~L., \&
  {Erb}, D.~K. 2006, \apj, 651, 688

\bibitem[{{Shapley} {et~al.}(2016){Shapley}, {Steidel}, {Strom},
  {Bogosavljevi{\'c}}, {Reddy}, {Siana}, {Mostardi}, \& {Rudie}}]{Shapley2016}
{Shapley}, A.~E., {Steidel}, C.~C., {Strom}, A.~L., {et~al.} 2016, \apjl, 826,
  L24

\bibitem[{{Siana} {et~al.}(2007){Siana}, {Teplitz}, {Colbert}, {Ferguson},
  {Dickinson}, {Brown}, {Conselice}, {de Mello}, {Gardner}, \&
  {Giavalisco}}]{Siana2007}
{Siana}, B., {Teplitz}, H.~I., {Colbert}, J., {et~al.} 2007, \apj, 668, 62

\bibitem[{{Siana} {et~al.}(2010){Siana}, {Teplitz}, {Ferguson}, {Brown},
  {Giavalisco}, {Dickinson}, {Chary}, {de Mello}, {Conselice}, {Bridge},
  {Gardner}, {Colbert}, \& {Scarlata}}]{Siana2010}
{Siana}, B., {Teplitz}, H.~I., {Ferguson}, H.~C., {et~al.} 2010, \apj, 723, 241

\bibitem[{{Siana} {et~al.}(2015){Siana}, {Shapley}, {Kulas}, {Nestor},
  {Steidel}, {Teplitz}, {Alavi}, {Brown}, {Conselice}, {Ferguson}, {Dickinson},
  {Giavalisco}, {Colbert}, {Bridge}, {Gardner}, \& {de Mello}}]{Siana2015}
{Siana}, B., {Shapley}, A.~E., {Kulas}, K.~R., {et~al.} 2015, \apj, 804, 17

\bibitem[{{Steidel} {et~al.}(2018){Steidel}, {Bogosavljevi{\'c}}, {Shapley},
  {Reddy}, {Rudie}, {Pettini}, {Trainor}, \& {Strom}}]{Steidel2018}
{Steidel}, C.~C., {Bogosavljevi{\'c}}, M., {Shapley}, A.~E., {et~al.} 2018,
  \apj, 869, 123

\bibitem[{{Steidel} {et~al.}(2001){Steidel}, {Pettini}, \&
  {Adelberger}}]{Steidel2001}
{Steidel}, C.~C., {Pettini}, M., \& {Adelberger}, K.~L. 2001, \apj, 546, 665

\bibitem[{{Tanvir} {et~al.}(2019){Tanvir}, {Fynbo}, {de Ugarte Postigo},
  {Japelj}, {Wiersema}, {Malesani}, {Perley}, {Levan}, {Selsing}, \&
  {Cenko}}]{Tanvir2019}
{Tanvir}, N.~R., {Fynbo}, J.~P.~U., {de Ugarte Postigo}, A., {et~al.} 2019,
  \mnras, 483, 5380

\bibitem[{{Teplitz} {et~al.}(2013){Teplitz}, {Rafelski}, {Kurczynski}, {Bond},
  {Grogin}, {Koekemoer}, {Atek}, {Brown}, {Coe}, {Colbert}, {Ferguson},
  {Finkelstein}, {Gardner}, {Gawiser}, {Giavalisco}, {Gronwall}, {Hanish},
  {Lee}, {de Mello}, {Ravindranath}, {Ryan}, {Siana}, {Scarlata}, {Soto},
  {Voyer}, \& {Wolfe}}]{Teplitz2013}
{Teplitz}, H.~I., {Rafelski}, M., {Kurczynski}, P., {et~al.} 2013, \aj, 146,
  159

\bibitem[{{Trebitsch} {et~al.}(2017){Trebitsch}, {Blaizot}, {Rosdahl},
  {Devriendt}, \& {Slyz}}]{Trebitsch2017}
{Trebitsch}, M., {Blaizot}, J., {Rosdahl}, J., {Devriendt}, J., \& {Slyz}, A.
  2017, \mnras, 470, 224

\bibitem[{{Trebitsch} {et~al.}(2018){Trebitsch}, {Volonteri}, {Dubois}, \&
  {Madau}}]{Trebitsch2018}
{Trebitsch}, M., {Volonteri}, M., {Dubois}, Y., \& {Madau}, P. 2018, \mnras,
  478, 5607

\bibitem[{{van der Wel} {et~al.}(2012){van der Wel}, {Bell}, {H{\"a}ussler},
  {McGrath}, {Chang}, {Guo}, {McIntosh}, {Rix}, {Barden}, {Cheung}, {Faber},
  {Ferguson}, {Galametz}, {Grogin}, {Hartley}, {Kartaltepe}, {Kocevski},
  {Koekemoer}, {Lotz}, {Mozena}, {Peth}, \& {Peng}}]{vanderWel2012}
{van der Wel}, A., {Bell}, E.~F., {H{\"a}ussler}, B., {et~al.} 2012, \apjs,
  203, 24

\bibitem[{{Vanzella} {et~al.}(2010{\natexlab{a}}){Vanzella}, {Siana},
  {Cristiani}, \& {Nonino}}]{Vanzella2010MNRAS}
{Vanzella}, E., {Siana}, B., {Cristiani}, S., \& {Nonino}, M.
  2010{\natexlab{a}}, \mnras, 404, 1672

\bibitem[{{Vanzella} {et~al.}(2005){Vanzella}, {Cristiani}, {Dickinson},
  {Kuntschner}, {Moustakas}, {Nonino}, {Rosati}, {Stern}, {Cesarsky}, {Ettori},
  {Ferguson}, {Fosbury}, {Giavalisco}, {Haase}, {Renzini}, {Rettura}, {Serra},
  \& {GOODS Team}}]{Vanzella2005}
{Vanzella}, E., {Cristiani}, S., {Dickinson}, M., {et~al.} 2005, \aap, 434, 53

\bibitem[{{Vanzella} {et~al.}(2006){Vanzella}, {Cristiani}, {Dickinson},
  {Kuntschner}, {Nonino}, {Rettura}, {Rosati}, {Vernet}, {Cesarsky},
  {Ferguson}, {Fosbury}, {Giavalisco}, {Grazian}, {Haase}, {Moustakas},
  {Popesso}, {Renzini}, {Stern}, \& {GOODS Team}}]{Vanzella2006}
---. 2006, \aap, 454, 423

\bibitem[{{Vanzella} {et~al.}(2008){Vanzella}, {Cristiani}, {Dickinson},
  {Giavalisco}, {Kuntschner}, {Haase}, {Nonino}, {Rosati}, {Cesarsky},
  {Ferguson}, {Fosbury}, {Grazian}, {Moustakas}, {Rettura}, {Popesso},
  {Renzini}, {Stern}, \& {GOODS Team}}]{Vanzella2008}
---. 2008, \aap, 478, 83

\bibitem[{{Vanzella} {et~al.}(2010{\natexlab{b}}){Vanzella}, {Giavalisco},
  {Inoue}, {Nonino}, {Fontanot}, {Cristiani}, {Grazian}, {Dickinson}, {Stern},
  {Tozzi}, {Giallongo}, {Ferguson}, {Spinrad}, {Boutsia}, {Fontana}, {Rosati},
  \& {Pentericci}}]{Vanzella2010}
{Vanzella}, E., {Giavalisco}, M., {Inoue}, A.~K., {et~al.} 2010{\natexlab{b}},
  \apj, 725, 1011

\bibitem[{{Vanzella} {et~al.}(2012){Vanzella}, {Guo}, {Giavalisco}, {Grazian},
  {Castellano}, {Cristiani}, {Dickinson}, {Fontana}, {Nonino}, {Giallongo},
  {Pentericci}, {Galametz}, {Faber}, {Ferguson}, {Grogin}, {Koekemoer},
  {Newman}, \& {Siana}}]{Vanzella2012}
{Vanzella}, E., {Guo}, Y., {Giavalisco}, M., {et~al.} 2012, \apj, 751, 70

\bibitem[{{Vanzella} {et~al.}(2014){Vanzella}, {Fontana}, {Pentericci},
  {Castellano}, {Grazian}, {Giavalisco}, {Nonino}, {Cristiani}, {Zamorani}, \&
  {Vignali}}]{Vanzella2014}
{Vanzella}, E., {Fontana}, A., {Pentericci}, L., {et~al.} 2014, \aap, 569, A78

\bibitem[{{Vanzella} {et~al.}(2015){Vanzella}, {de Barros}, {Castellano},
  {Grazian}, {Inoue}, {Schaerer}, {Guaita}, {Zamorani}, {Giavalisco}, {Siana},
  {Pentericci}, {Giallongo}, {Fontana}, \& {Vignali}}]{Vanzella2015}
{Vanzella}, E., {de Barros}, S., {Castellano}, M., {et~al.} 2015, \aap, 576,
  A116

\bibitem[{{Vanzella} {et~al.}(2016){Vanzella}, {de Barros}, {Vasei}, {Alavi},
  {Giavalisco}, {Siana}, {Grazian}, {Hasinger}, {Suh}, {Cappelluti}, {Vito},
  {Amorin}, {Balestra}, {Brusa}, {Calura}, {Castellano}, {Comastri}, {Fontana},
  {Gilli}, {Mignoli}, {Pentericci}, {Vignali}, \& {Zamorani}}]{Vanzella2016}
{Vanzella}, E., {de Barros}, S., {Vasei}, K., {et~al.} 2016, \apj, 825, 41

\bibitem[{{Vanzella} {et~al.}(2018){Vanzella}, {Nonino}, {Cupani},
  {Castellano}, {Sani}, {Mignoli}, {Calura}, {Meneghetti}, {Gilli}, {Comastri},
  {Mercurio}, {Caminha}, {Caputi}, {Rosati}, {Grillo}, {Cristiani}, {Balestra},
  {Fontana}, \& {Giavalisco}}]{Vanzella2018}
{Vanzella}, E., {Nonino}, M., {Cupani}, G., {et~al.} 2018, \mnras, 476, L15

\bibitem[{{Vanzella} {et~al.}(2020){Vanzella}, {Caminha}, {Calura}, {Cupani},
  {Meneghetti}, {Castellano}, {Rosati}, {Mercurio}, {Sani}, {Grillo}, {Gilli},
  {Mignoli}, {Comastri}, {Nonino}, {Cristiani}, {Giavalisco}, \&
  {Caputi}}]{Vanzella2019}
{Vanzella}, E., {Caminha}, G.~B., {Calura}, F., {et~al.} 2020, \mnras, 491,
  1093

\bibitem[{{Verhamme} {et~al.}(2015){Verhamme}, {Orlitov{\'a}}, {Schaerer}, \&
  {Hayes}}]{Verhamme2015}
{Verhamme}, A., {Orlitov{\'a}}, I., {Schaerer}, D., \& {Hayes}, M. 2015, \aap,
  578, A7

\bibitem[{{Verhamme} {et~al.}(2017){Verhamme}, {Orlitov{\'a}}, {Schaerer},
  {Izotov}, {Worseck}, {Thuan}, \& {Guseva}}]{Verhamme2017}
{Verhamme}, A., {Orlitov{\'a}}, I., {Schaerer}, D., {et~al.} 2017, \aap, 597,
  A13

\bibitem[{{Wise} {et~al.}(2014){Wise}, {Demchenko}, {Halicek}, {Norman},
  {Turk}, {Abel}, \& {Smith}}]{Wise2014}
{Wise}, J.~H., {Demchenko}, V.~G., {Halicek}, M.~T., {et~al.} 2014, \mnras,
  442, 2560

\bibitem[{{Worseck} {et~al.}(2014){Worseck}, {Prochaska}, {O'Meara}, {Becker},
  {Ellison}, {Lopez}, {Meiksin}, {M{\'e}nard}, {Murphy}, \&
  {Fumagalli}}]{Worseck2014}
{Worseck}, G., {Prochaska}, J.~X., {O'Meara}, J.~M., {et~al.} 2014, \mnras,
  445, 1745

\bibitem[{{Yue} {et~al.}(2018){Yue}, {Castellano}, {Ferrara}, {Fontana},
  {Merlin}, {Amor{\'\i}n}, {Grazian}, {M{\'a}rmol-Queralto}, {Micha{\l}owski},
  {Mortlock}, {Paris}, {Parsa}, {Pilo}, {Santini}, \& {Di
  Criscienzo}}]{Yue2018}
{Yue}, B., {Castellano}, M., {Ferrara}, A., {et~al.} 2018, \apj, 868, 115

\end{thebibliography}

\end{document}